%% file: final.tex
\documentclass[aip,pof,reprint]{revtex4-2}
\usepackage[utf8]{inputenc}
\usepackage{graphicx}% Include figure files
\usepackage{amsmath}
\usepackage{dcolumn}% Align table columns on decimal point
\usepackage{bm}% bold math
\usepackage{hyperref}% add hypertext capabilities
\hypersetup{
    colorlinks=true,
    linkcolor=red,
    filecolor=magenta,      
    urlcolor=blue,}
\usepackage{url}
\usepackage{xcolor}
\usepackage{titlesec}
\usepackage[labelformat=simple]{subcaption}

\usepackage{multirow}
\usepackage{newtxtext,newtxmath}
\usepackage{etoolbox}
\usepackage{float}
\usepackage[normalem]{ulem}
\DeclareUnicodeCharacter{03B2}{\ensuremath{\beta}}

\usepackage{geometry}
\usepackage{subdepth}
\usepackage{color}
\input{macros}

\titlespacing*{\section}
{0pt}{4ex}{2ex}
\titlespacing*{\subsection}
{0pt}{4ex}{2ex}
\titlespacing*{\subsubsection}
{0pt}{4ex}{2ex}
\begin{document}

\title{X-CAL: Explaining latent causality in physical space for fluid mechanics}

\author{Marcial Sanchis-Agudo$^{\text{1}}$, Andrés Cremades$^{\text{2,1}}$, Alvaro Martinez-Sanchez$^{\text{3}}$, Adrian Lozano-Duran$^{\text{4,3}}$ and Ricardo Vinuesa$^{\text{5,1}}$}
\affiliation{1: FLOW, Engineering Mechanics, KTH Royal Institute of Technology, SE-100 44 Stockholm, Sweden.\\
2: Instituto Universitario de Matemática Pura y Aplicada, Universitat Politècnica de València, Valencia, 46022, Spain\\
3: Department of Aeronautics and Astronautics, Massachusetts Institute of Technology, Cambridge, MA, USA.\\
4: Graduate Aerospace
Laboratories, California Institute of Technology, Pasadena, CA 91125, USA.\\
5: Department of Aerospace Engineering, University of Michigan, Ann Arbor, MI 48109, USA.
}
\begin{abstract}
\noindent We present X-CAL, a pipeline that combines a $\beta$-variational autoencoder ($\beta$-VAE) with the synergistic-unique-redundant decomposition (SURD)~\cite{surd} approach for causality analysis to interpret low-dimensional latent representations of turbulent fluid flows. Combining $\beta$-VAE compression with SURD and SHAP (SHapley Additive exPlanations) yields interpretable latent representations and structure-level attributions in physical space, offering a general methodology for causal analysis of high-dimensional flows. Using direct numerical simulation (DNS) data of the flow around a wall-mounted square cylinder at $Re_h=2000$, we (i) learn a compact latent space with near-orthogonal variables, (ii) quantify directed information flows among these variables via the SURD approach, and (iii) map latent-space causality back to physical space through gradient-SHAP fields . By means of percolation analysis of the SHAP fields, we extract the coherent, time-resolved structures that most influence each latent variable. The analysis connects coherent structures with latent variables which are in turn associated with wake-boundary-layer interactions. This method enables translating the insight obtained through causal analysis in the latent space into interpretable phenomena in physical space.
\end{abstract}

\maketitle

\section{Introduction}

Turbulence is characterized by its high dimensionality and complexity, governed by the nonlinear Navier–Stokes equations. This inherent complexity necessitates the development of data-driven, reduced-order modeling techniques capable of efficiently capturing the dominant flow dynamics. A key feature of turbulent flow is its organization by coherent structures such as vortices and shear layers which dictate the fundamental mechanisms of energy transport and mixing. While the existence and energetic contribution of these structures are known, a central challenge is understanding their specific dynamic interactions and causal dependence over time. Identifying not only what structures exist, but how and why they influence the subsequent flow evolution, is crucial for achieving robust predictive capability and effective control. The inherent non-linearity and high-dimensionality of the system demand an analysis grounded in information theory to quantify the directed exchange of information among flow mechanisms.

\noindent To address the complexities of causal analysis in turbulence, the present study focuses on the canonical configuration of the flow around a wall-mounted square cylinder, using direct numerical simulation (DNS) data. This geometry isolates complex wake-wall interactions and features a rich set of mechanisms relevant to engineering applications, particularly in urban design $\cite{vinuesa2015direct}$.

\noindent The flow field features four critical vortical structures that govern transport and mixing in the near wake $\cite{Sakamoto1986ArchtypeVF,WangZhou2009FiniteLength}$: the tip (or roof) vortex, the base vortex (a streamwise vortex formed near the cylinder base), the spanwise (primary shedding) vortex, and the horseshoe vortex (which forms around the obstacle). Even in this simpler flow case, these arch-vortices have a key role in phenomena like pollutant transport $\cite{vinuesa2015direct}$. The ability to characterize the dynamic, causal interactions of these structures is paramount for developing control strategies that target specific mechanisms, thereby facilitating the physical interpretation of the system for future engineering applications $\cite{WangZhou2009FiniteLength}$.

\noindent Conventional state-of-the-art reduced-order modeling primarily relies on linear techniques such as Proper Orthogonal Decomposition (POD) and Dynamic Mode Decomposition (DMD) $\cite{POD_org,Schmid2010}$. While these tools reveal dominant energetic or kinematic structures, their linear formulation presents critical limitations when pursuing an explicit, physically interpretable causal view of nonlinear dynamics $\cite{Eiximeno_Sanchis-Agudo_Miro_Rodriguez_Vinuesa_Lehmkuhl_2025,Schmid2010}$.

\noindent The primary deficiency is that, since turbulence is fundamentally a nonlinear phenomenon, these methods can only encode nonlinear dependencies implicitly. This results in highly coupled temporal coefficients, which obscure the physical meaning and explicit causal links of individual modes. Consequently, while POD/DMD efficiently identify what dominant structures exist, they are mathematically incapable of quantifying the directed causality or explicit information exchange among these structures. A robust, physically meaningful ROM requires capturing the flow's dynamics in a framework that explicitly addresses causality and explainability.

\noindent The methodological foundation for applying information theory to turbulent flows is rooted in the dual legacy of Andrey Kolmogorov, a key figure in the study of complex systems $\cite{kolmogorov1965,kolmogorovLegacy}$.

\noindent Kolmogorov’s contributions to turbulence theory established the statistical laws governing the energy cascade, including the famous $-5/3$ power law $\cite{kolmogorov1965}$ and its corrections. Simultaneously, his work in information theory and chaos introduced concepts for the quantification and interpretation of chaotic systems. This led to the development of the Kolmogorov–Sinai (K-S) entropy, a measure quantifying the rate of information generation within a dynamical system $\cite{kolmogorov1965,millionshchikov1970}$. Theoretical relations connect the Lyapunov exponents with the K-A entropy and the K-S entropy $\cite{millionshchikov1970}$, thereby linking the rate of orbital divergence in phase space with the fundamental informational content of the system. This connection justifies the use of informational path measures, such as the informational exchange introduced by Wiener $\cite{wiener1956}$, as a robust metric for quantifying directed causality, which moves the analysis beyond simple correlation $\cite{millionshchikov1970}$.

\noindent We present X-CAL (Explainable causal analysis for latent representations), a novel pipeline that explicitly addresses the critical need for interpreting nonlinear flow dynamics through an explicitly causal lens. X-CAL achieves this by integrating three core concepts:

\begin{enumerate}
\item Nonlinear Compression ($\beta$-VAE): We learn a compact, nonlinear latent space using a $\beta$-Variational Autoencoder ($\beta$-VAE) $\cite{SoleraRico2024,EIVAZI2022117038}$. The $\beta$-regularization term encourages the latent variables to be near-orthogonal and disentangled, minimizing redundancy and maximizing the interpretability of the latent manifold.\\
\item Explicit Causality Analysis (SURD): We rigorously quantify directed information flow among the disentangled latent variables using the synergistic–unique–redundant (SURD) decomposition of causality $\cite{surd}$. This method uses the informational path as the causality measure, decomposing mutual information into unique, redundant, and synergistic contributions, and can be extended to a state-dependent analysis $\cite{surd}$.\\
\item Physical Attribution (Gradient-SHAP and Percolation): We project the latent causality back to the physical space via Gradient-SHAP (SHapley Additive exPlanations) $\cite{Cremades2024,lundberg2017unified}$. This yields spatially explicit SHAP fields that quantify how each grid point contributes to the encoding of a specific latent variable. By applying percolation analysis to these fields, we extract the coherent, time-resolved structures that are most instrumental in driving the dynamics $\cite{Jimenez_2018}$, yielding structure-level attributions.
\end{enumerate}
\noindent The complete framework involves data compression, causal analysis, and explainability. Incorporating causality and explainability into machine learning (ML) models is necessary to increase trust in these architectures $\cite{vinuesa2025decodingcomplexitymachinelearning}$. This makes X-CAL an excellent tool for flow control and scientific discovery purposes in fluid mechanics and, more generally, in high-dimensional chaotic systems.

\noindent The remainder of this paper is structured as follows. Section~\ref{sec:framework} introduces the Framework conceptually and the corresponding tools; $\beta$-VAE architectures, SURD and the Gradient-SHAP method. \S~\ref{method} introduces in detail the methods leveraged on the X-CAL framework, by defining notation and operations. Following, in \S~\ref{val} we inspect the X-CAL methodology conceptually and numerically on controlled synthetic cases, specifically the 2D Torus and the Lorenz system, to ensure the $\beta$-VAE successfully preserves or learns the key causal signatures. \S~\ref{obstacle} presents the results from the application of X-CAL to the flow around the wall-mounted obstacle, analyzing latent causality through unique/redundant contributions and mapping these causal events to physical structures. Finally, Section~\ref{conclusions} summarizes the conclusions and discusses the implications of the identified causal mechanisms for the understanding and control of turbulent wakes. The application of X-CAL for other fluid mechanic analysis or chaotic systems discovery is also mentioned.

\section{Framework overview}
\label{sec:framework}
We will introduce the framework both conceptually and numerically by studying two simple cases to validate and further understand the proposed method.
As mentioned already, the objective of this investigation is to gain further insight into the physical mechanisms through causal artificial intelligence (AI).
% designing a $\vartheta-machine$, $\beta-VAE$, which is causally closed, given that the coarse-grained latent space inherits memoryless or stochastic dynamics from the fluid. Even more, such strong lump-able systems, systems which after coarse-graining are still dependable on transition probabilities, are information-ally closed, which allows for the observer to predict the macro states given information of the micro-world only, latent space.
Due to the causally closed nature of the system, we apply the synergistic, unique and redundant (SURD) decomposition~\cite{surd} identifying all causal relationships among latent variables. Furthermore, we interpret such causal relations in the physical space by identifying the most important physical structures for each latent variable applying SHAP~\cite{Cremades2024}, which concepts will be later introduced in \S~\ref{method}. The causal framework, summarized in Fig.~\ref{fig:frame}, is applied as follows:
\begin{figure}
    \centering
    \includegraphics[width=1\linewidth]{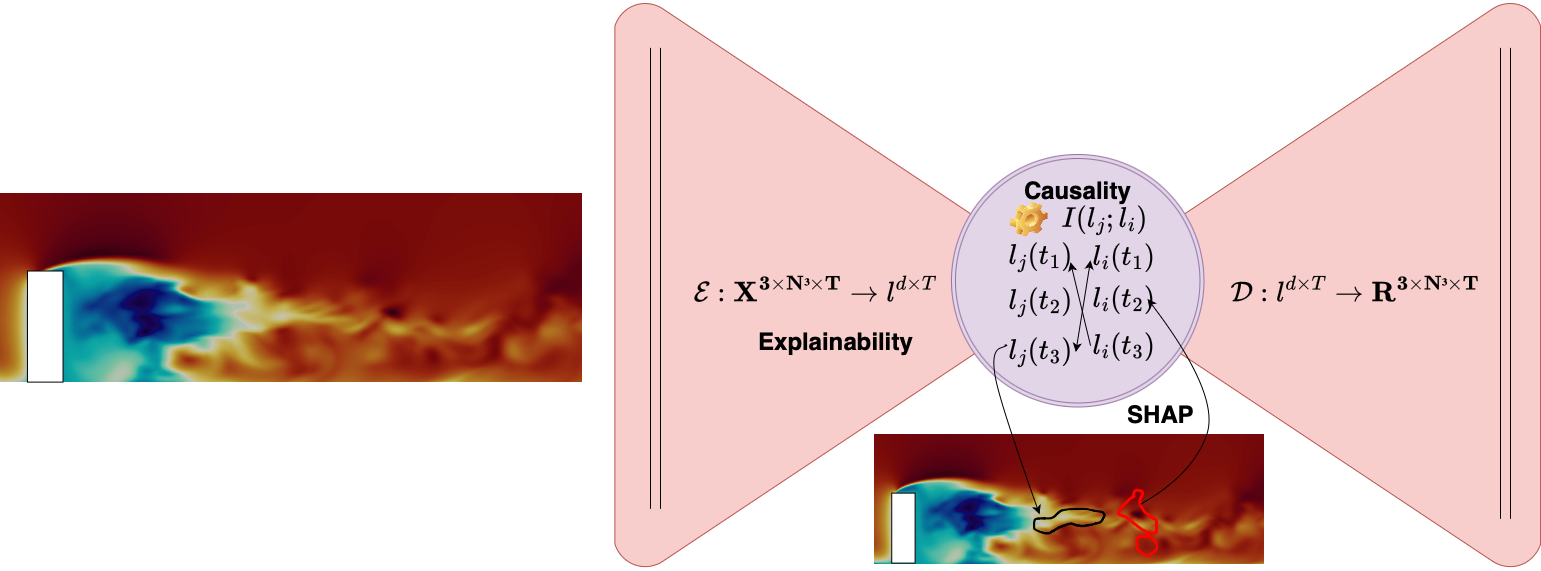}
        \caption{Causal artificial-intelligence (AI) framework visualization. (Left) Instantaneous flow field of the case under study. (Right) Schematic representation of the method, where 1) we encode the flow field into a latent space with a $\beta$-VAE; 2) we analyze causality among the latent variables, identifying key causal mechanisms; 3) we perform SHAP analysis to identify the most important flow structures associated with each of the latent variables. This enables formulating the causal relations identified in the latent space in terms of coherent structures in physical space. }
    \label{fig:frame}
\end{figure}

\begin{enumerate}
    \item Use the $\beta$-VAE in Fig.~\ref{fig:frame} to encode physical fields into latent variables (analogous to temporal coefficients in classical linear-decomposition methods).
    \item Perform causality analysis on the latent space, e.g. $l_j \rightarrow l_i$ (where $l$ denotes a latent variable).
    \item For each time step, apply SHAP to identify the points in physical domain contributing to encode latent variable $l_j$ and $l_i$.
\item Percolate and perform statistics on those physical structures associated with the causal relation $l_j \rightarrow l_i$.
\end{enumerate}

\section{Methodology}
\label{method}
\subsection{Numerical database}

The database was obtained via direct numerical simulation (DNS) of the flow around a wall-mounted square cylinder, as documented in Ref.~\cite{Yousif2023}. The DNS data set was produced using the open-source code Nek5000, which employs the spectral-element method to solve the incompressible Navier--Stokes equations.

% \begin{subequations} \begin{equation} \nabla \cdot \mathbf{u} = 0, \end{equation} 
% \begin{equation} \frac{\partial \mathbf{u}}{\partial t} + \mathbf{u} \cdot \nabla \mathbf{u} = - \nabla p + \nu \nabla^2 \mathbf{u}. \end{equation} \end{subequations}

\noindent The following variables will used along the rest of the study; \( \mathbf{u} = (u,v,w) \) is the velocity vector field, representing the instantaneous velocity components in the three directions: streamwise (\( x \)), wall-normal (\( y \)) and spanwise (\( z \)).

\noindent The simulation involves a square obstacle with a width-to-height ratio ($b/h$) of 0.25, and all dimensions are normalized by the obstacle height, $h$. The flow is simulated at a Reynolds number of 2000 based on this height, with a laminar inflow boundary layer. To accurately capture the flow dynamics, a spectral-element mesh with 21.8 million grid points is used. A subset of this mesh is extracted focusing on the domain near the obstacle with boundaries of $-1 \leq x/h \leq 5$, $0 \leq y/h \leq 2$, and $-1.5 \leq z/h \leq 1.5$. The data is then interpolated onto a uniform grid with resolution $(N_x, N_y, N_z) = (300, 100, 150)$. Temporal parameters are given in convective time units, defined as the ratio of the free-stream velocity $U_\infty$ to the obstacle height $h$, with a constant time step, $\Delta t_s = 0.005$, ensuring sufficient resolution to capture both the low- and high-frequency flow phenomena.

\noindent To develop an accurate reduced-order model, it is essential to efficiently capture the dominant flow dynamics while minimizing computational complexity. We focus on extracting statistically stationary fields by excluding transient periods, resulting in a dataset of 30,000 snapshots corresponding to 150 convective time units. This dataset is divided into training and testing sets in a 5:1 ratio, maintaining the temporal sequence to simulate a real-world prediction scenario. This approach ensures that the model can predict future flow states based on past observations, assuming that the turbulent-flow dynamics remain statistically consistent over time.

\noindent In this work we analyze a two-dimensional (2D) section at $z/h = 0$, which corresponds to the mid-span of the obstacle. This reduction to a 2D plane is intended to balance computational efficiency and model complexity, facilitating performance evaluation.

\subsection{Space-time Compression}
\label{space}

Turbulence is a paradigmatic high-dimensional complex system, with millions of degrees of freedom governed by the nonlinear Navier--Stokes equations. Since no closed-form analytical solution exists, progress relies on a synergy between theoretical models, numerical simulations, and data-driven approaches. A central challenge is to construct reduced representations that are both interpretable and capable of capturing causal dependencies among flow structures. Our approach is to encode the fluctuating velocity fields into a disentangled low-dimensional manifold, where the number of observables is drastically reduced while preserving the dominant mechanisms of the flow. In this latent space, causal relations can be analyzed more clearly.

\noindent To achieve this, we employ a $\beta$-variational autoencoder ($\beta$-VAE), which learns a probabilistic mapping between the input flow fields and a compressed latent representation. Let $X^t$ denote an input snapshot and $\ell_i^t$ a latent variable. The encoder learns an approximate posterior distribution $q_\phi(\ell_i^t \mid X^t),$
which is a variational approximation to the true but intractable posterior. The prior distribution over latents is taken as
$
p(\ell^t)\sim \mathcal{N}(0, I),
$
a standard multivariate Gaussian. The decoder then reconstructs the input via the conditional likelihood
$
p_\theta(\tilde{X}^t \mid \ell_i^t),
$
where $\tilde{X}^t$ is the reconstruction of $X^t$.

\noindent The $\beta$-VAE objective balances reconstruction fidelity and latent regularization. Specifically, it minimizes the variational free energy
\begin{equation}
\begin{split}
  \mathcal{F}_\theta
  &= \mathcal{L}_{\beta\text{-VAE}} \\[4pt]
  &= \underbrace{-\mathbb{E}_{q(\ell_i^t\mid X^t)}\bigl[\log p_\theta(\tilde{X}^t\mid \ell_i^t)\bigr]}_{\text{Reconstruction error}}
     \;+\; \beta
     \underbrace{\mathrm{D_{KL}}\bigl[q_\phi(\ell_i^t\mid X^t)\,\|\,p(\ell_i^t)\bigr]}_{\text{Prior regularization}} \\[6pt]
\end{split}
\label{eq:beta_vae_decomp}
\end{equation}

\noindent The loss function combines two main terms: a \textbf{reconstruction error} and a \textbf{regularization term}. The first term penalizes the error between the input data ($X^t$) and its reconstruction from the latent representation ($\ell^t$), ensuring the latent variables retain enough information to recover the original flow. The second term, scaled by a hyperparameter $\beta$, is a regularization that forces the approximate posterior distribution of the latents, $q_\phi(\ell \mid X^t)$, to be close to a predefined prior, $p(\ell)$, which is typically a standard Gaussian. This structure encourages the latent variables to form a disentangled, interpretable manifold, which is crucial for the physical interpretation of the latent space.

\subsection{Observational causality with SURD}
\label{SURD}

% Introduction
For quantifying causal interactions, we adopt the definition of causality proposed in Martínez-Sánchez et al.\cite{surd}, implemented through SURD. Consider the collection of $N$ input variables evolving in
time given by the vector $\bQ = [Q_1(t),Q_2(t),$
  $\dots,Q_N(t)]$. For example, $Q_i$ may represent the time
coefficients associated with each of the latent variables extracted from $\beta$-VAE. The components of $\bQ$ are the input variables and are treated
as random variables. Our objective is to quantify the causal influence of $\bQ$ on the future of a target variable $Q_j^+$, denoted by $Q_j^+ = Q_j(t+\Delta T)$, where $\Delta T>0$ is an arbitrary time increment.

% Definition of causality in SURD
In the SURD framework, this causality is quantified as the increase in information about the future output $Q_j^+$ that is gained by observing individual or groups of past inputs $\bQ$. The information content in $Q_j^+$ is measured using Shannon entropy~\cite{shannon1948}, denoted as $H(Q_j^+)$, which reflects the average level of unpredictability or expected surprise associated with the outcomes of the random variable $Q_j^+$. 

% Step 2
Next, the information in $H(Q_j^+)$ is decomposed into a
sum of information increments contributed by distinct types of
interactions from $\bQ$ namely, redundant, unique, and synergistic
components using the principle of forward-in-time propagation of information~\cite{surd}:
\begin{equation}
\label{eq:surd}
    H(Q_j^+) = \sum_{\bi \in \mathcal{C}} \Delta I ^ R _ {\bi\to j} + \sum_{i=1}^N \Delta I ^ U _ {i\to j} + \sum_{\bi \in \mathcal{C}} \Delta I ^ S _ {\bi\to j} + \Delta I_{{\rm{leak}}\to j},
\end{equation}
where the terms $\Delta I^R_{\bi
  \rightarrow j}$, $\Delta I^U_{i \rightarrow j}$, and $\Delta
I^S_{\bi \rightarrow j}$ denote redundant, unique, and synergistic
causalities, respectively, from $\bQ$ to $Q_j^+$, and $\Delta I_{{\rm{leak}}\to j}$ is the causality from unobserved variables, referred to as the causality leak. Unique causalities
are associated with individual components of $\bQ$, while redundant
and synergistic causalities emerge from interactions among groups of
variables. The set $\mathcal{C}$ includes all subsets of indices with
cardinality greater than one, i.e., $\mathcal{C} = \{ \bi \subseteq
\{1, \dots, N\} \mid |\bi| > 1 \}$. For instance, for $N=2$, Eq. \ref{eq:surd} reduces to $H(Q_j^+) = \Delta I ^ R _ {12\to j} + \Delta I ^ U _ {1\to j} + \Delta I ^ U _ {2\to j} + \Delta I ^ S _ {12\to j} + \Delta I_{{\rm{leak}}\to j}$.
The formal definitions of
causality can be found in Ref.\cite{surd}. Here, we offer
an interpretation of each term:
\begin{itemize}
    \item \textit{Redundant causality} from a subset $\bQ_{\bi} = \{
      Q_{i_1}, Q_{i_2}, \dots \} \subseteq \bQ$ to $Q_j^+$, denoted by
      $\Delta I^R_{\bi \rightarrow j}$, is the information about the
      output that is identically present in all variables within the
      group $\bQ_{\bi}$. Redundant causality arises when each variable
      in the group individually contains the same information about
      the target. 
      % Therefore, forecasting models for $Q_O^+$ can be
      % optimized by selecting the most convenient variable from the
      % redundant set and disregarding the rest.
    
    \item \textit{Unique causality} from an individual variable $Q_i$
      to $Q_j^+$, denoted by $\Delta I^U_{i \rightarrow j}$, is the
      information about the output that is available exclusively
      through $Q_i$ and cannot be recovered from any other single
      variable. Unique causality indicates that $Q_i$ provides
      critical information not found elsewhere in the set of
      individual variables. 
      % Therefore, forecasting models for $Q_O^+$
      % should always retain $Q_i$ as input, since its information
      % cannot be found in any other variable alone.
    
    \item \textit{Synergistic causality} from a subset $\bQ_{\bi} = \{
      Q_{i_1}, Q_{i_2}, \dots \} \subseteq \bQ$ to $Q_j^+$, denoted by
      $\Delta I^S_{\bi \rightarrow j}$, corresponds to the information
      that can only be accessed when all variables in the group are
      considered jointly. Synergy captures higher-order interactions,
      where the collective observation of variables reveals
      information that is absent when they are observed
      individually. 
      % Therefore, it is crucial for models to incorporate
      % all variables in $\bQ_{\bi}$ as inputs to ensure accurate
      % forecasts.
    \item \textit{Causality leak} represents the effect from unobserved variables
   that influence $Q_j^+$ but are not contained in $\bQ$.
   {This is the amount of information missing that would be
     required to unambiguously determine the future of $Q_j$ after
     considering all observable variables collectively}.
\end{itemize}

% Specific mutual information
To quantify the causal components in Eq. \ref{eq:surd}, SURD relies on the concept of specific mutual
information \cite{DeWeese1999} between a specific value $q_j^+$ of the target variable $Q_j^+$. This quantity can be
mathematically described as:
\begin{equation}
  \label{eq:specific_mutual_info}
   \Is(q_j^+;\bQ) = \sum_{\bq\in \bQ} \frac{p(q_j^+,\bq)}{p(q_j^+)}
   \log_2\left( \frac{p(q_j^+,\bq)}{p(q_j^+)p(\bq)} \right) \geq 0.
\end{equation}
where $p(q_j^+, \bq_\bi)$, $p(q_j^+)$, and $p(\bq_\bi)$ denote the joint and
marginal probability density functions of the output and input
variables, respectively, and $q_j^+$ and $\bq_\bi$ represent particular
values of the output and input variables. The specific mutual information measures
for each of the values of the target variable $q_j^+$ how different the joint probability density function $p(q_j^+, \bq_\bi)$
is from the hypothetical distribution $p(q_j^+)p(\bq_\bi)$, where $q_j^+$
and $\bq_\bi$ are assumed to be independent. For instance, if $q_j^+$ and
$\bq_\bi$ are not independent, then $p(q_j^+, \bq_\bi)$ will differ
significantly from $p(q_j^+)p(\bq_\bi)$. Hence, SURD assess causality by
examining how the probability of $Q_j^+$ changes when accounting for
$\bQ_\bi$ for each $q_j^+$.

% Definitions
Then, SURD quantifies the
increments in specific information $\Delta \Is$ about $q_j^+$ obtained
by observing an individual or groups of components from $\bQ$. 
 For a given state $q_j^+$
    of the target variable $Q_j^+$, the specific causalities $\Is$
    are computed for all the possible combinations of past
    variables. These components are organized in ascending order,
    which allows to assign the redundant, unique, and
    synergistic  causalities. The
quantities in Equation (\ref{eq:surd}) are then obtained
as the expectation of their corresponding values:
\begin{equation}
    \Delta I_{\bi \rightarrow j}^R = \sum_{q_j^+} p(q_j^+) \Delta \Is_{\bi}^R(q_j^+),
    \quad\quad\quad
    \Delta I_{i \rightarrow j}^U = \sum_{q_j^+} p(q_j^+) \Delta \Is_{i}^U(q_j^+),
    \quad\quad\quad
    \Delta I_{\bi \rightarrow j}^S = \sum_{q_j^+} p(q_j^+) \Delta \Is_{\bi}^S(q_j^+).
\end{equation}

The portion of information in 
$Q^+_j$ that remains unexplained by the source variables $\bQ$ is 
accounted for in the causality leak term. This can be quantified in closed 
form as the conditional Shannon information $H(Q_j^+|\bQ)$ \cite{shannon1948}:
\begin{equation}
    H(Q_j^+\vert\bQ) = \sum_{q_j^+\in Q_j^+} \sum_{\bq\in \bQ} -p(q_j^+,\bq) \log_2 \left[p(q_j^+\vert\bq)\right]  \geq 0.
\end{equation}

Finally, we also examine the contributions of each of the values $\bq$ of the source variables $\bQ$ to the terms in Eq. \ref{eq:surd} using the definition proposed in Ref.\cite{states2025}. Therefore, the resulting causal maps will be a function of both source and target states. The reader is referred to Ref.\cite{states2025} for further details about the formulation of this approach and its implementation.

\subsection{Explainability}
\label{SHAP_method}
As proposed by~\citet{Cremades2024,cremades2024classically}, we use SHAPley Additive exPlanation (SHAP) values~\cite{SHAPley1953,lundberg2017unified} to quantify, for each latent variable, how each grid point of the flow field contributes to its encoding. Concretely, let
\[
\mathbf{X}^t \in \mathbb{R}^{N\times N\times N\times 3}
\]
be a single three-dimensional (3D) snapshot of the flow, let $p(\mathbf{X}_r)$ denote a baseline distribution over reference snapshots, and let $f_j$ be the model's mapping from an input snapshot to the $j$-th latent variable.

\noindent
We adopt the Expected Gradients formulation of~\citet{Erion2021}, which is a gradient-based estimator of SHAP values. For a fixed latent index $j$ and a feature index $p$ (here $p$ stands for the multi-index $(i,j,k,c)$ on the grid), the Expected Gradients attribution is defined as
\begin{equation}
  \phi^{\mathrm{EG}}_{p,j}(\mathbf{X}^t)
  \;=\;
  \mathbb{E}_{\mathbf{X}_r\sim p(\mathbf{X}_r),\,\alpha\sim U(0,1)}
  \Bigl[
    \bigl(\mathbf{X}^t_p - \mathbf{X}_{r,p}\bigr)\,
    \partial_{\mathbf{X}_p}
    f_j\bigl(\mathbf{X}_r + \alpha(\mathbf{X}^t - \mathbf{X}_r)\bigr)
  \Bigr].
  \label{eq:EG-def}
\end{equation}
Collecting all feature indices $p$ into a tensor yields the attribution map
\(\mathbf{\Phi}_j(\mathbf{X}^t)\in\mathbb{R}^{N\times N\times N\times 3}\), with
\(\mathbf{\Phi}_j(\mathbf{X}^t)_{i j k c} = \phi^{\mathrm{EG}}_{p,j}(\mathbf{X}^t)\)
after identifying $p\equiv(i,j,k,c)$.

\noindent
Expected Gradients satisfies a completeness property analogous to integrated gradients~\cite{Sundararajan2017,Erion2021}: summing attributions over all features recovers the difference between the model output at $\mathbf{X}^t$ and its expected value over baselines,
\begin{equation}
  \label{eq:gradSHAP-main}
  \sum_{i,j,k,c} \mathbf{\Phi}_j(\mathbf{X}^t)_{i j k c}
  \;=\;
  f_j(\mathbf{X}^t)
  \;-\;
  \mathbb{E}_{\mathbf{X}_r\sim p(\mathbf{X}_r)}\!\bigl[f_j(\mathbf{X}_r)\bigr].
\end{equation}

\medskip
\noindent
\textbf{Connection to integrated gradients.}
For a fixed baseline $\mathbf{X}_r$, define
\[
\Delta\mathbf{X} \;=\; \mathbf{X}^t - \mathbf{X}_r.
\]
By the fundamental theorem of calculus for line integrals, the integrated-gradients identity for $f_j$ along the straight path
$\mathbf{X}(\alpha) = \mathbf{X}_r + \alpha\,\Delta\mathbf{X}$, $\alpha\in[0,1]$, reads
\begin{equation}
  f_j(\mathbf{X}^t) - f_j(\mathbf{X}_r)
  \;=\;
  \int_0^1
  \Bigl\langle
    \Delta\mathbf{X},\,
    \nabla_{\mathbf{X}} f_j\bigl(\mathbf{X}_r + \alpha\,\Delta\mathbf{X}\bigr)
  \Bigr\rangle
  \,\mathrm{d}\alpha,
  \label{eq:IG-continuous}
\end{equation}
where
\(\langle A, B\rangle = \sum_{i,j,k,c} A_{i j k c}\,B_{i j k c}\)
denotes the inner product over spatial indices \((i,j,k)\) and channel $c$.

\noindent
In practice, we approximate the integral in~\eqref{eq:IG-continuous} by a Riemann sum over $m$ points $\alpha_k = k/m$, $k=1,\dots,m$. For each baseline $\mathbf{X}_r$ we define
\[
\overline{\mathbf{J}}_j(\mathbf{X}^t,\mathbf{X}_r)
\;=\;
\frac{1}{m}\sum_{k=1}^m
 \nabla_{\mathbf{X}}\,
 f_j\!\Bigl(\mathbf{X}_r + \tfrac{k}{m}\,\Delta\mathbf{X}\Bigr)
\;\in\;\mathbb{R}^{N\times N\times N\times 3},
\]
i.e. the average Jacobian of the mapping $f_j$ along the straight path from $\mathbf{X}_r$ to $\mathbf{X}^t$. Substituting this into~\eqref{eq:IG-continuous} yields the discrete integrated-gradients identity
\begin{equation}
f_j(\mathbf{X}^t) - f_j(\mathbf{X}_r)
 \;\approx\;
 \bigl\langle
 \Delta\mathbf{X},\,\overline{\mathbf{J}}_j(\mathbf{X}^t,\mathbf{X}_r)
 \bigr\rangle.
 \label{eq:IG-discrete}
\end{equation}

\medskip
\noindent
\textbf{Practical Gradient-SHAP / Expected Gradients estimator.}
In practice, we approximate the expectations in~\eqref{eq:EG-def} and~\eqref{eq:gradSHAP-main} by Monte Carlo sampling. We draw a finite set of baselines
\(\mathcal{R} = \{\mathbf{X}_r^{(1)},\dots,\mathbf{X}_r^{(|\mathcal{R}|)}\}\)
from \(p(\mathbf{X}_r)\), and approximate the integral over
\(\alpha\in[0,1]\) by the Riemann sum in~\eqref{eq:IG-discrete}.
This yields the following estimator of the attribution map:
\begin{equation}
 \mathbf{\Phi}_j(\mathbf{X}^t)
 \;\approx\;
 \frac{1}{|\mathcal{R}|}
 \sum_{\mathbf{X}_r\in\mathcal{R}}
 \left[
 \Delta\mathbf{X}
 \;\odot\;
    \left(
      \frac{1}{m}\sum_{k=1}^m
      \nabla_{\mathbf{X}} f_j\bigl(\mathbf{X}_r + \tfrac{k}{m}\,\Delta\mathbf{X}\bigr)
    \right)
 \right],
 \label{eq:EG-monte-carlo}
\end{equation}
where $\odot$ denotes the element-wise (Hadamard) product. Equation~\eqref{eq:EG-monte-carlo} is exactly the “expected gradients” form used by GradientSHAP implementations~\cite{Erion2021,captumGradientShap}, i.e. the expected value of
\(\Delta\mathbf{X} \odot \nabla_{\mathbf{X}} f_j\)
over random baselines and random points along the path.

\noindent
By construction, this estimator preserves the completeness property in expectation. Summing~\eqref{eq:EG-monte-carlo} over all grid points and channels gives
\begin{equation}
 \sum_{i,j,k,c} \mathbf{\Phi}_j(\mathbf{X}^t)_{i j k c}
 \;\approx\;
 f_j(\mathbf{X}^t)
 \;-\;
 \frac{1}{|\mathcal{R}|}\sum_{\mathbf{X}_r\in\mathcal{R}} f_j(\mathbf{X}_r),
\end{equation}
which is the Monte Carlo approximation of~\eqref{eq:gradSHAP-main}. In other words, the sum of the feature-wise attributions approximately recovers the difference between the latent activation at $\mathbf{X}^t$ and its expected value over baselines.

\noindent
SHAP values, grounded in cooperative game theory, thus arise here as Expected Gradients (Gradient-SHAP) attributions for each latent variable $f_j$: each grid point $(i,j,k)$ and channel $c$ is assigned the expected contribution of its offset $\Delta\mathbf{X}_{i j k c}$, weighted by the average directional derivative of $f_j$ along the straight-line paths from the baselines $\mathbf{X}_r$ to the target snapshot $\mathbf{X}^t$.

\noindent Practically, the gradient-SHAP method calculates this attribution by assessing the average sensitivity of the model outputs along a linear interpolation between a baseline (reference state) and the actual input. This results in spatially explicit importance fields that reveal precisely which regions in the input most significantly influence the latent representations.

\noindent In conclusion, the SHAP values constitute a scalar field which classifies by importance each of the pixels composing the input of our $\beta$-VAE. The importance is measured as previously presented. Extending the study of SHAP values to classical theories of coherent structures~\cite{Jimenez_2018}, it is possible to apply a percolation study to the SHAP fields, $\phi(\mathbf{X}^t) = \mathbf{S}^t(\mathbf{X})$ in order to retrieve organized structures on the data. In our study, we consider 2D fluid flows, where each velocity component has a SHAP field, stream-wise and vertical fluctuations respectively: $\mathbf{S^t(u)},\mathbf{S^t(v)}$. Before diving into the details of this approach, it is important to note that the SHAP values are in fact a sensitivity analysis which quantifies the importance locally of each pixel when applying an operator:
\begin{equation}
    M: I \rightarrow O
\end{equation}
where $I$ and $O$ do not need to have the same dimensions.
The percolation study is conducted as follows; let $\mathbf{u},\mathbf{v} \subset I$ where $ I \in \mathbb{R}^{2 \times N \times N \times T} $, then:
\begin{equation}
    \sqrt{\bigl(\mathbf{S^t(u)}\bigr)^2 + \bigl(\mathbf{S^t(v)}\bigr)^2}
    > H\sqrt{\bigl(\overline{\mathbf{S(u)}\bigr)^2} + \bigl(\overline{\mathbf{S(v)}\bigr)^2}},
    \quad \mathbf{S^t(u)},\overline{\mathbf{S(v)}^2} \in \mathbb{R}^{ N \times N}, \quad t\in T
\end{equation}
This equation ensures that only the points with high importance with respect to the square of the root-mean-square (rms) importance value of the field are considered. Additional constraints can be included, if we want the SHAP structures to satisfy known constraints, such as volumetric limitations imposed by the boundaries. Returning to our particular case, we use an injection operator (submersion), which reduces the dimensionality of the fluid-flow problem around the obstacle:
\begin{equation}
    \mathcal{E}: I \rightarrow O,\quad
    I \in \mathbb{R}^{2 \times N \times N \times T},\;
    O \in \mathbb{R}^{d \times T},\;
    N \times N > d
\end{equation}
By doing so we can inspect how each element of the input contributes to the encoding of each of the latent variables, obtaining a temporal sequence of SHAP fields for each of such latent variables; $\mathbf{S^t_i(u)},\mathbf{S^t_i(v)}, \forall i \in d$, where $d$ is the size of the latent space. After computing the SHAP fields, we can proceed to percolate independently for each element in $O$ the SHAP fields:
\begin{align}
  P_i^t(x,y)
  &= \sqrt{\bigl(\mathbf{S^t_i}(\mathbf{u}(x,y))\bigr)^2 + \bigl(\mathbf{S^t_i}(\mathbf{v}(x,y))}\bigr)^2,
    \quad x,y \in N, \\
  || \mu_S||_2
  &= \sqrt{\bigl(\overline{\mathbf{S}(\mathbf{u}(x,y))\bigr)^2} + \bigl(\overline{\mathbf{S}(\mathbf{v}(x,y))\bigr)^2}}, \\
  P_i^t(x,y)
  &=
    \begin{cases}
      1, & P_i^t(x,y) > H_i || \mu_S||_2,\\
      0, & P_i^t(x,y) \le H_i || \mu_S||_2,
    \end{cases}
\label{percenc}
\end{align}
After applying the percolation independently for each latent variable and later the boolean mask for each time step, we obtain a temporal sequence of a spatial subset of the SHAP fields, $\tilde{\mathbf{S^t_i}} \subset \mathbf{S^t_i}$. Note that the field $\tilde{\mathbf{S^t_i}}$ contains the SHAP-based structures at time $t$ for latent variable $i$.

\section{Framework Validation: Explainability and Causality}
\label{val}
\subsection{X-CAL on a 2D Torus}
The causal representation of latent variables learned by a $\beta$-VAE remains a challenging topic, particularly when the underlying dynamical system exhibits strong coupling among its degrees of freedom. To rigorously assess how well a $\beta$-VAE captures known causal structure, we consider the \emph{coupled torus}, a toy model whose causal mechanisms are fully specified and whose attractor dimension matches its embedding dimension.

\noindent We define two signals $f_1(t)$ and $f_2(t)$ sampled at $N$ equally spaced times over $[0,T]$:
\begin{subequations}
\begin{align}
t_n &= \frac{n}{N-1},T, \quad n=0,1,\dots,N-1, \\
f_1(t_n) &= \cos(\omega_1 t_n),\\
f_2(t_0) &= \cos(\omega_2 t_0), \quad
f_2(t_n) = \cos(\omega_2 t_n) + \alpha f_1(t_{n-1}), \quad n\ge1.
\end{align}
\label{eq:torus}
\end{subequations}

Here $\omega_1 = 1$, $\omega_2 = \sqrt{2}$, and $\alpha$ controls a pure lag-1 influence $f_1\to f_2$, $\alpha = 0.7$. This dimension of the attractor of this system is two, spanned by $(f_1,f_2)$. Regarding the input of the $\beta$-VAE, we lift up the temporal coefficients presented in Eq.~(\ref{eq:torus}) into a spatio-temporal field, $F$ which is defined as follows:

\begin{equation}
F(x,z,t_n) = f_1(t_n) \sin(\pi x)\sin(\pi z) + f_2(t_n) \cos(\pi x)\sin(\pi z)
\label{Field_torus}
\end{equation}

\begin{figure}[H]
    \centering
    \begin{subfigure}{0.3\linewidth}
        \centering
        \includegraphics[width=1\linewidth]{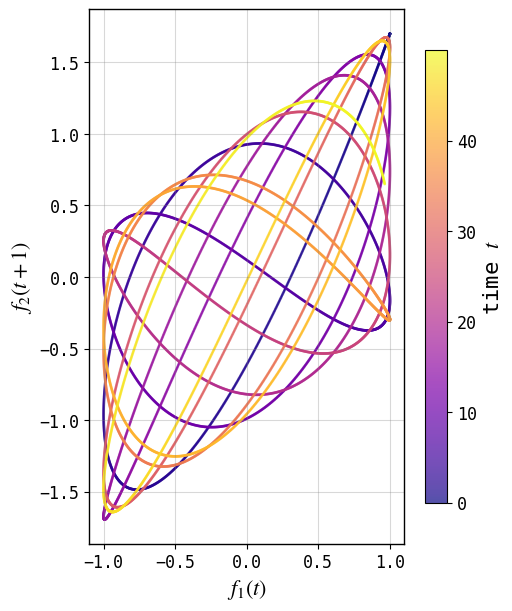}
    \end{subfigure}
    \quad
    \begin{subfigure}{0.40\linewidth}
        \centering
        \includegraphics[width=1\linewidth]{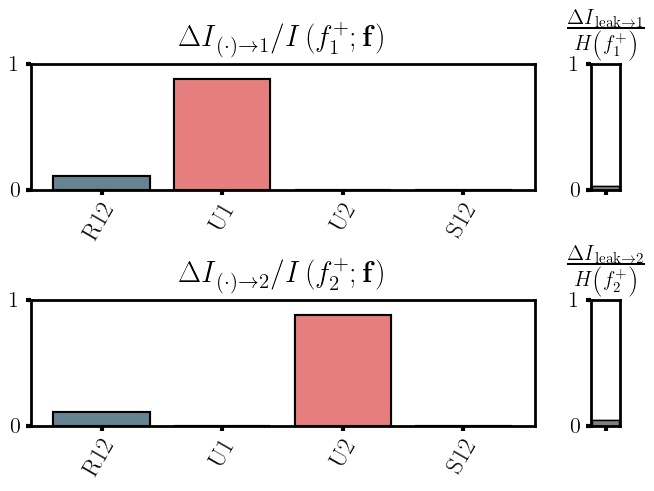}
    \end{subfigure}
    \caption{(Left) First lag phase plot between temporal coefficients of the system and (Right) the SURD decomposition for the temporal sequences $f_1,f_2$. }
    \label{torus_info}
\end{figure}

\noindent Given the spatio-temporal field in Eq.~(\ref{Field_torus}), we train four $\beta$-VAEs which have different $\beta$ coefficients, inspecting how the latter affects the causality present in the data. However, before we dive into the analysis of the different cases, it is important to note that regardless of the coupling presented in Eq.~(\ref{eq:torus}) the causal decomposition is dominated by self-unique terms and small redundancy between $f_1$ and $f_2$ depicted in Fig.~\ref{torus_info}. It is also noticeable the first lag phase plot between both temporal signals in Fig.~\ref{torus_info}, which will also serve as a measure to assess the performance of the autoencoders.

\begin{figure}[H]
    \centering
    \begin{subfigure}{0.45\linewidth}
        \centering
        \includegraphics[width=1\linewidth]{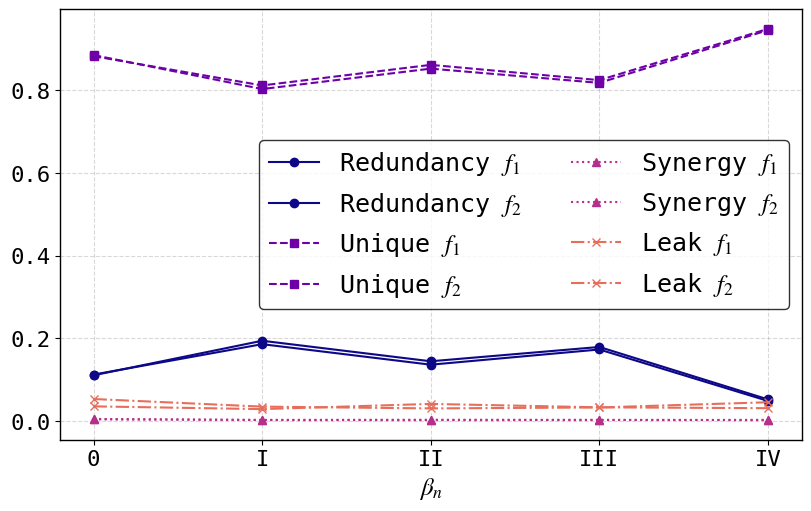}
    \end{subfigure}
    \quad
    \begin{subfigure}{0.30\linewidth}
        \centering
        \includegraphics[width=1.1\linewidth]{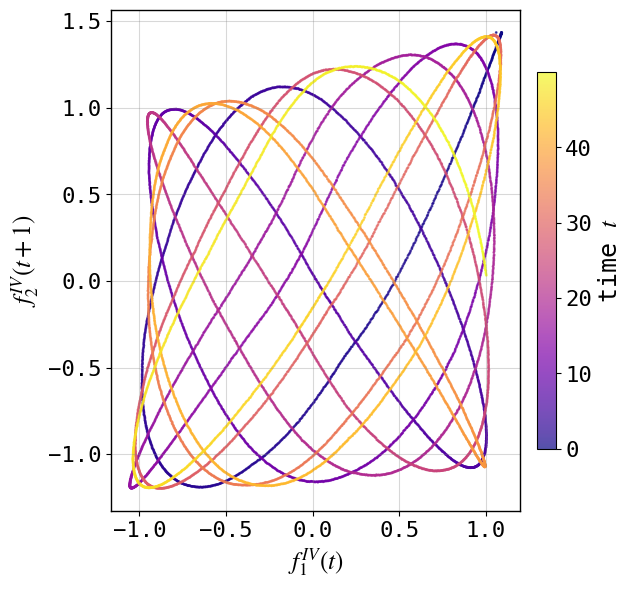}
    \end{subfigure}
        \caption{(Left) SURD decomposition analysis for all cases presented in Table~\ref{torus_tab} and the reference temporal signal in Fig~\ref{torus_info}, the different terms in the decomposition are plotted with their respective magnitude as a function of the $\beta_n$, where the numbers on the $x$-axis refer to the study cases and (Right) First lag phase plot between latent temporal coefficients of Case IV.  }
    \label{cases_torus}
\end{figure}

\noindent Once we have the dynamics and causality of our simple 2-D torus, we will compare the causal performance of all cases to assess the ability of $\beta$-VAE for the identification of causal mechanisms. In Fig.~\ref{cases_torus} one can observe the evolution of the SURD causal terms as a function of $\beta$. Note that the ground truth signal is characterized by the values at $\beta_n = 0$, and only Case $IV$ is capable of improving the unique causality magnitude while also reducing the redundancy present in the original system. In terms of causality, the main mechanisms remain unchanged, as $U_1$ and $U_2$ dominate the causal space. The increase in unique causality and the decrease in redundancy is visible on the right plot in Fig.~\ref{torus_info}, where the geometry of the phase plot has been rotated, reducing the redundant coupling between both temporal signals.

\begin{table}[ht]
\centering
\begin{tabular}{l|c|c|c|c|c}
\textbf{Case} 
  & \textbf{Orthogonality} 
  & \textbf{Reconstruction} 
  & \textbf{Latent dim} 
  & \textbf{Epochs}
  & \textbf{$\beta_n$} \\ \hline 
I  & 88.53\% &  98.98\% & 2 & 300&$5\times10^{-5}$ \\  
II   & 91.14\% & 99.92\% & 2 & 300&$5\times10^{-4}$ \\  
III  & 99.69\% &  99.89\% & 2 & 300&$5\times10^{-3}$ \\  
IV   & 95.79\% &  99.89\% & 2 & 800&$5\times10^{-3}$ \\
\end{tabular}
\caption{Training results for the three Lorenz‐based cases with different hyper-parameters.}
\label{torus_tab}
\end{table}
When the attractor dimension already matches the number of observables (i.e.\ the torus), there is no new causal structure to discover. Increasing $\beta$ simply collapses redundancy into one latent, while also reducing the information-leak and leaves the original causality partially intact. It is important to mention that the information-leak could also increase due to reconstruction regularization induced by the $\beta$-VAE.

\subsection{X-CAL on the Lorenz System}
\label{Lorenz_sec}
To further validate our $\beta$-VAE framework in a controlled setting, we consider two variants of the classic Lorenz system~\cite{Lorenz1963}, each embedded into a low-dimensional latent space of size $d\in\{3,2\}$.  We generate time series of length $N=10{,}000$ by integrating
\begin{subequations}
\begin{align}
  \frac{{\rm d}x}{{\rm d}t} &= \sigma\,(y - x), \\
  \frac{{\rm d}y}{{\rm d}t} &= x\,(\rho - z) - y, \\
  \frac{{\rm d}z}{{\rm d}t} &= x\,y - \beta\,z,
\end{align}
\label{eq:lorenz}
\end{subequations}
\noindent with the standard parameters $\sigma=10$, $\rho=28$, and $\beta=8/3$.  These three real-valued trajectories $(X,Y,Z)$ are then projected onto a set of spatial basis functions
\begin{equation}
  \phi_1(x,z) = \sin(\pi x)\sin(\pi z),\quad
  \phi_2(x,z) = \cos(\pi x)\sin(\pi z),\quad
  \phi_3(x,z) = \sin(2\pi z)
\end{equation}
via
\begin{equation}
  F(x,z,t) = x_t\,\phi_1(x,z)
  + y_t\,\phi_2(x,z)
 + z_t\,\phi_3(x,z),
\end{equation}
so that each snapshot is a 2D field in $(x,z)\in[0,1]^2$.

\begin{figure}[H]
  \centering
  \includegraphics[width=1.05\linewidth]{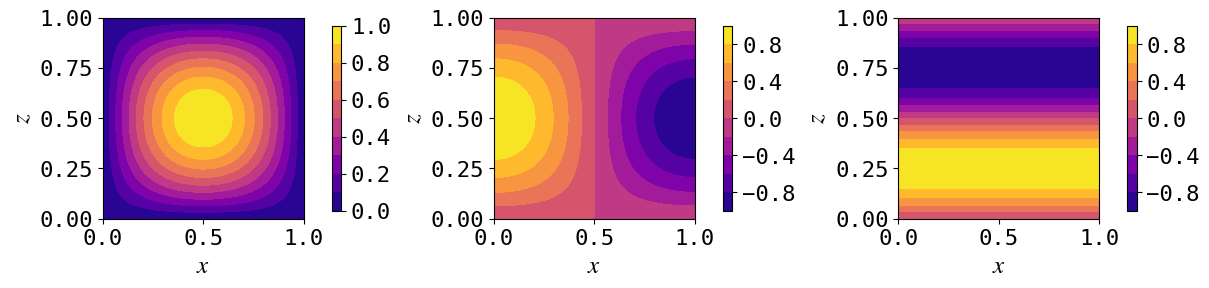}
  \caption{Spatial basis functions used to lift the Lorenz trajectories into 2D fields: (a) $\phi_1$, (b) $\phi_2$, (c) $\phi_3$.}
  \label{fig:lorenz_bases}
\end{figure}

\noindent The resulting data is encoded and decoded by a convolutional $\beta$-VAE with four hidden layers (two in the encoder, two in the decoder), each using $4\times4$ kernels and ELU activations.  A fully connected layer of size 128 projects into the $d$-dimensional latent space, with regularization weight $\beta$ as indicated in Table~\ref{tab_train_val}.  We train each model for 300 epochs with a batch size of 64 and learning rate $10^{-5}$.  The training metrics, i.e. latent‐variable orthogonality (measured via $\det(\mathrm{cov}(\mathcal L))$), reconstruction accuracy and the chosen $\beta$, are reported in Table~\ref{tab_train_val}.

\begin{table}[ht]
\centering
\begin{tabular}{l|c|c|c|c}
\textbf{Case} 
  & \textbf{Orthogonality} 
  & \textbf{Reconstruction} 
  & \textbf{Latent dim.} 
  & \textbf{$\beta$} \\ \hline
I   & 79.44\% & 100\% & 3 & $5\times10^{-4}$ \\  
II  & 97.04\% &  99\% & 2 & $5\times10^{-4}$ \\  
\end{tabular}
\caption{Training results after 300 epochs for the three Lorenz‐based cases.}
\label{tab_train_val}
\end{table}

\noindent Note that projecting the Lorenz ordinary differential equation (ODE) into a spatial field via these sinusoidal modes is mathematically equivalent to a three‐mode Galerkin projection of the underlying dynamics.  This simple yet non‐trivial testbed lets us assess how well the $\beta$-VAE disentangles and reconstructs known chaotic behavior before tackling the full 2D urban‐flow direct numerical simulation (DNS) data.  

\begin{figure}[H]
    \centering
    \begin{subfigure}{0.28\linewidth}
        \centering
        \includegraphics[width=1\linewidth]{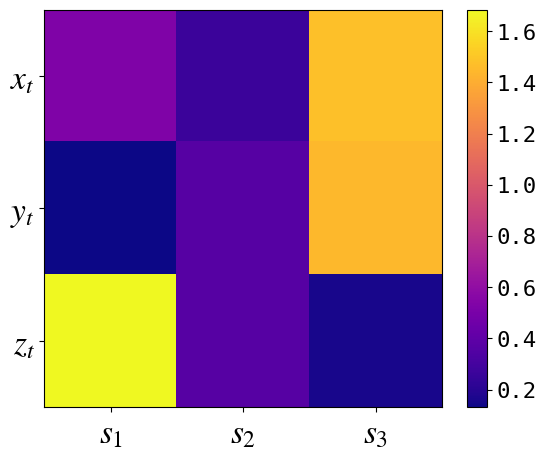}
    \end{subfigure}
    \quad
    \begin{subfigure}{0.28\linewidth}
        \centering
        \includegraphics[width=1\linewidth]{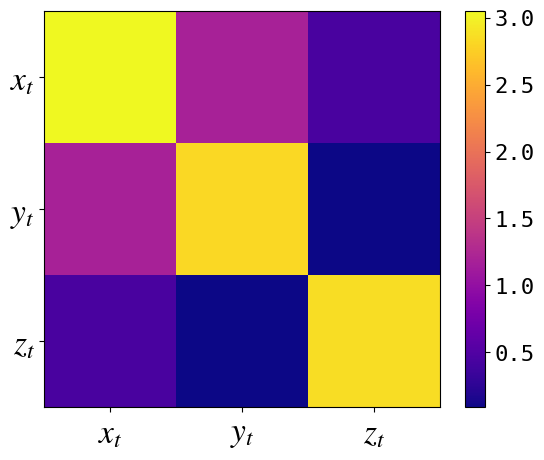}
    \end{subfigure}
    \begin{subfigure}{0.28\linewidth}
        \centering
        \includegraphics[width=1\linewidth]{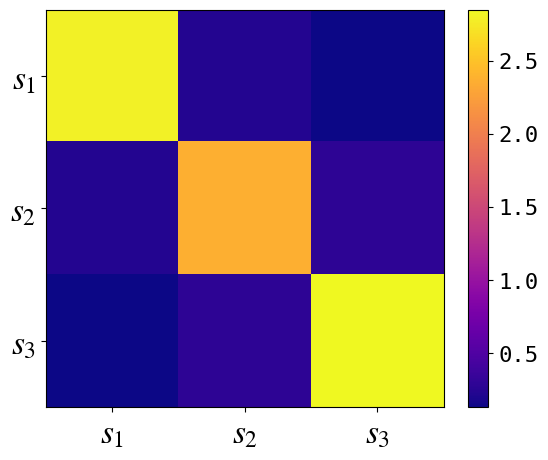}
    \end{subfigure}
    \caption{First we inspect the Mutual information between the original temporal coefficients of the Lorenz System $x_t,y_t,z_t$ and the latent coefficients for Case I, $s_1,s_2,s_3$ (left), secondly we inspect the mutual information in between Lorenz coefficients (middle) and finally we compute the mutual information between the latent elements of Case I (right). }
    \label{mutual}
\end{figure}
\medskip
\noindent
In our analysis we will compare the latent space encoded by the two autoencoders against the original attractor spanned by the Lorenz system ODE, specifically the 10,000 temporal sequence used along training. Before we continue the study, it is important to mention that all variables $x_t,y_t,z_t,s_1,s_2,s_3,l_1$ and $l_2$ will be treated as random variables from now on, leaving behind their deterministic origin, making possible the usage of the mutual information to quantify causality. In Fig.~\ref{mutual}, one can observe on the left plot how $x_t$ and $y_t$ are mainly captured by $s_3$ in terms of mutual information while $z_t$ is principally sharing the information with $s_1$. The latter stays consistent with the second plot, as variables $x_t$ and $y_t$ have a high cross mutual information, suggesting a strong dynamical coupling. Finally on the last plot of Fig.~\ref{mutual}, one can see how the latent space with $d=3$, has a better disentangling on the mutual information as the dominant elements are the diagonal terms.

\begin{figure}[h!]
    \centering
    \begin{subfigure}{0.48\linewidth}
        \centering
        \includegraphics[width=1\linewidth]{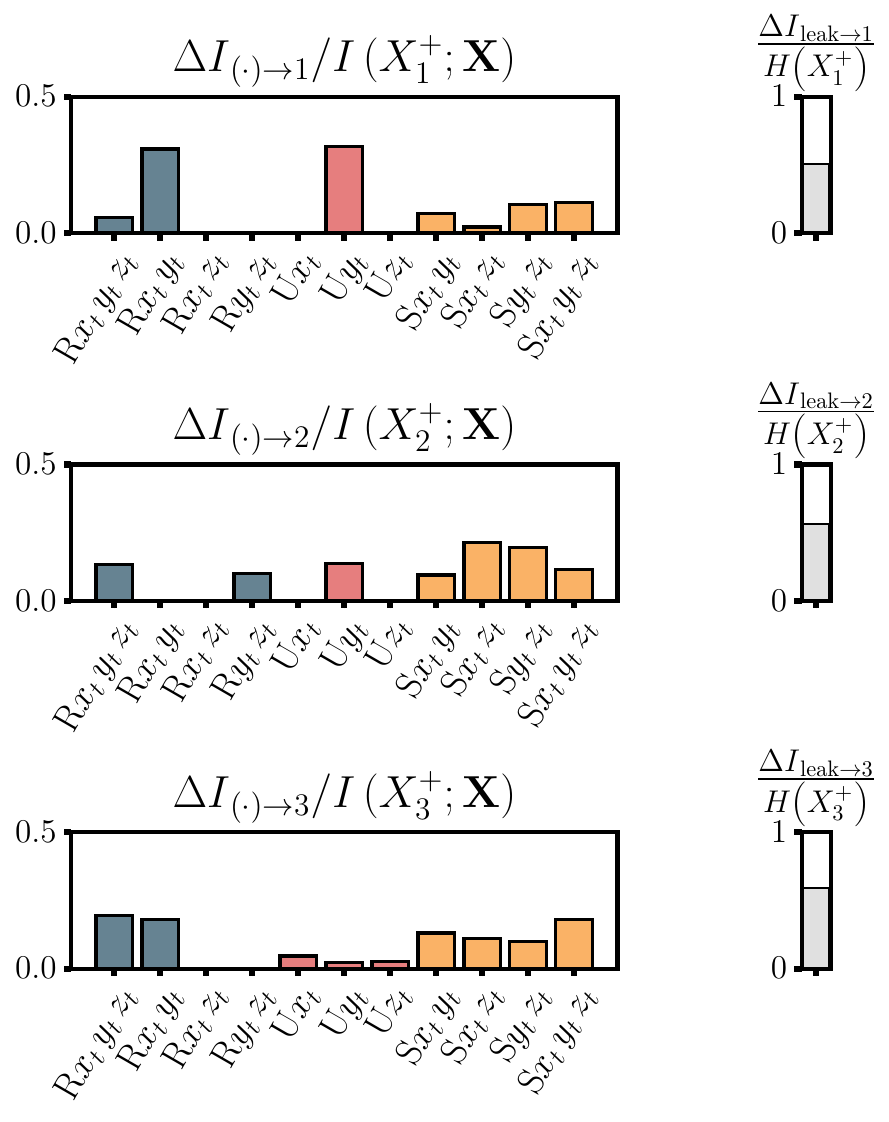}
    \end{subfigure}
    \quad
    \begin{subfigure}{0.48\linewidth}
        \centering
        \includegraphics[width=1\linewidth]{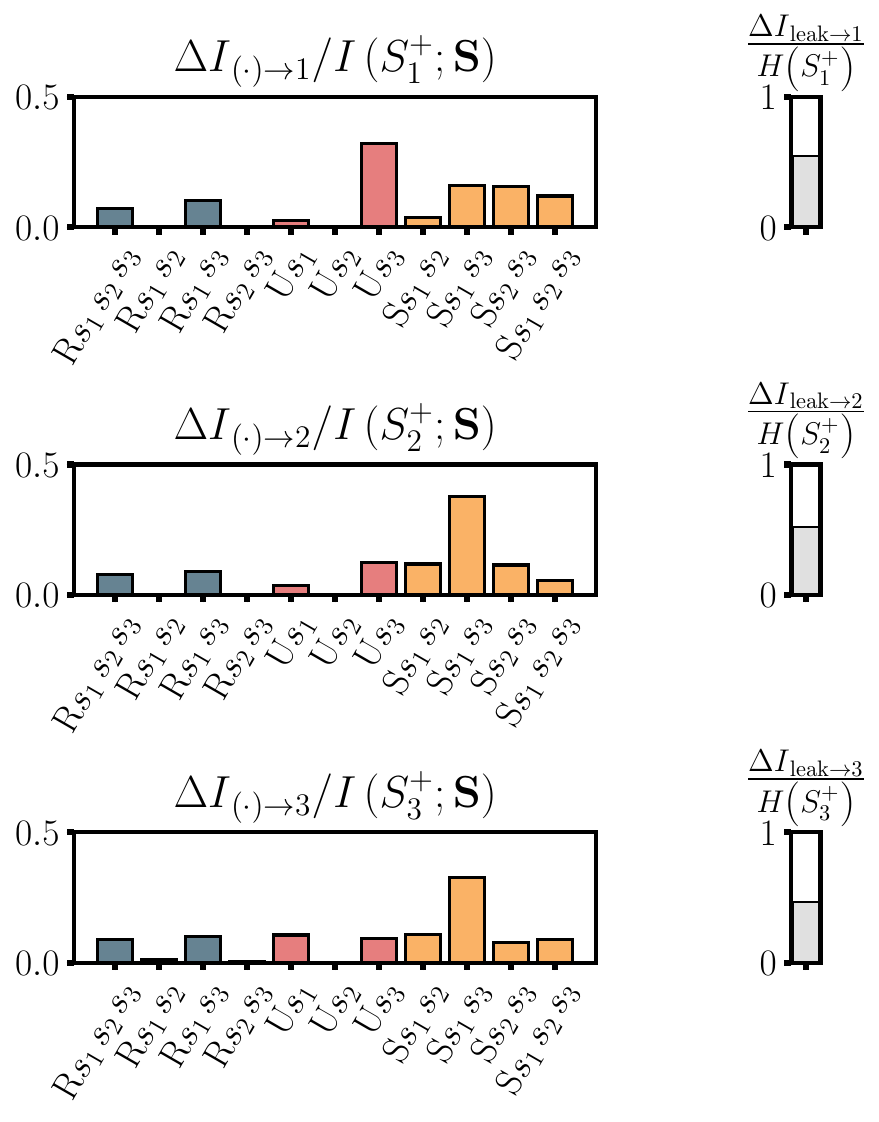}
    \end{subfigure}
    \caption{SURD analysis conducted with 10 bins and a lag of 61 on the original temporal sequence of the Lorenz ODE (left) where $\boldsymbol{X}$ is the probability dual of $(X,Y,Z)$ with respective duals $X_i$ $\in [1,2,3]$ are the respective random variables. On the (right) the latent temporal sequence of Case I where $S_1,S_2,S_3$ are the non-deterministic duals of $s_1,s_2,s_3$ respectively. }
    \label{histos_lor}
\end{figure}

When inspecting the SURD histograms depicted in Fig.~\ref{histos_lor}, one can observe how the main causal relationships extracted at lag 61 are similar for both the original Lorenz system and the latent attractor when considering $s_3 \sim y_t$ and $s_1 \sim z_t$. These perspective on causality still remains limited as the the causality imposed by the geometry of the attractor is unknown. In order to extend the analysis, it is needed to introduce the study of causality by states~\cite{states2025}, where one can observe how the different states of a source temporal sequence affects the states of target sequence. Specifically we will focus on the unique and synergistic terms of the SURD decomposition. 

\begin{figure}[h!]
    \centering
    \begin{subfigure}{0.23\linewidth}
        \centering
        \includegraphics[width=1\linewidth]{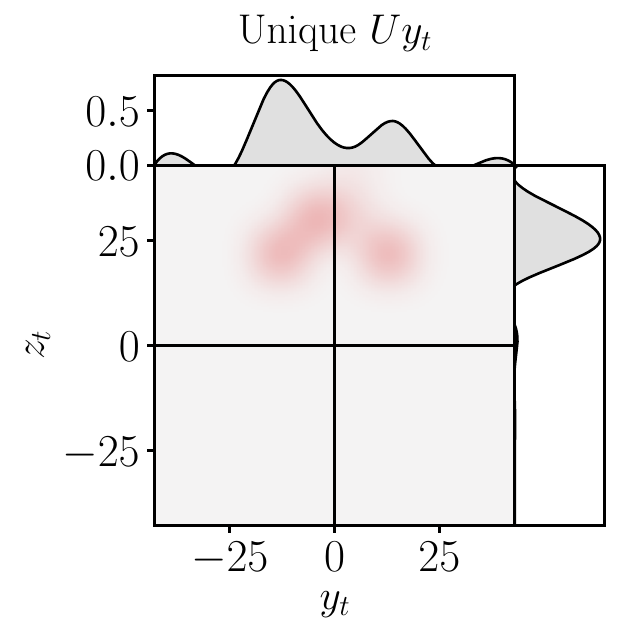}
    \end{subfigure}
    \quad
    \begin{subfigure}{0.23\linewidth}
        \centering
        \includegraphics[width=1\linewidth]{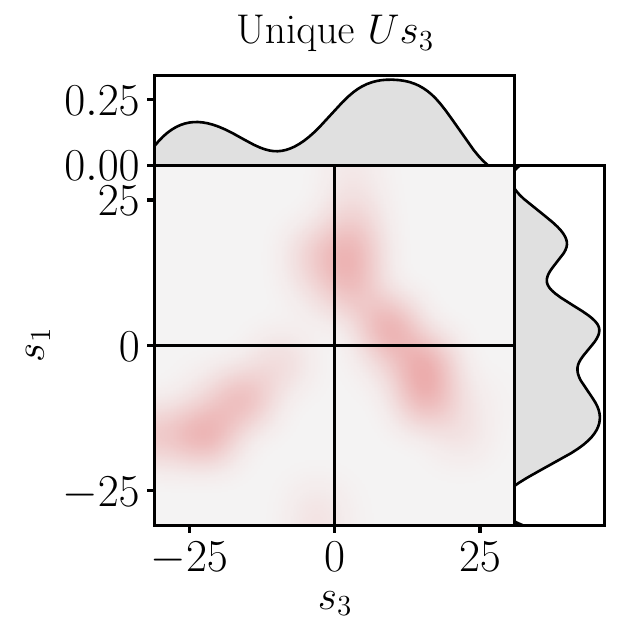}
    \end{subfigure}
    \begin{subfigure}{0.23\linewidth}
        \centering
        \includegraphics[width=1\linewidth]{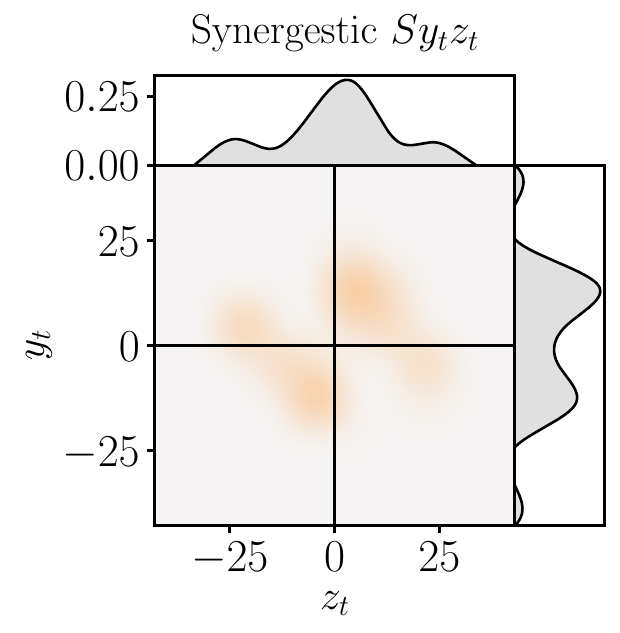}
    \end{subfigure}
    \quad
    \begin{subfigure}{0.23\linewidth}
        \centering
        \includegraphics[width=1\linewidth]{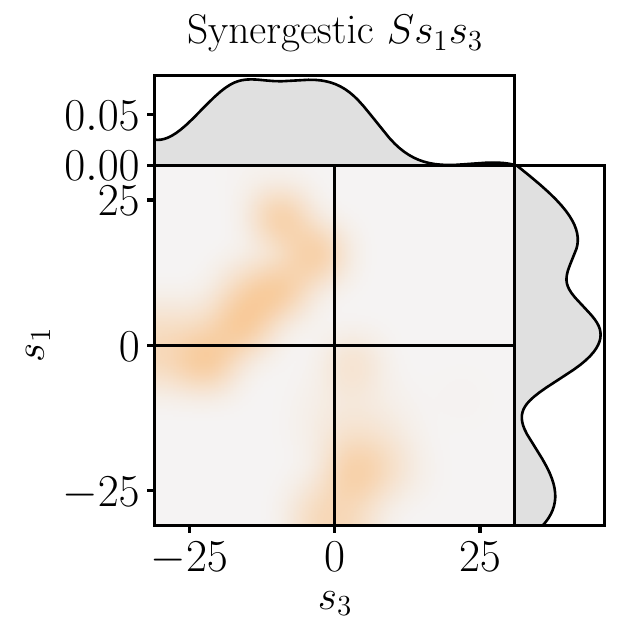}
    \end{subfigure}
    \caption{Unique and synergistic state analysis, representing from (left) to (right) the causal maps for the original Lorenz ODE and Case I presented in Fig.~\ref{histos_lor}. On the $x$ axis we plot the source and on the $y$ axis the target.}
    \label{states_lor}
\end{figure}

\begin{figure}[H]
    \centering
    \begin{subfigure}{0.20\linewidth}
        \centering
        \includegraphics[width=1\linewidth]{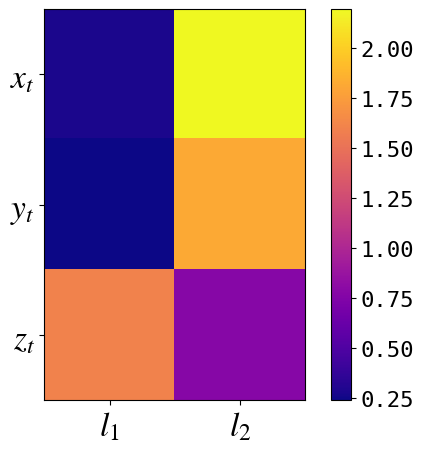}
    \end{subfigure}
    \quad
    \begin{subfigure}{0.30\linewidth}
        \centering
        \includegraphics[width=1\linewidth]{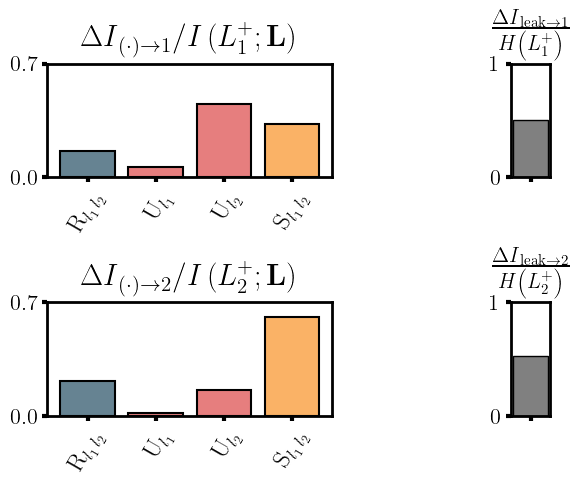}
    \end{subfigure}
        \begin{subfigure}{0.21\linewidth}
        \centering
        \includegraphics[width=1\linewidth]{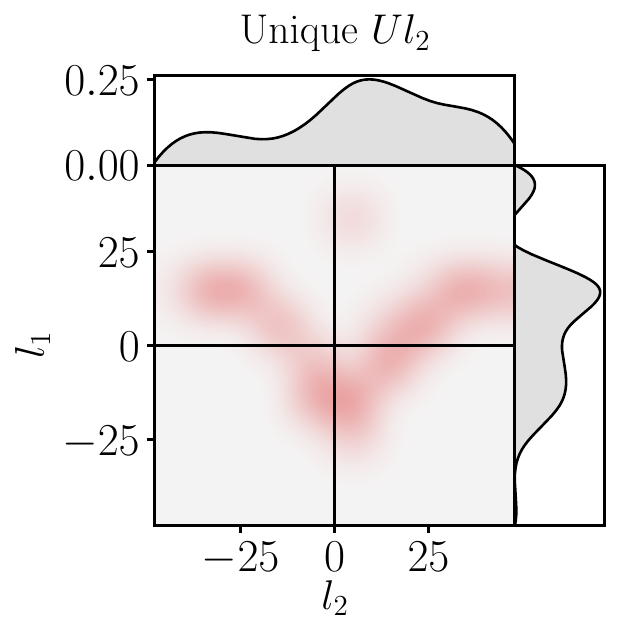}
    \end{subfigure}
    \quad
    \begin{subfigure}{0.21\linewidth}
        \centering
        \includegraphics[width=1\linewidth]{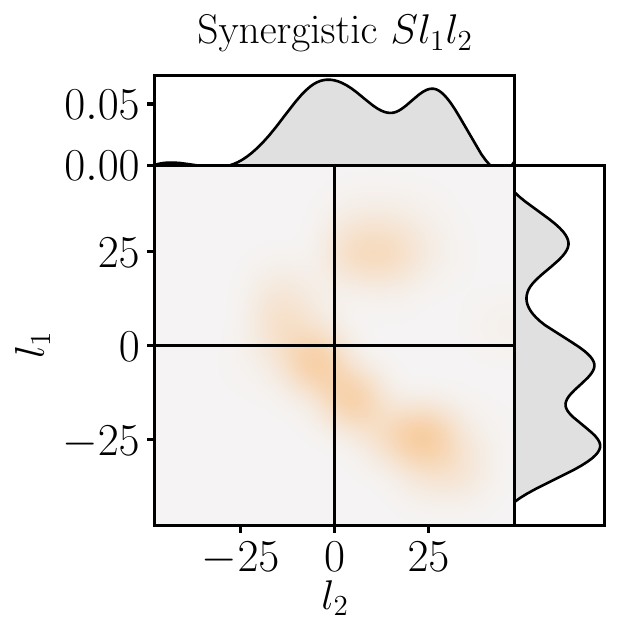}
    \end{subfigure}
    \caption{From left to right; the Mutual Information between $x_t,y_t,z_t$ and $l_1,l_2$ remarking that $l_1\sim z_t$ and $l_2 \sim x_t,y_t$. SURD analysis conducted with 10 bins and a lag of 61 on the latent temporal sequence of Case II and SURD causal state study on the unique and synergistic terms.   }
    \label{l2}
\end{figure}

\noindent Furthermore, Fig.~\ref{states_lor} shows that the latent space spans a broader causal domain when compared to the original system, specially the unique term $Us_3$ where the flow of information connecting both lobes of the attractor is captured. Even-more $Ul_2$, depicted in Fig.~\ref{l2}, seems to capture an information circle around $(0,25)$ resembling that of the saddle point of the attractor. 

\noindent Given the causal structure presented in Figs.~\ref{states_lor},\ref{l2}, it is sensitive to focus our attention on the study of Case II. If one has the objective of controlling the system, Case II seems the most appealing as the information is better disentangled, spanning an a priori interpretable causal space compared with the original system, while reducing the redundant contribution, Fig.~\ref{l2}. It is important to mention that one cannot study the causal structure of the original Lorenz system through these latent dynamical systems, as the informational content of each variable is intrinsically different. Given so, when applying causal techniques based on the mutual information or other dynamical metrics, one can not expect to obtain the same causal mechanisms across the different systems.

\noindent To gain a further insight into the models latent space, we will inspect the mean SHAP fields for a given latent variable and its mean decodification at the causal time steps depicted in Figs.~\ref{l2}\ref{states_lor}. Specifically we will focus our study on the unique causal states, which we refer to as causal time steps too as each state have a corresponding position in the temporal sequence.
\begin{figure}[H]
  \centering
  \includegraphics[width=1.05\linewidth]{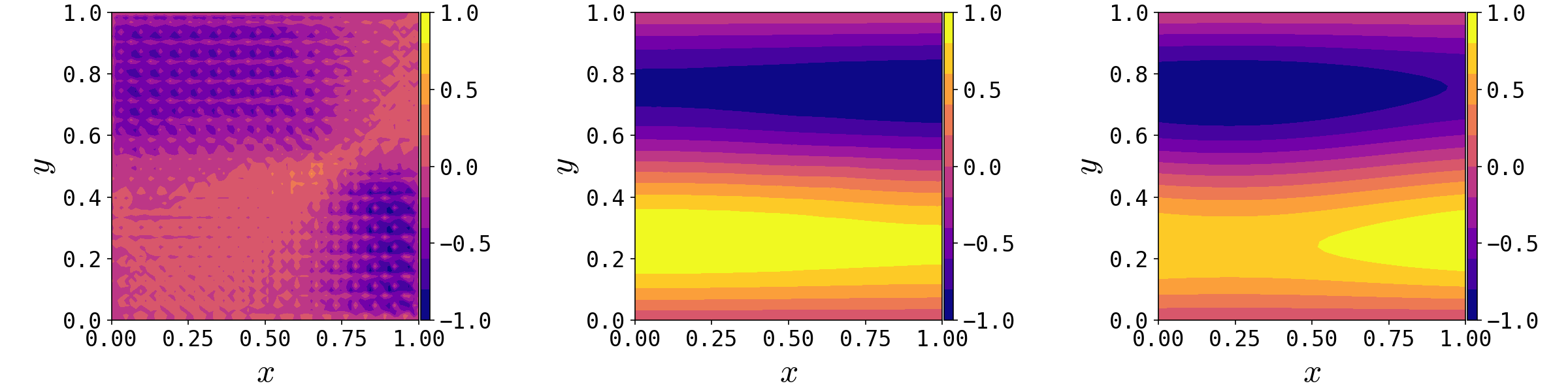}
  \caption{Mean SHAP field for the causal states in Fig.~\ref{states_lor}, $\overline{S_{s_1}(F)}$ , particularly the unique contribution. Secondly we show the Mean decoded field for $s_1$, $\overline{{\cal D}(s_1)}$. Finally, the last plot represents the Mean field for the selected causal states.}
  \label{fig:lorenz_bases_l3}
\end{figure}
\begin{figure}[H]
  \centering
  \includegraphics[width=1.05\linewidth]{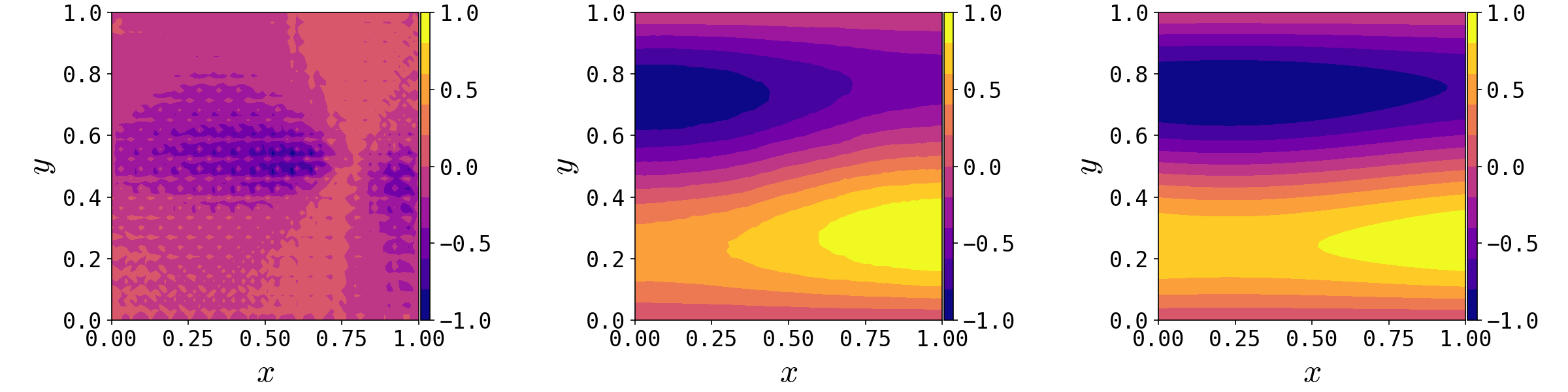}
  \caption{As above, we first find the Mean SHAP field for the causal states in Fig.~\ref{states_lor}, $\overline{S_{s_3}(F)}$ , particularly the unique contribution. Secondly we show the Mean decoded field for $s_3$, $\overline{{\cal D}(s_3)}$. Finally, the last plot represents the Mean field for the selected causal states.}
  \label{fig:lorenz_bases_l3_2}
\end{figure}

\begin{figure}[h!]
  \centering
  \includegraphics[width=1.05\linewidth]{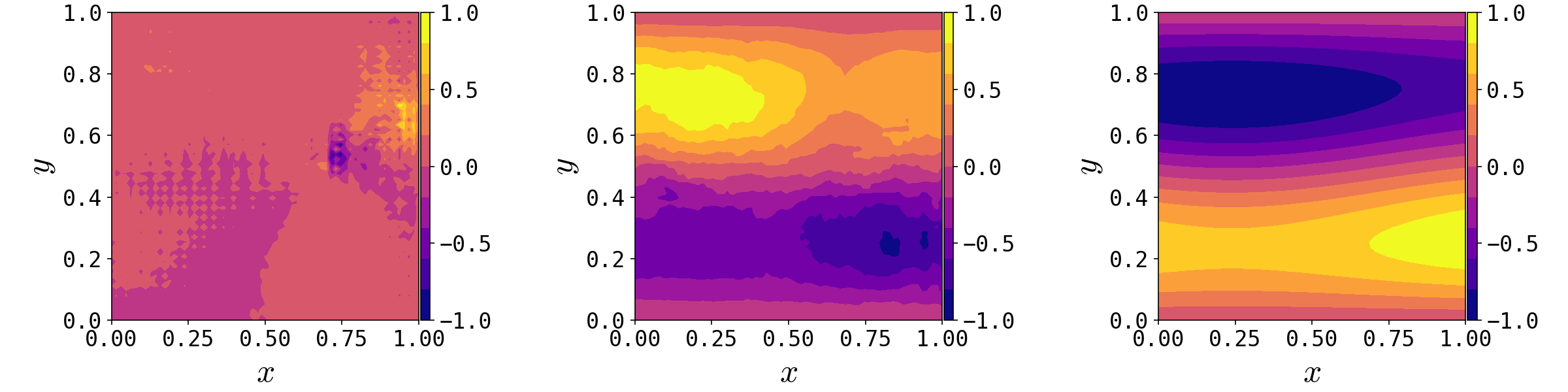}
  \caption{First, Mean SHAP field for the causal states in Fig.~\ref{states_lor}, $\overline{S_{l_1}}$ , particularly the unique contribution. Secondly, we depict the Mean decoded field for $l_1$, $\overline{{\cal D}(l_1)}$. Finally, the Mean field for the selected causal states.}
  \label{fig:lorenz_bases_l2}
\end{figure}

\begin{figure}[h!]
  \centering

  \includegraphics[width=1.05\linewidth]{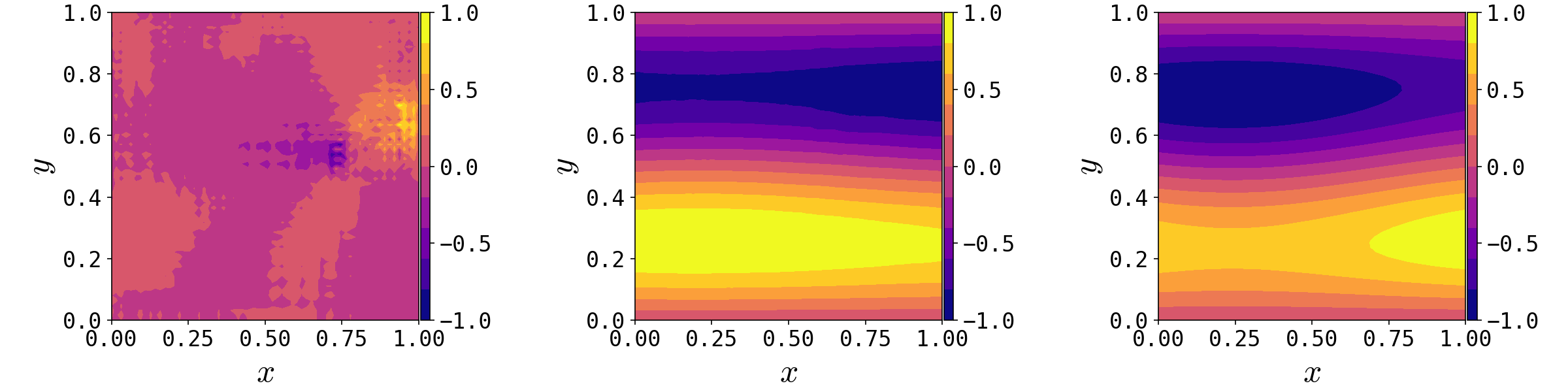}
  \caption{As above, we first show the Mean SHAP field for the causal states in Fig.~\ref{states_lor}, $\overline{S_{l_2}}$ , particularly the unique contribution. Secondly, we depict the Mean decoded field for $l_2$, $\overline{{\cal D}(l_2)}$. Finally, the Mean field for the selected causal states.}
  \label{fig:lorenz_bases_l2_2}
\end{figure}

\noindent When inspecting Figs.~\ref{fig:lorenz_bases_l3},~\ref{fig:lorenz_bases_l3_2} and \ref{fig:lorenz_bases_l2},\ref{fig:lorenz_bases_l2_2}, it can be observed that the individual mean decodings $\overline{{\cal D}(l_2)}$ and $\overline{{\cal D}(s_3)}$ produce non-linear representations of the original basis functions of the Lorenz system, presented in Fig.~\ref{fig:lorenz_bases}. Such behavior is expected as the $\beta$-VAE objective is to find the optimal non-linear basis that represents the original system. We know that the basis functions proposed by Lorenz~\cite{Lorenz1963} are the best linear representation of the system, but not the most informationally rich, clearly visible on the redundant terms depicted in Fig.~\ref{histos_lor}. The different latent variables yield similar configurations to that of the original basis, spanning geometrical SHAPs resembling a non-linear mix of the basis functions. On the other hand, the mean SHAP fields for the same temporal window of causal states appears to capture a \emph{butterfly} pattern which follows the symmetry change present on $F(x,y,t)$. Given the fact that the \emph{butterfly} SHAP is clearly visible in Figs.~\ref{fig:lorenz_bases_l3} and~\ref{fig:lorenz_bases_l3_2},  where $\overline{S_{s_1}(F)}$ and $\overline{S_{s_3}(F)}$ capture difference instances of the \emph{butterfly} oscillations. In order to make further claims, one would need to study the symmetries governing the dynamics of the system and how the critical points determine the transitions between dynamical regimes (symmetry transitions).

\noindent To conclude this section, we also investigate the decodification and SHAP fields at the unique causal temporal window for states presented in Fig.~\ref{l2}. Once again the $\overline{{\cal D}(l_1)}$ and $\overline{{\cal D}(l_2)}$ reconstruct non-linear representations of the original linear basis, while the mean SHAP fields $\overline{S_{l_1}(F)}$ and $\overline{S_{l_2}(F)}$ show a fragmented \emph{butterfly} pattern which could be attributed to a smaller temporal window (more data needed to converge the SHAP fields) or to the incapability of the model to create a latent space below the fractal dimension, making the model incapable of learning both symmetries and transitions completely.
\section{Results for the flow around a wall-mounted obstacle}
\label{obstacle}
\subsection{Spatial compression through $\beta$-VAE}
After introducing and validating the framework, here we analyze the $z=0$ plane of the flow around the wall-mounted obstacle. First, we reduce the dimensionality of the problem as follows:
\begin{align}
    {\cal E} &: {\cal U}^t \rightarrow \mathbf{{\cal L}}^t, \quad {\cal U}^t \subset \mathbf{{\cal O}},\\
     {\cal D} &:  \mathbf{{\cal L}}^t \rightarrow{ \tilde{\cal U}}^t, \quad \mathbf{{\cal L}}^t \in \mathbb{R}^{3}.
\end{align}

where $\mathbf{{\cal L}}^t$ is a time instant of the latent space corresponding to snapshot $t$ while ${\cal E}$ and ${\cal D}$ are the encoding and decoding operators. As introduced in \S.\ref{space}, the $\beta$-VAE has the capability of achieving accurate reconstructions while disentangling the latent space when regularizing the loss function through the Kullback--Leibler divergence by means of a hyperparameter $\beta$. Assuming gaussianity of the latent space, Eq.~\ref{eq:beta_vae_decomp} simplifies to:
\begin{equation}
    \mathcal{F}_\theta = \mathcal{L}_{\text{$\beta$-VAE}} =  \underbrace{{\lVert \tilde{\cal U} - \cal U \rVert}_2}_{\text{Accuracy}} - \frac{\beta}{2} \underbrace{\sum_{i=1}^{|{\cal L}|} 1 + \log(\sigma_i ^2) - \mu_i - \sigma_i^2}_{\text{KL divergence}}; \quad \beta = 5 \times 10^{-8} |{\cal L}| = 3
    \label{loss}
\end{equation}

Our database ${\cal O}$ is composed of 25,000 snapshots, ${\cal O} \in \mathbb{R}^{25,000 \times 288 \times 96 \times 2}$, used for training and 5000 more for testing, ${\cal D}_v$. Specifically we have the stremawise fluctuations field and the vertical fluctuations ${\cal U},{\cal V} \in \mathbb{R}^{25,000 \times 288 \times 96} $.   Regarding the training, we train for a total of 300 epochs, with a learning rate of $5 \times 10^{-4}$ with a batch size of 64. The training process takes around 2,050 s in an NVIDIA A100 graphics-processing unit (GPU) with 40GB RAM. After the process is concluded, we first study the orthogonality on the latent space, by means of the following metric, $\det(\mathrm{cov}(\mathcal L))$, while on the physical space the reconstructive capability will be tested by $E = \frac{\sum ({ \tilde{\cal U}} - {\cal{U}})^2}{\sum {\cal {U}}^2}$. 
% All the performance analysis will be conducted on ${\cal D}_v$.

\begin{figure}[H]
    \centering
    \begin{subfigure}{0.28\linewidth}
        \centering
        \includegraphics[width=1\linewidth]{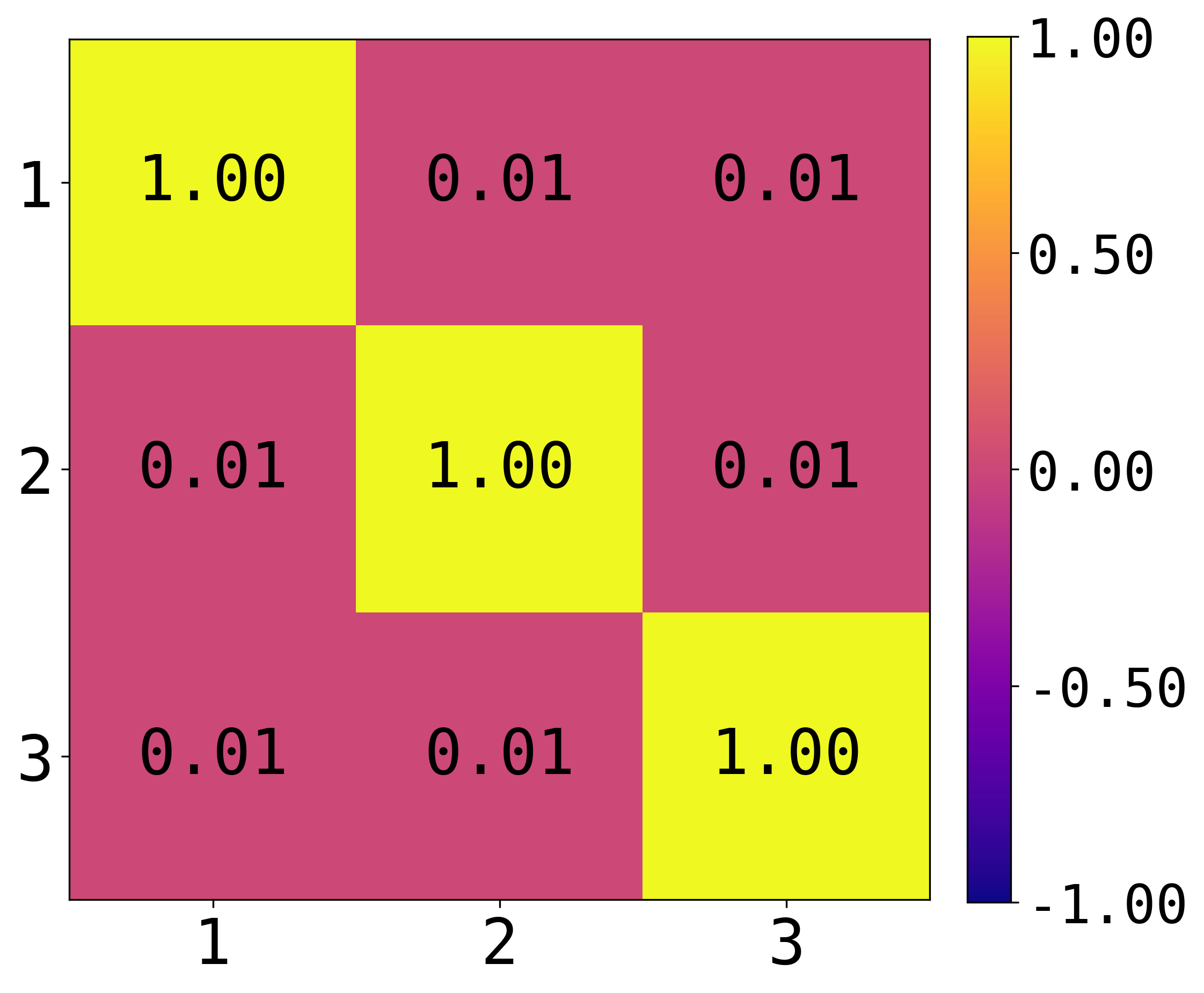}
        \caption{Correlation matrix for the model latent space, where $\det(\mathrm{cov}(\mathcal L))\approx 0.99$}
    \end{subfigure}
    \quad
    \begin{subfigure}{0.48\linewidth}
        \centering
        \includegraphics[width=1\linewidth]{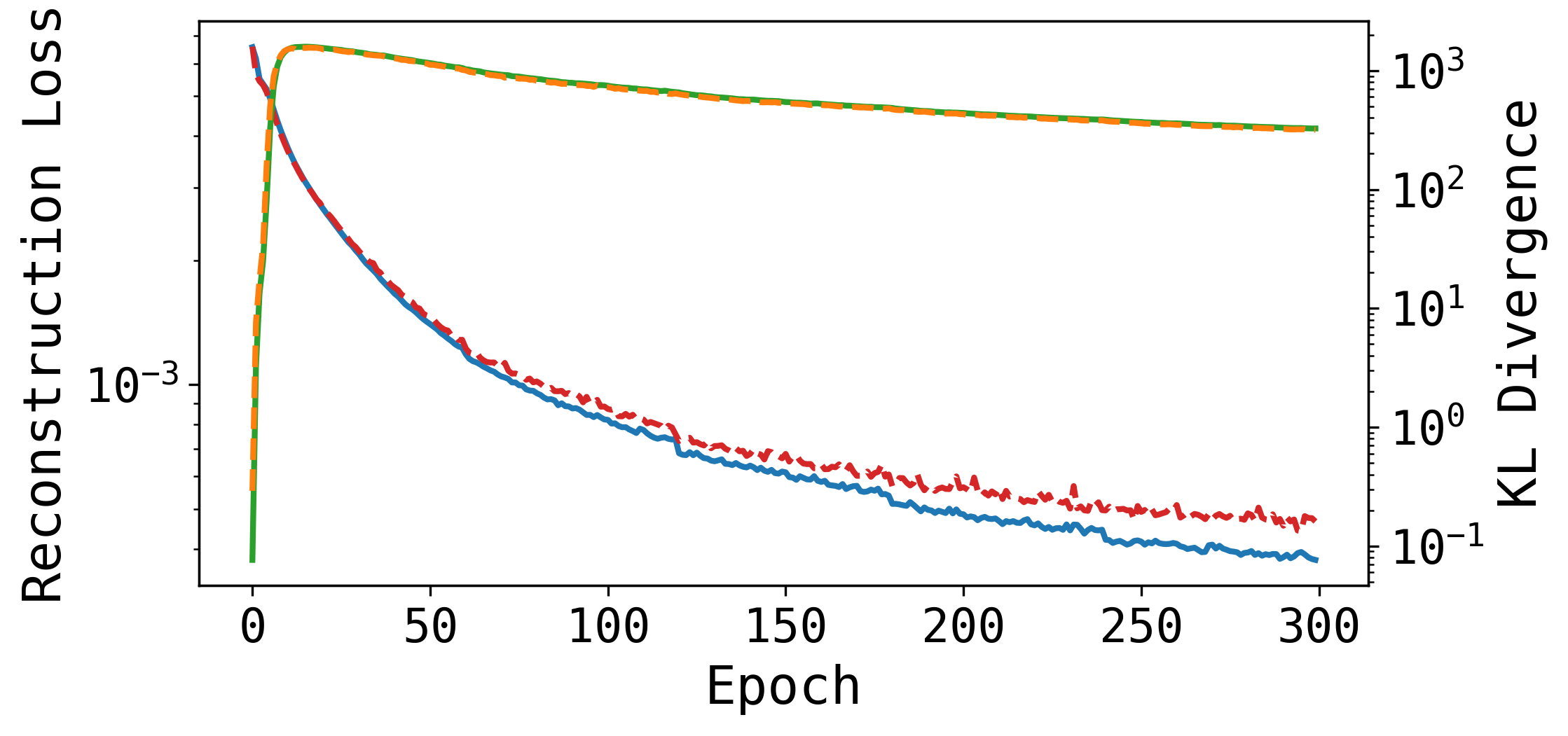}
        \caption{Loss and disentanglement convergence along the training process.}
    \end{subfigure}
    \caption{Performance matrix on the latent space and evolution of the loss along the the training and validation. Dashed line (yellow and red) for validation and straight (green and blue) for training.}
    \label{train}
\end{figure}

\begin{figure}[h!]
  \centering
  \includegraphics[width=1.05\linewidth]{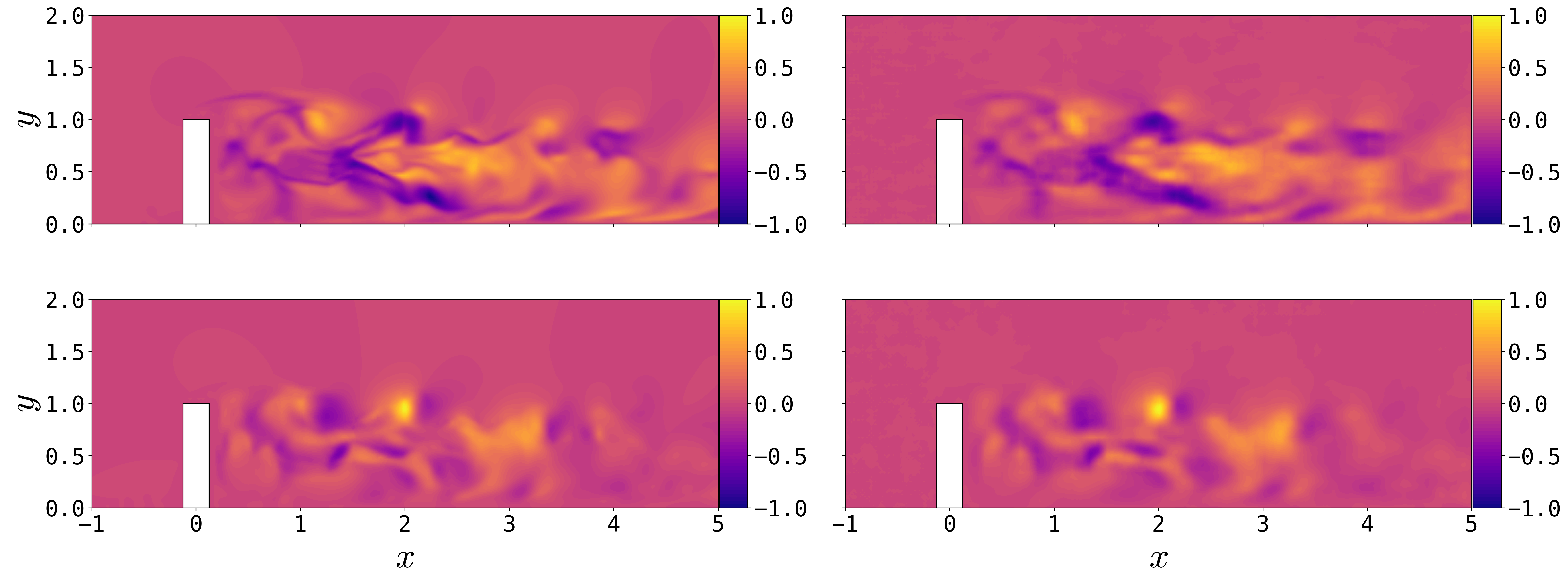}
  \caption{Total reconstructed energy on ${\cal D}_v$ for both channels: stream-wise (top) and vertical (bottom) fluctuations, depicting both ground truth (left) and reconstruction (right). }
  \label{ener}
\end{figure}

\noindent As observed in Fig.~\ref{train}, the orthogonality between latent variables is approximately $Tr(\mathrm{cov}({\cal L}))\approx 3$, where $\mathrm{cov}({\cal L})$ refers to the covariance matrix between latent variables, while the reconstruction loss begins to converge after 100 epochs to $10^{-4}$. In Fig.~\ref{ener}, we assess the performance of the reconstruction in physical space, achieving $93.5\%$ and $94.4\%$ of the reconstruction of the streamwise and vertical fluctuations, respectively. 
%\begin{figure}[H]
%    \centering
%    \begin{subfigure}{0.4\linewidth}
%%        \centering
%        \includegraphics[width=1\linewidth]{3D_VAE_3/Snap_stocha-deterministic_v5_25000n_3d_0e-4beta_originalconv_6Nf_512Fdim_256Ldimeluconvact_elu_10e-5LR_0e-5Wd64bs_300epoch_FalseES_0P_0.png}
%        \caption{Stream-wise fluctuations}
%    \end{subfigure}
%    \quad
%    \begin{subfigure}{0.4\linewidth}
%        \centering
%        \includegraphics[width=1\linewidth]{3D_VAE_3/Snap_stocha-deterministic_v5_25000n_3d_0e-4beta_originalconv_6Nf_512Fdim_256Ldimeluconvact_elu_10e-5LR_0e-5Wd64bs_300epoch_FalseES_0P_1.png}
%        \caption{Vertical fluctuations }
%    \end{subfigure}
%    \caption{Total reconstructed energy on ${\cal D}_v$ for both channels, ground truth (up) and reconstruction (down).}
%    \label{ener}
%\end{figure}
Next, we will enhance the understanding of each latent variable by ranking the elements of the latent space in terms of their contribution to the reconstruction,  as proposed in Ref.~\cite{EIVAZI2022117038}. The methodology uses the non-linear decoding of the $\beta$-VAE when ranking the latent vectors as follows:
\begin{align}
\tilde{{\cal L}}_1 &=  \sup_{i \in |{\cal L}|} \{E({\cal D}({\cal L}_i))\} \quad \tilde{{\cal L}}_1 \in \tilde{{\cal L}} \\
\tilde{{\cal L}}_2 &=  \sup_{i \in |\tilde{{\cal L}}/ c|} \{ E({\cal D}(\tilde{{\cal L}},{\cal L}_i))\}\\
\tilde{{\cal L}}_n &=  \sup_{i \in |\tilde{{\cal L}}/ c|} \{ E({\cal D}(\tilde{{\cal L}},{\cal L}_i))\} \quad  \forall n \in [1, |{\cal L}|],
\label{rank}
\end{align}
where $|\tilde{{\cal L}}| = 0$ when ranking the first mode, then as we rank each latent vector the magnitude of the latter increases. Finally, when completed the iterative procedure stated in Eq.~(\ref{rank}) both $\tilde{{\cal L}}$ and ${\cal L}$ will be equivalent, however permuted.
\begin{figure}
    \centering
    \begin{subfigure}{1.05\linewidth}
        \centering
        \includegraphics[width=1\linewidth]{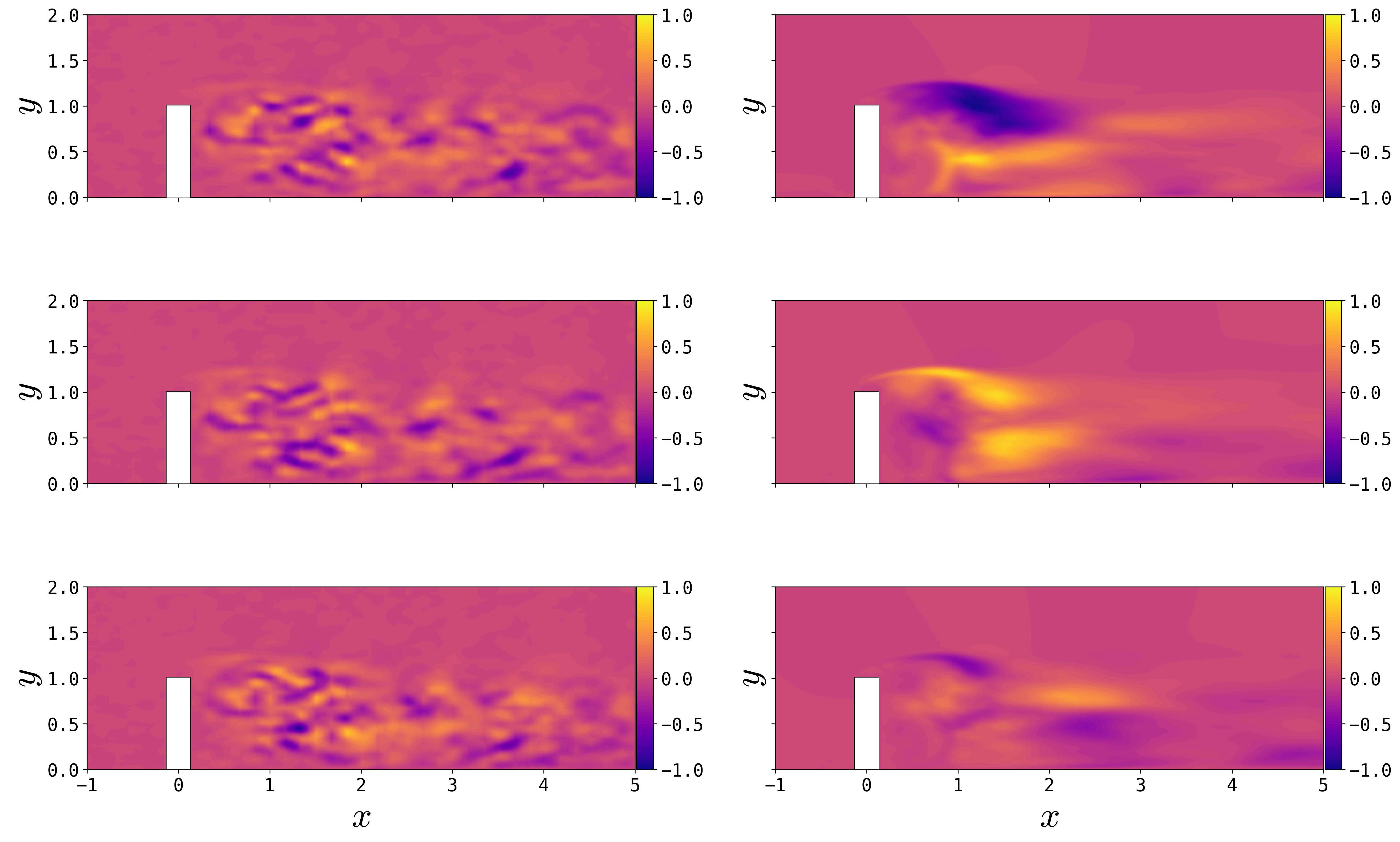}
        \caption{Stream-wise Modes}
        \label{stream-mode}
    \end{subfigure}
    \quad
    \begin{subfigure}{1.05\linewidth}
        \centering
        \includegraphics[width=1\linewidth]{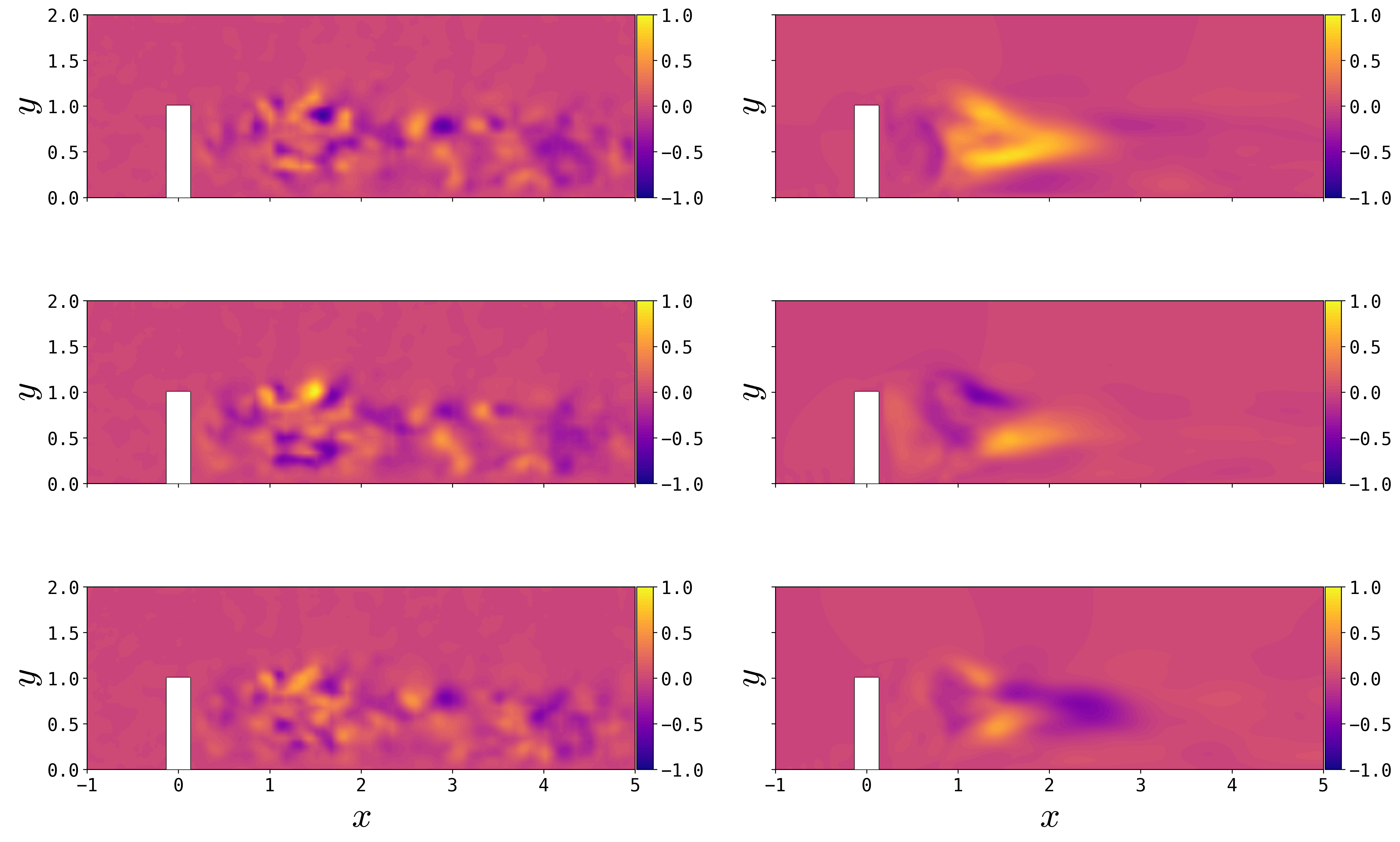}
        \caption{Vertical Modes}
    \end{subfigure}
    \caption{Mode comparison between $\beta$-VAE and POD for both channels.}
    \label{ranking}
\end{figure}

\noindent As observed in Fig.~\ref{ranking}, we represent side by side the decoded latent vectors against their corresponding POD modes, which by construction are also ranked from higher to lower contribution to the flow reconstruction. While the relationship between both POD and $\beta$-VAE modes is not straightforward, it is clear how the latter modes have richer dynamics than the former, due to the non-linear capability that the latent vectors have to capture non-linear dynamics, therefore the latent vectors exhibit a wider range of scales when decoded. In past works, the resemblances between non-linear and POD modes have been explored, but in this study we will focus only on the $\beta$-VAEs modes~\cite{SoleraRico2024,WANG2024109254} and their causality.

\subsection{Latent Causality}
To enhance our analysis, we conduct a causal study on the temporal signals of the latent space, understanding that the low-dimensional manifold behaves as a stochastic dynamical system. In order to do so, we will investigate different lags to assess causality.

\noindent We compute a causal study on the latent space searching for the most informative lag. In doing so, we investigate how the different terms of the SURD decomposition evolve as a function of lag,  where the lag denotes the temporal shift between a source variable’s past state and a target variable’s present state: in SURD, one quantifies how information from a source uniquely, redundantly, or synergistically contributes to predicting a target variable once all other sources are accounted for.  Equivalently, the lag measures the delay between eliminating the influence of the source at time $t-\rm{lag}$ and assessing its impact on the target at time $t$, thus tracing the temporal scale over which causal interactions unfold. In order to bound the lag search we use prior physical knowledge of our system, i.e. the maximal vortex-shedding frequency in the wake given by the Strouhal number, $St \approx 0.064$, which in simulation steps is approximately $3500$ time steps.

%It is interesting as these lag is a quarter of the vortex shedding, $St=0.1$, reported in the bibliography for a similar flow case~\cite{vinuesa2015direct}.
\begin{figure}[H]
    \centering
    \begin{subfigure}{0.48\linewidth}
        \centering
        \includegraphics[width=1\linewidth]{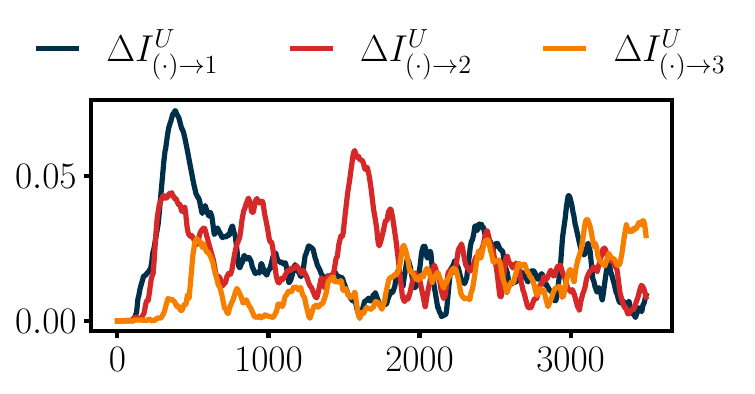}
    \end{subfigure}
    \quad
    \begin{subfigure}{0.48\linewidth}
        \centering
        \includegraphics[width=1\linewidth]{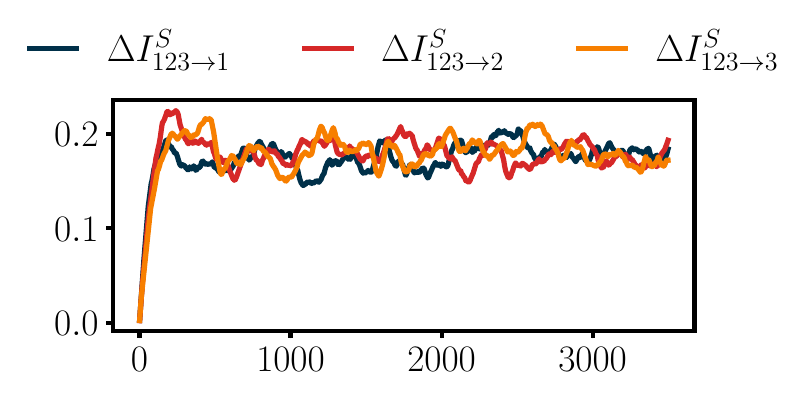}
    \end{subfigure}
    \caption{Total unique (left) and Synergistic (right) contributions for each latent variable as a function of lag.}
    \label{syni}
\end{figure}
\begin{figure}[H]
    \centering
    \begin{subfigure}{0.48\linewidth}
        \centering
        \includegraphics[width=1\linewidth]{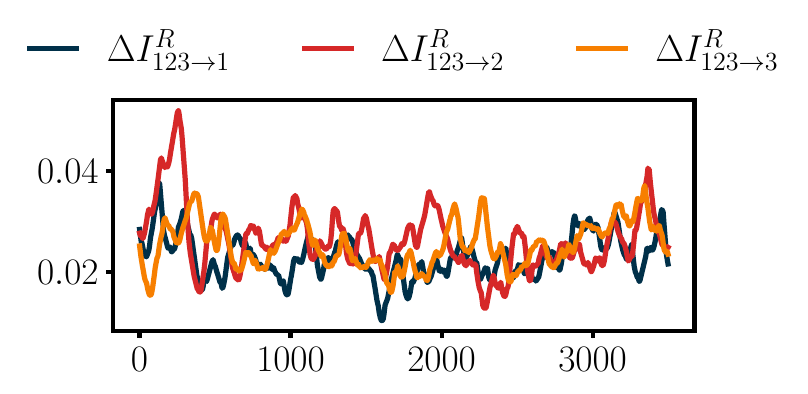}
    \end{subfigure}
    \quad
    \begin{subfigure}{0.48\linewidth}
        \centering
        \includegraphics[width=1\linewidth]{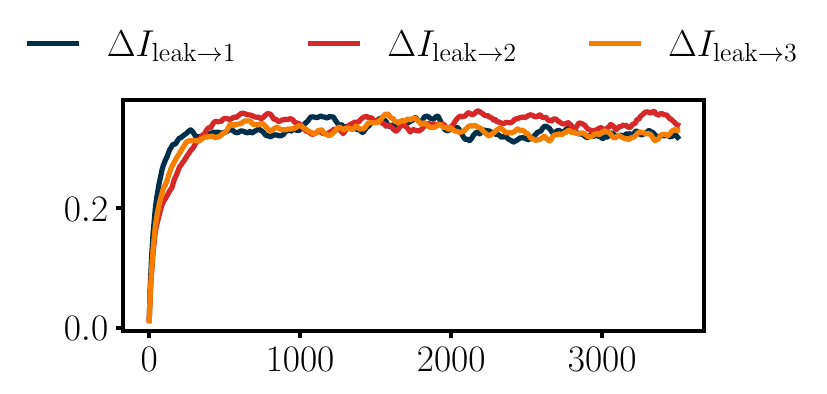}
    \end{subfigure}
    \caption{Total redundant contribution (left) and leak (right)  for each latent variable as a function of lag.}
    \label{redu}
\end{figure}

\noindent It is important to recall that SURD decomposes the mutual information into three distinct contributions: unique (\( \Delta I^U \)), redundant (\( \Delta I^R \)) and ynergistic (\( \Delta I^S \)), as depicted in Figs.~\ref{syni},\ref{redu}. Unique contributions indicate information exclusive to a single variable, redundant contributions identify information commonly shared by multiple variables and synergistic contributions describe cooperative information that only emerges from combined groups of variables, and cannot be attributed individually to any single variable.

\noindent As depicted in Figs.~\ref{syni},\ref{redu}, the leak and synergistic contributions converge to two different values, which in principle are bounded by the informational nature of the system, in this case the invariant set spanned by the statistical properties of the flow. Therefore, we focus on understanding the unique and redundant contributions when interpreting the physical space through the latent one. In particular, we investigate the most informative lags for each latent variable when considering the latter contributions. We will repeat this analysis for the redundant contribution also and inspect these lags in the physical space through the SHAP structures.

\begin{figure}[H]
    \centering
    \begin{subfigure}{0.48\linewidth}
        \centering
        \includegraphics[width=1\linewidth]{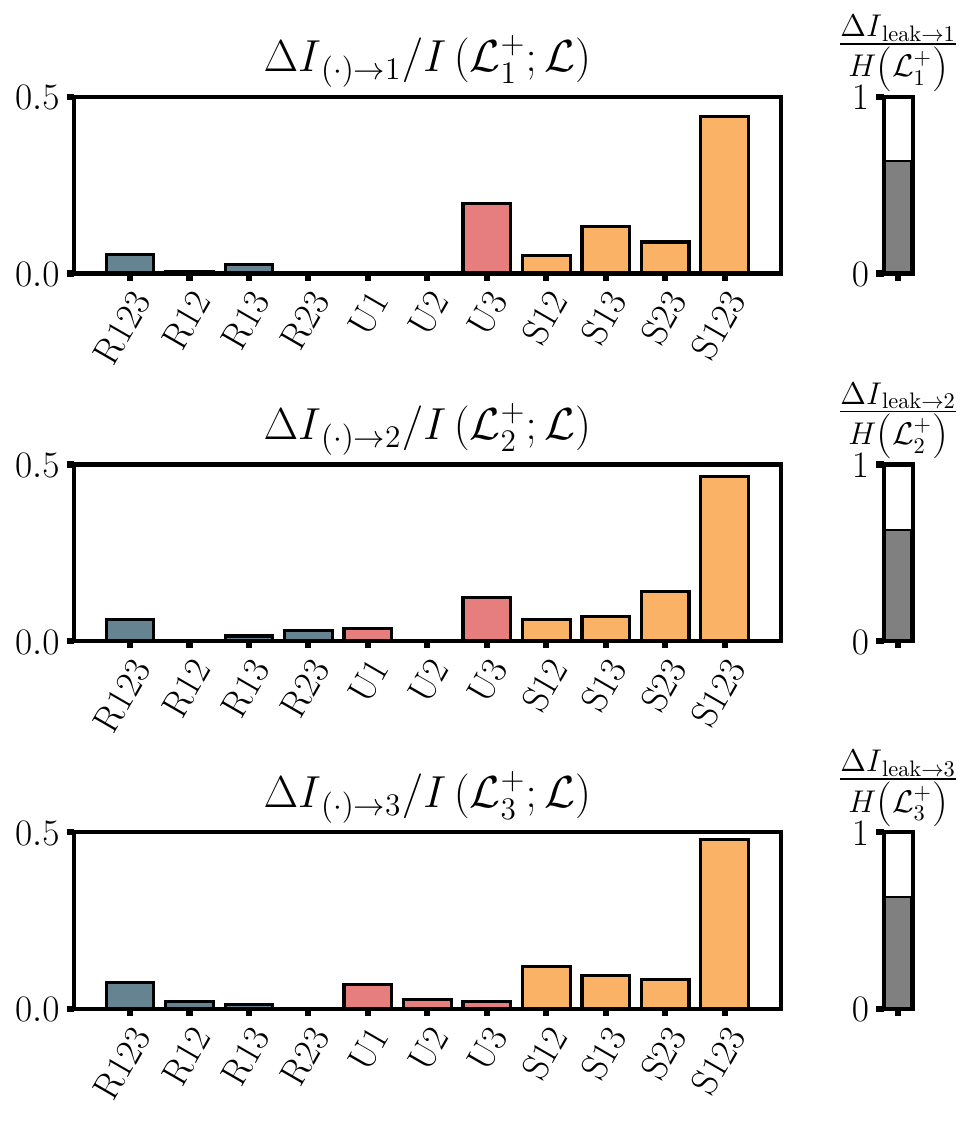}
    \end{subfigure}
    \quad
    \begin{subfigure}{0.48\linewidth}
        \centering
        \includegraphics[width=1\linewidth]{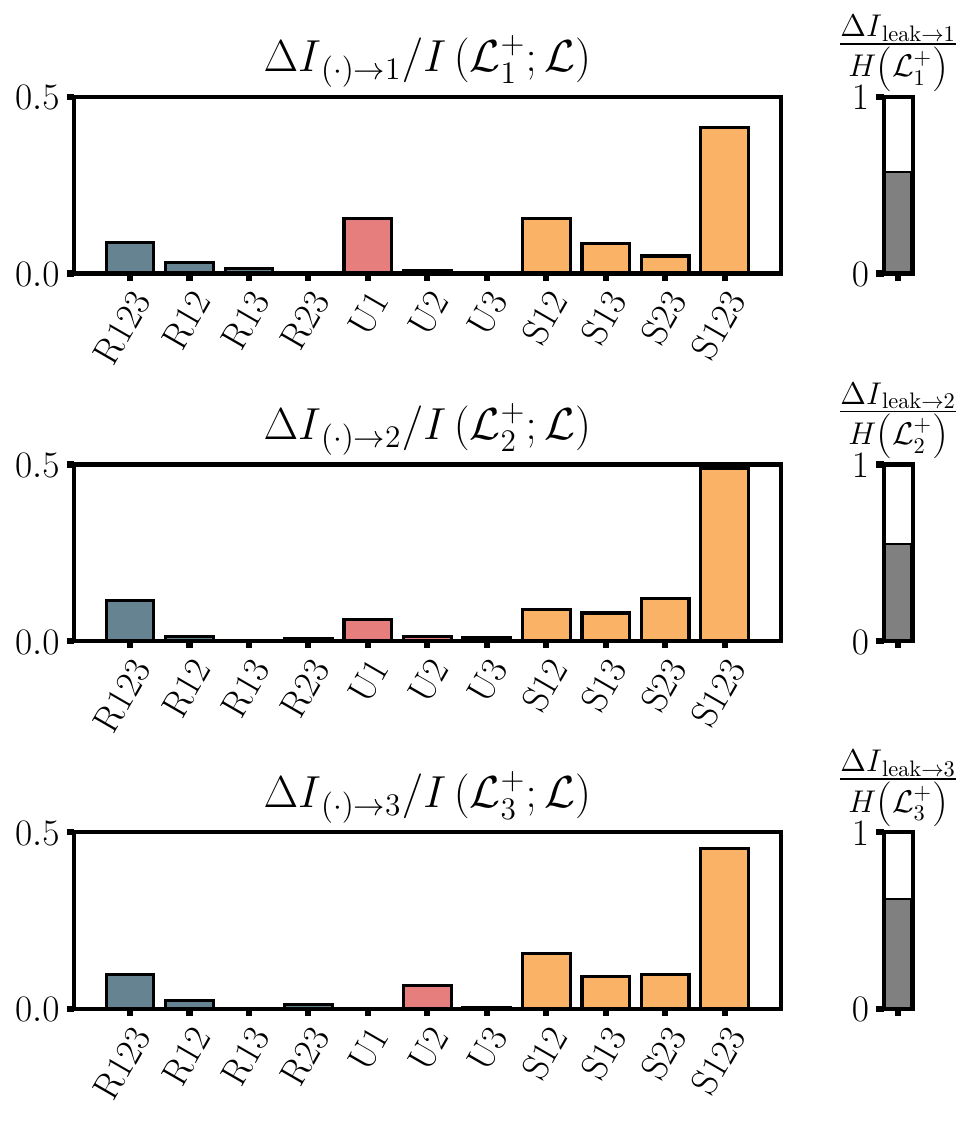}
    \end{subfigure}
    \caption{SURD analysis conducted with 10 samples per bin for the most informative lags. unique contribution maximization at lags 381, 1569, 3103 (left) and redundant contribution maximization at lags 128, 257 and 3344 (right) from to bottom}
    \label{histos}
\end{figure}

\noindent When studying the unique contributions, one can observe in the left histograms of Fig.~\ref{histos} the contribution from latent variable $\mathbf{\cal{L}}_3$, $U3$ as the dominant one for both $\mathbf{\cal{L}}_3$ and $\mathbf{\cal{L}}_3$ respectively. On the other hand the mutual information of variable $\mathbf{\cal{L}}_3$ has similar contributions from all elements of the latent space $\cal{L}$. This analysis is repeated for the redundant contribution in Fig.~\ref{histos}. For the first two latent variables the dominant unique contribution is coming from $\mathbf{\cal{L}}_1$ while for $\mathbf{\cal{L}}_3$ the unique contribution arises from latent variable $\mathbf{\cal{L}}_2$. Regarding Synergectic and redundant contributions, both $S123$ and $R123$ remain as the most dominant terms in their respective categories. The latter is expected, since the the high-dimensional 2D flow is being compressed into $\cal{L}$ where each latent variable learns different regions of the flow, and they will collaborate in time to sustain the turbulent flow and its coherent motions. Therefore, we will further investigate in the following section the unique and redundant contributions at their most informative lags respectively with the usage of the SHAP (encoder) and $\beta$-VAE (decoder) fields .

\subsection{Explainability}
\begin{figure}
    \centering
    \includegraphics[width=1\linewidth]{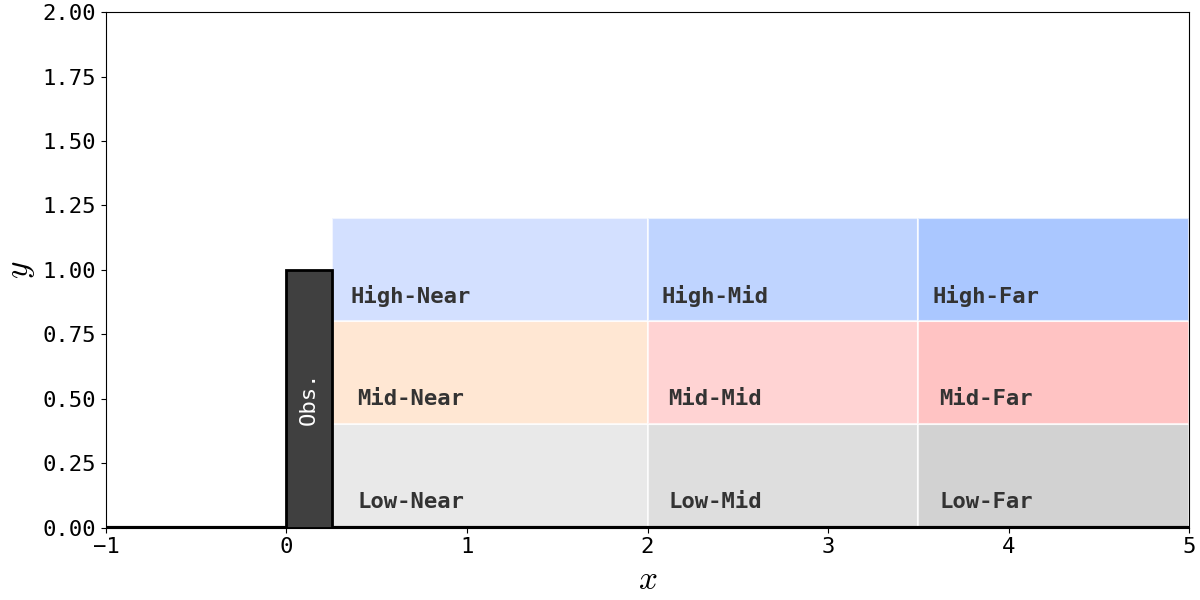 }
        \caption{Wake subdomain labeling: in $x$ (near,mid,far) and $y$ (low,mid,high) directions.}
    \label{wake}
\end{figure}
\label{IE}

\noindent To facilitate the interpretation of the spatial structures identified by the network, we introduce a domain segmentation strategy illustrated in Fig.~\ref{wake}. Characterizing the wake behind the obstacle through such zonal decomposition is critical, as distinct physical phenomena dominate different regions of the flow; for instance, the near-wake is typically governed by strong recirculation and shear-layer instabilities, whereas the far-wake is characterized by viscous decay and structure recovery. By discretizing the domain into streamwise (near, mid, far) and wall-normal (low, mid, high) subdomains, we can more precisely attribute the learned latent representations to specific localized flow features rather than global statistics. In the following subsections, we utilize this spatial framework to analyze the interpretability of the model: first by examining the generative capabilities of the \emph{\textbf{Decoder}} to reconstruct energetic modes, and subsequently by applying feature-selection techniques to the \emph{\textbf{Encoder}} to identify the causal regions driving these representations.

\subsubsection{Decoder}
Here we investigate the physical mechanisms driving the flow over the obstacle through both the encoder, $\cal{E}$, and decoder, $\cal{D}$. Specifically, using the SHAP values~\cite{Cremades2024}, it is possible to interpret the latent space by categorizing the importance of each pixel in the input of the $\beta-VAE$ when constructing each latent variable $\mathbf{\cal{L}}^t$ $\forall t\in T$, where $T = 25,000$. However, we will start our investigation by decoding each latent variable independently, $\cal{D} :$ $ \mathbf{\cal{L}}_i^t \rightarrow $ $\mathbf{\cal{U}}_i^t$, where $\mathbf{\cal{U}}_i^t$ is the physical mode at time $t$ for latent variable $i$. It is important to mention that the de-decoding is performed for the snapshot at time $t$ after re-parametrization, setting the rest of the latent values to $0$. The main objective of this section is to relate the causal events detected in the latent space back to the physical domain. Once the main structures or subdomains are identified one could design a control strategy, based on the causality of the low-dimensional manifold. We will focus on the energy domain, specifically on the instantaneous turbulent kinetic energy (TKE) as both $\beta$-VAE input channels are considered; $TKE_i^t =$ $(\mathbf{\cal{U}}_i^t)^2/2 $ $+$ $(\mathbf{\cal{V}}_t)^2/2 $. It is interesting how the concentration of energy differs between all three latent variables once we compute the average along the whole database, localizing the information along different sections of the boundary layer.

\begin{figure}[H]
    \centering
    \begin{subfigure}{0.6\linewidth}
        \centering
        \includegraphics[width=1\linewidth]{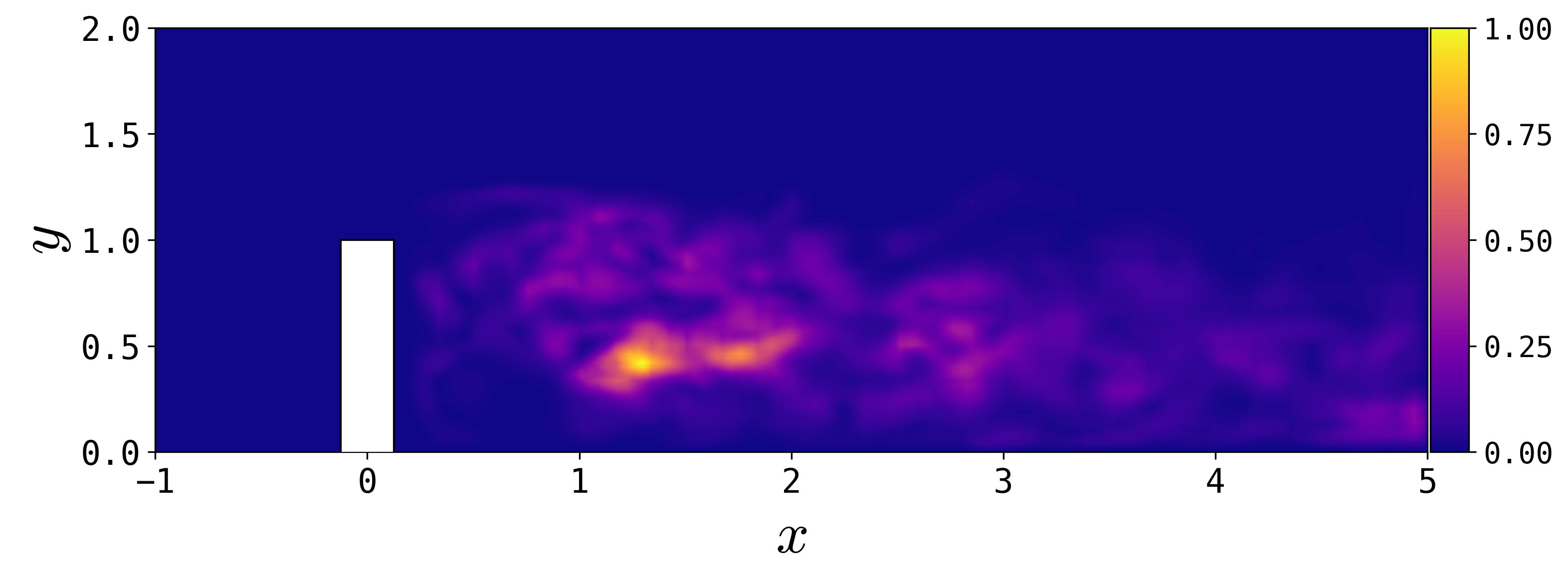}
    \end{subfigure}
    \quad
    \begin{subfigure}{0.6\linewidth}
        \centering
        \includegraphics[width=1\linewidth]{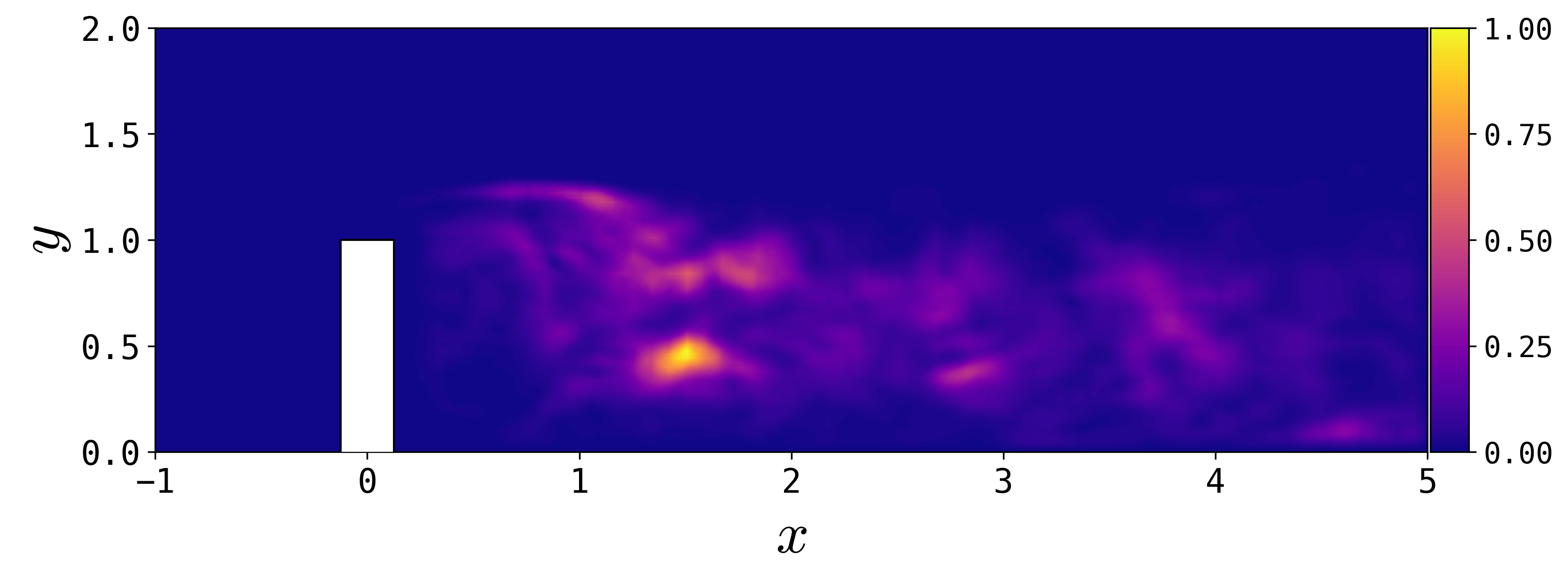}
    \end{subfigure}
    \quad
    \begin{subfigure}{0.6\linewidth}
        \centering
        \includegraphics[width=1\linewidth]{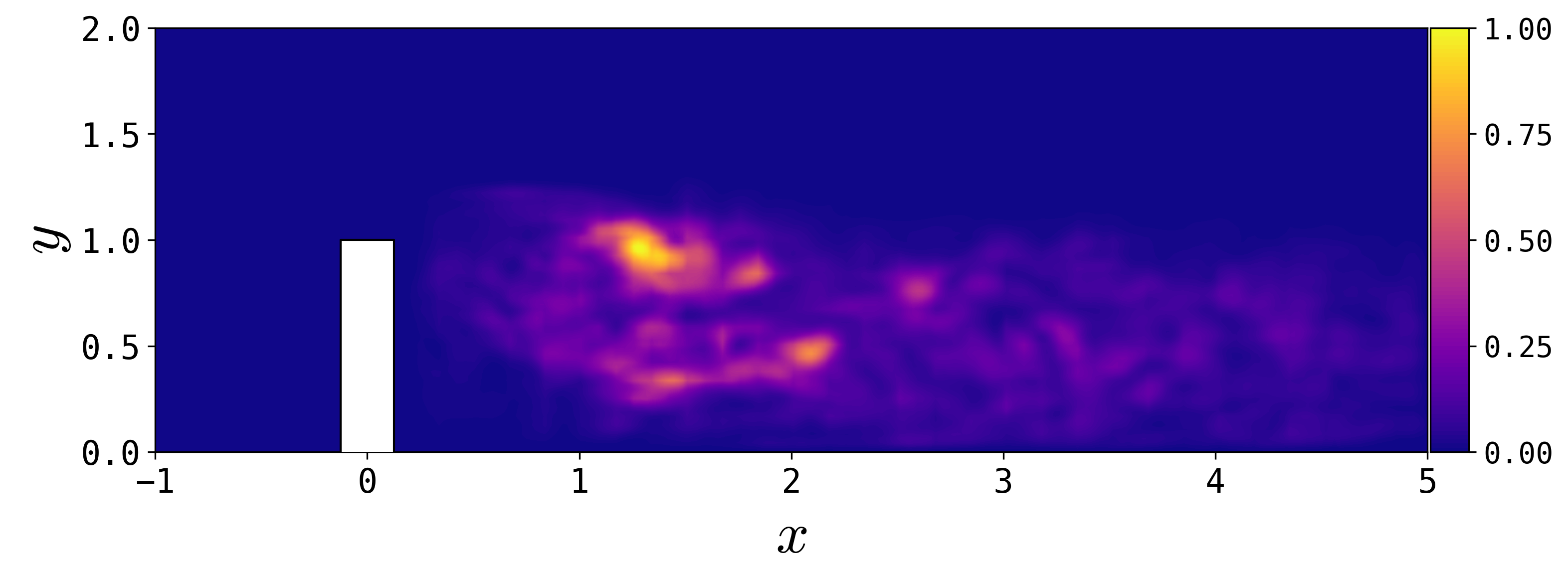}
    \end{subfigure}
    \caption{Mean Reconstructed $TKE$ for $\mathbf{\cal{L}}_1$ (top),  $\mathbf{\cal{L}}_2$ (mid) and  $\mathbf{\cal{L}}_3$ (bottom)}
    \label{ener1}
\end{figure}
%\begin{figure}[H]
%%    \centering
 %   \includegraphics[width=0.8\linewidth]{final_comp/meanphy_v3.png}
%    \caption{Mean Reconstructed $TKE$ for $\mathbf{\cal{L}}_3$ }
%    \label{ener2}
%\end{figure}

\noindent In Fig.~\ref{ener1} it it can be observed that the energy of the decoded latent variables concentrates along the wake in $x\in[1,4]$ for $\mathbf{\cal{L}}_1$ and $\mathbf{\cal{L}}_3$. Their spatial location along the vertical direction differs, the former is localized mainly on the lower-mid section of the wake while 
the latter exhibits higher energy on the top of the wake close to the obstacle height $y/h=1$. Furthermore, when inspecting the mean decoded turbulent kinetic energy for $\mathbf{\cal{L}}_3$, $\overline{TKE_3}$, one can observe a structure related to the three classical tip vortices arising from the interaction of the flow with the obstacle and the spanwise vortices. The end of this structure where the main vortical mechanisms arise are mainly identified in the upper wake of $\overline{TKE}_3$, a fact that suggests on the other hand that both $\mathbf{\cal{L}}_2$ and $\mathbf{\cal{L}}_1$ are indeed capturing the spanwise~\cite{Monnier2018BLM} vortex arising from the bottom of the obstacle,  (although they may be connected with different subregions of this vortical structure.)

\subsubsection{Encoder}
To extend the interpretation, one can use the $\mathcal{E}$ network to identify the most important features on the input when constructing each of the latent variables. This approach provides a different perspective to the decoder $\mathcal{D}$, however, it relies on further tools to localize the most informative regions when constructing the latent space. Such tools are known as \textit{feature-selection methods}. Among these, SHAP values \cite{lundberg2017unified}, particularly gradient SHAP~\cite{Cremades2024}, are particularly powerful as they combine integrated gradients with SHAPley value theory to provide smooth, consistent attributions. Other widely-used methods include Integrated Gradients \cite{sundararajan2017axiomatic}, which compute the path-integrated gradients between a baseline and the input, and occlusion sensitivity analysis \cite{zeiler2014visualizing}, a model-agnostic approach that perturbs input regions to assess their effect. On this investigation, the SHAP values introduced on \S~\ref{SHAP_method} will be the selected tool to interpret the construction of the latent space. We compute the SHAP fields for each of the latent variables and for both streamwise and vertical fluctuations. The reference for the gradient SHAP is a subset of 7000 snapshots which we use as baseline when calculating the SHAPs~\cite{Cremades2024}. Once the feature-selection method is used we perform a percolation analysis as shown in Eq.~(\ref{percenc}) where the percolation factor is set as $H=[1.14,1.36.1.14]$ for each latent variable respectively. Note that $H$ is set to maximize the number of structures obtained after the percolation. Once the percolated SHAP fields are obtained, one can investigate $STKE_i^t = (\mathbf{\tilde{S_i^t}(u)}^2)/2 + (\mathbf{\tilde{S_i^t}(v)}^2)/2 $ which is defined as the structured turbulent kinetic energy. In particular, we first inspect the unique $STKE_i^t$ for each latent variable retaining all energy points which lie outside the intersection: $STKE_1^t \cap STKE_2^t \cap STKE_3^t $.
\begin{figure}[h!]
    \centering
    \begin{subfigure}{0.6\linewidth}
        \centering
        \includegraphics[width=\linewidth]{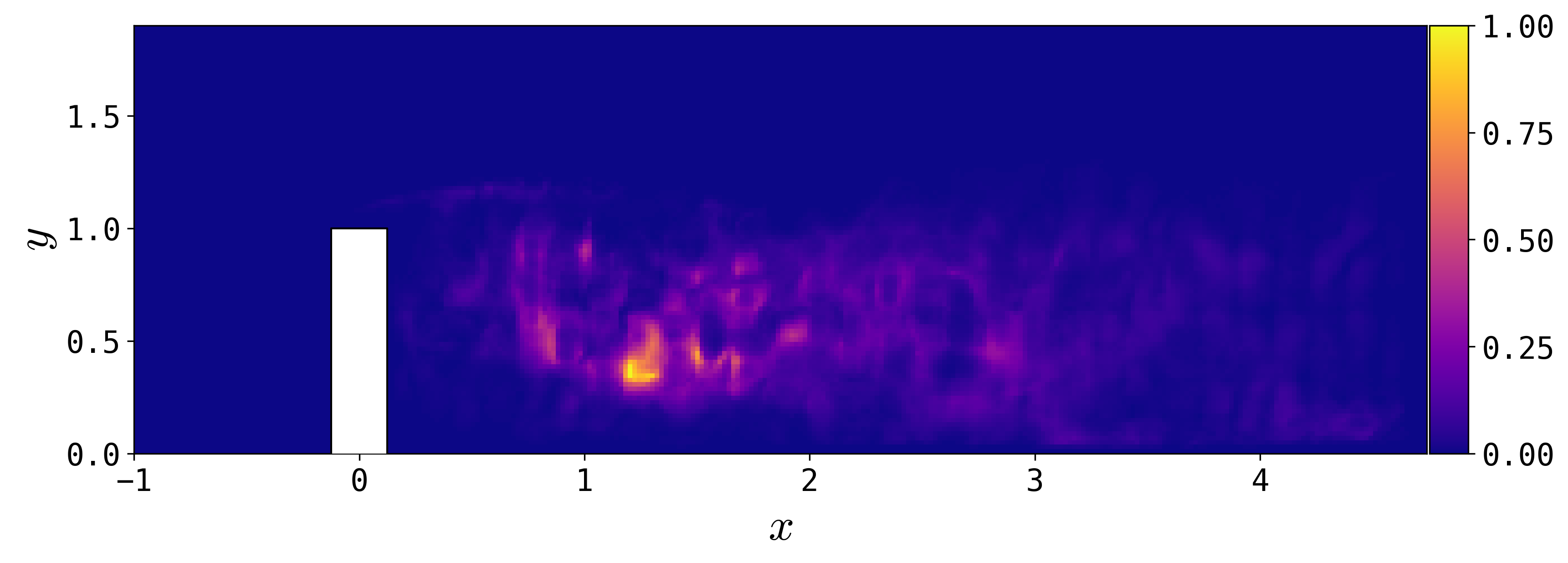}
        \caption{}\label{fig:ener_a}
    \end{subfigure}
    \quad
    \begin{subfigure}{0.6\linewidth}
        \centering
        \includegraphics[width=\linewidth]{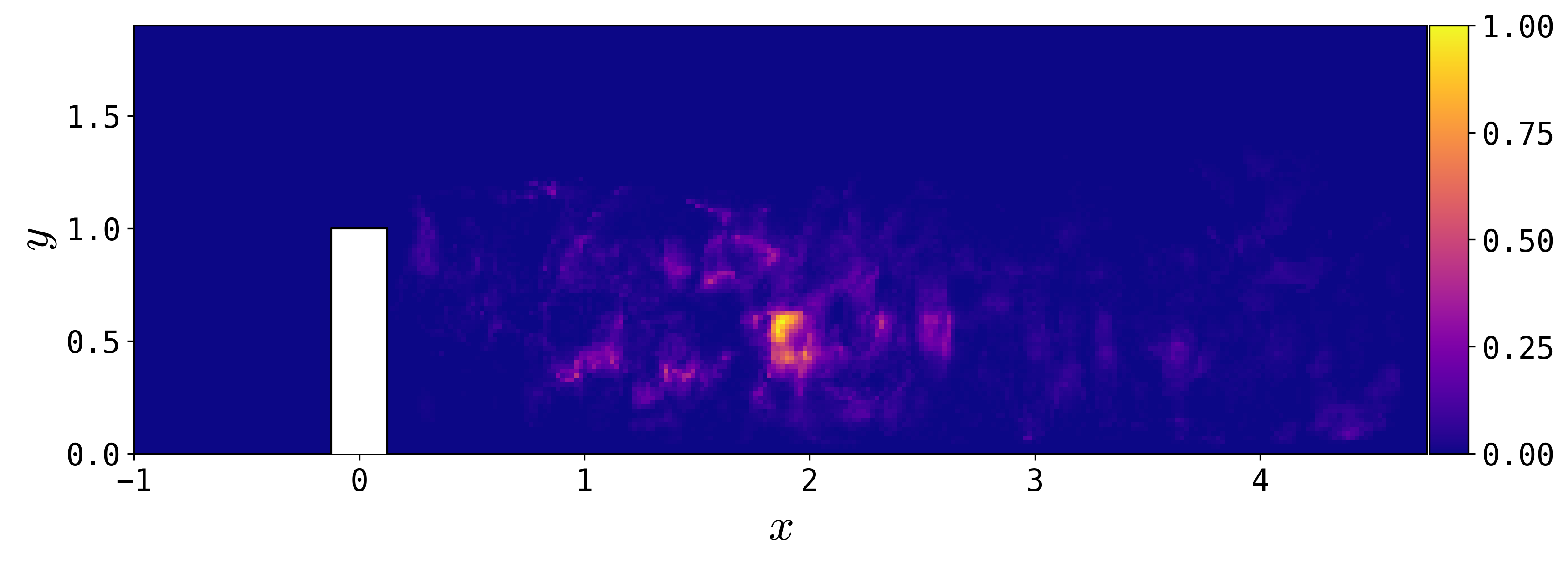}
        \caption{}\label{fig:ener_b}
    \end{subfigure}

    \begin{subfigure}{0.6\linewidth}
        \centering
        \includegraphics[width=\linewidth]{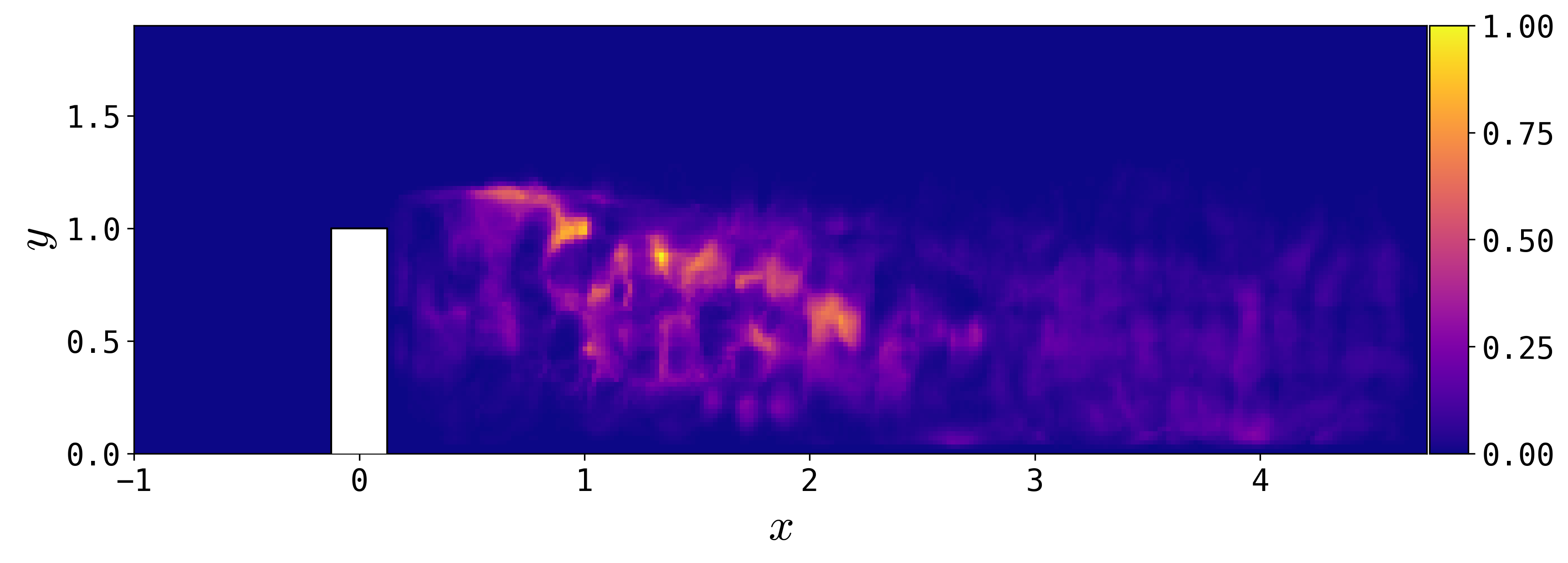}
        \caption{}\label{fig:ener_c}
    \end{subfigure}
    \begin{subfigure}{0.6\linewidth}
        \centering
        \includegraphics[width=\linewidth]{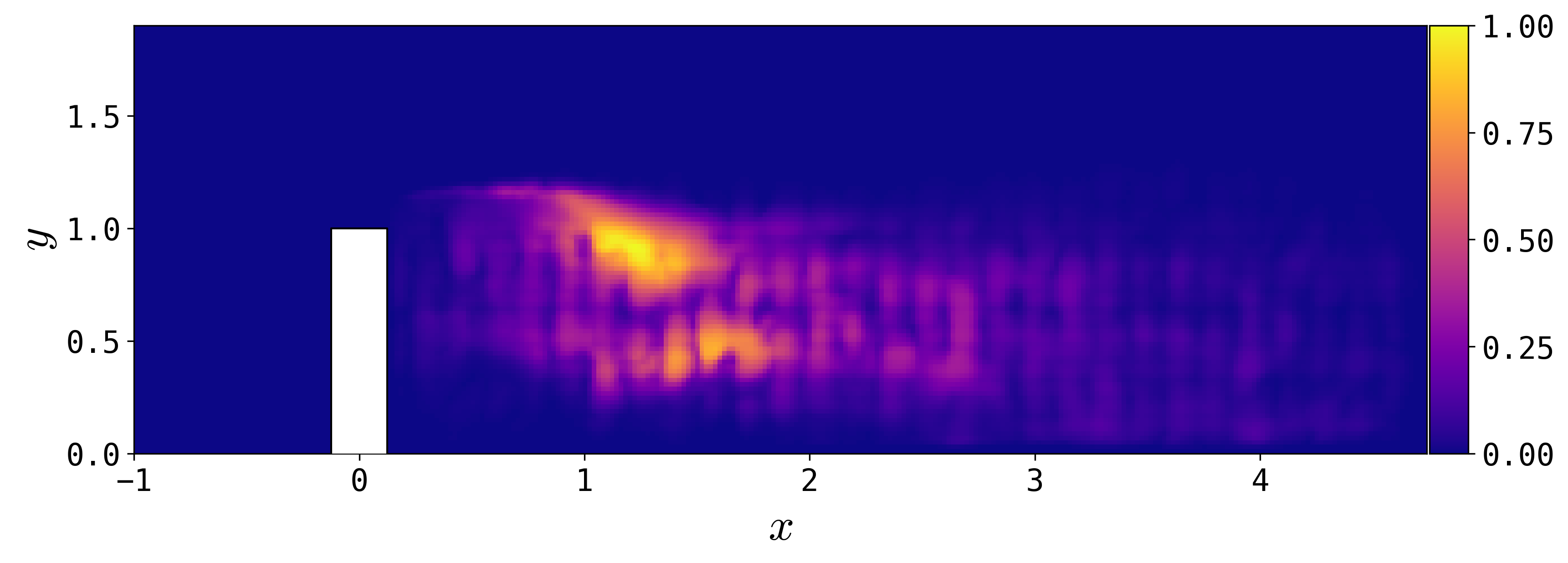}
        \caption{}\label{fig:ener_d}
    \end{subfigure}

    \caption{Mean unique reconstructed $\overline{STKE}$ for $\mathbf{\mathcal{L}}_1$ (a), 
    $\mathbf{\mathcal{L}}_2$ (b) and $\mathbf{\mathcal{L}}_3$ (c). Lastly (d),
    the mean reconstructed $\overline{STKE}_1 \cup \overline{STKE}_2 \cup \overline{STKE}_3$.}
    \label{eneruni1}
\end{figure}

%\begin{figure}[H]
%    \centering
%    \includegraphics[width=0.4\linewidth]%{Results_SHAP/energy_one_2.png}
%    \caption{Mean unique reconstructed $STKE$ for $\mathbf{\cal{L}}_3$. }
%    \label{eneruni2}
%\end{figure}

\noindent By assessing the unique $\overline{STKE_i}$ for each latent variable, we observe that all three latent variables depicted in Fig.~\ref{eneruni1} are located in different regions of the wake. Similarly to the results introduced above and generated by the decoder, the latent space segments the spatial domain into six sections: Low, Mid and High relative to the bottom wall combined with Near, Mid and Far relative to the obstacle. One can identify the top wake and the tip vortex at $\overline{STKE_2}$, while also capturing the evolution of this vortical structure into the mid-high wake of the flow. On the other hand $\overline{STKE_1}$ and $\overline{STKE_3}$ are concentrated on the low-near wake and the low-mid wake respectively, arising as a vortex from the base of the object to the top~\cite{WangZhou2009FiniteLength}. One can clearly conclude that all three latent variables are associated with different sections of the near-obstacle energetic subdomains; to gain a further insight into the dynamical behavior of the flow one could investigate the temporal evolution of the SHAP fields. It is well-known that when studying boundary layers and wake flows with different roughness levels or pressure gradients, the main particularities arise in the near-wake region~\cite{Gery,Waleffe2009Exact}, where the flow structures are supposed to be strongly correlated with the obstacle.

\noindent It is interesting to observe that $STKE_1^t \cup STKE_2^t \cup STKE_3^t$ is concentrated on the near-mid wake of the obstacle, clearly capturing the like tip-vortex structure on the high-wake and the energetic concentration on the mid-low wake.

\subsubsection{X-CAL on 2D Obstacle}
\begin{figure}[H]
    \centering
    \includegraphics[width=0.5\linewidth]{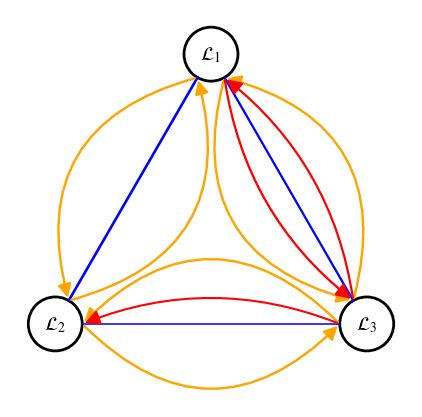}
    \caption{Global X-CAL graph, $\mathbf{\cal{G}}_c$, based on the most dominant contributions when maximizing the unique contribution. We sketch the main mechanisms identified on the histograms depicted in Fig.~\ref{histos}, where the main dominant unique contributions are depicted in red.}
    \label{graph}
\end{figure}

In this section we will explore the physical mechanisms the X-CAL framework highlights as relevant in the physical or input space. To do so, we will combine both causal tools SURD and SHAP. For this analysis, we will focus on the unique and synergistic causality depicted in Fig.~\ref{histos}. First, we will explore the dominant unique contribution for each latent variable, by inspecting the temporal windows were the variable producing that unique contribution is maximum, more details on the window analysis will be found on the following section. Secondly, the same unique term will be studied by splitting the latter in causal and non-causal, producing a temporal window of causal unique events. Note that both approaches are complementary and the same causal mechanisms through different perspectives
\paragraph{Windows analysis of unique causality:}
\label{study}
Firstly, the bridge between physical and latent spaces will be explored by studying the most informative lags presented in Table~\ref{tab:lags}, it is important to notice that each lag has been allocated a spatial location in the wake domain. We will study the unique lags through both; the SHAP and decoder fields as introduced in \S~\ref{IE}, specially exploring the unique causal relations depicted in Fig.~\ref{graph}. Note the visible similarities between $TKE$ and $STKE$ fields, which will provide an image of the mean spatial location for each latent variable before we study the most informative lags. Both $\overline{STKE_1}$ and $\overline{TKE_1}$ identify high-energy structures at $x/h \approx1$, near to the obstacle bottom around the bottom-mid section of the wake, a consequence of the base-vortex lift-up and mixing with the flow from the tip vortex in the mid-wake. On the other hand, $\overline{TKE_2}$ clearly captures the leading section of the base-vortex structure arising from the obstacle tip, including a mid energetic structure near those reconstructed by $\mathbf{\cal{L}}_1$, and the main shedding downstream in the mid-far wake. Comparing with $\overline{STKE_2}$, only the trail of the bottom-mid wake energetic structure present in Fig.~\ref{eneruni1} is captured, with sporadic energy concentrations identified in the high and bottom mid wake. It is noticeable in Fig.~\ref{ener1} (bottom) that this energetic cluster is captured in the mid-wake around $(x/h,y/h) \approx (2,0.5)$. The trail of the tip vortex, crucial for the vortex-shedding mechanism, is also present in $\overline{TKE}_3$. This same region is identified by $\overline{STKE}_3$, which is the richest of all latent variables in the $\overline{STKE}$ domain (Fig.~\ref{eneruni1}). The extensive domain of $\overline{STKE}_3$ from the near-high to mid wake compensates for the bottom wake concentration of $\overline{STKE}_2$ and $\overline{STKE}_1$ and constructs the energetic footprint from the tip vortex to the regions dominated by wall dynamics.
\begin{table}[h!]
\centering
\begin{tabular}{ll|l|l|l|l|l|}
                       & redundant & Location & unique & Location & Latent Time \\ \cline{1-6}
\multicolumn{1}{l|}{1} & 128  & far bottom &  381  & high-far shedding & [14792,14842]   \\ \cline{1-6}
\multicolumn{1}{l|}{2} & 257 & high-mid shedding &  1569  & mid-near shedding & [6611,6661]  \\ \cline{1-6}
\multicolumn{1}{l|}{3} & 3344 & mid-far shedding & 3103 & mid-far shedding & [14058,14108]
\end{tabular}
\caption{Most informative lags when studying the total redundant and unique contributions for each latent variable respectively. Location relates the corresponding lag with a domain shedding frequency on the wake. Latent time sets the studied range in \S.~\ref{study}, obtained by taking a 50-step temporal window after the $\lVert \mathbf{\cal{L}}_i(t) \lVert_{\infty}$ for each latent variable, respectively}
\label{tab:lags}
\end{table}

%\begin{figure}[H]
%    \centering
%    \includegraphics[width=1\linewidth]{compare/graph.png}
%    \caption{Causal graph, $\mathbf{\cal{G}}_i$, based on the most dominant contributions when maximizing the unique contribution. From left to right, causal graphs for histograms depicted in Fig.~\ref{histos}, where the main dominant unique contribution is depicted in red. From left to right, $\mathbf{\cal{G}}_1,\mathbf{\cal{G}}_2 $ and $\mathbf{\cal{G}}_3$.}
%\label{graph}
%\end{figure}

\noindent Once the energetic patterns produced by both encoder and decoder have been introduced, and how the latent variables allocate spatially along the wake. One needs to extract the most informative mechanisms in Fig.~\ref{histos}, specifically when maximizing the unique contribution, to do so we will filter the terms in the SURD histograms, and retain only the dominant contributions; unique, synergistic and redundant, for all latent variables. The global X-CAL graph in Fig.~\ref{graph}, depicts the latter, by describing the global synergy $S123$ (yellow), dominant in all latent variables, the unique cycle produced by the interchange of unique (red) information between $\mathcal{L}_1$ and $\mathcal{L}_3$, which also contributes to the informational nature of $\mathcal{L}_2$, giving rise to the co-founder variable~\cite{surd} fo this system $\mathcal{L}_3$. One can study the X-cal graph through different perspectives, however on this study we will focus on the two approaches introduced at the beginning of this section. First, we inspect the temporal signal for each of the latent variables, and extract a temporal window, after their respective maximums, defined as latent time in Tab.~\ref{tab:lags}. This latent time is responsible for the cause of the event while the latent time plus the lag is the effect of the event. In summary, what we propose with this first study, is an event driven analysis by cause and effect temporal windows, where the temporal window, $T_i$, is set by the source latent variable $i$, in charge of the studied informational mechanism, and the lag, $lag_j$, by the target variable $j$. We will inspect both cause and effect windows through the decoder and the encoder (SHAP) simultaneously, by computing ensemble means of the respective windows. As the latter are measured in the energetic domain, a complementary view is offered by depicting the same window for both vertical and stream-wise fluctuations. Inspecting the graph through different lens, we expect to obtain a better dynamical understanding of our X-cal graph $\mathcal{G}_c$.

\begin{figure}[H]
    \centering
    \includegraphics[width=1\linewidth]{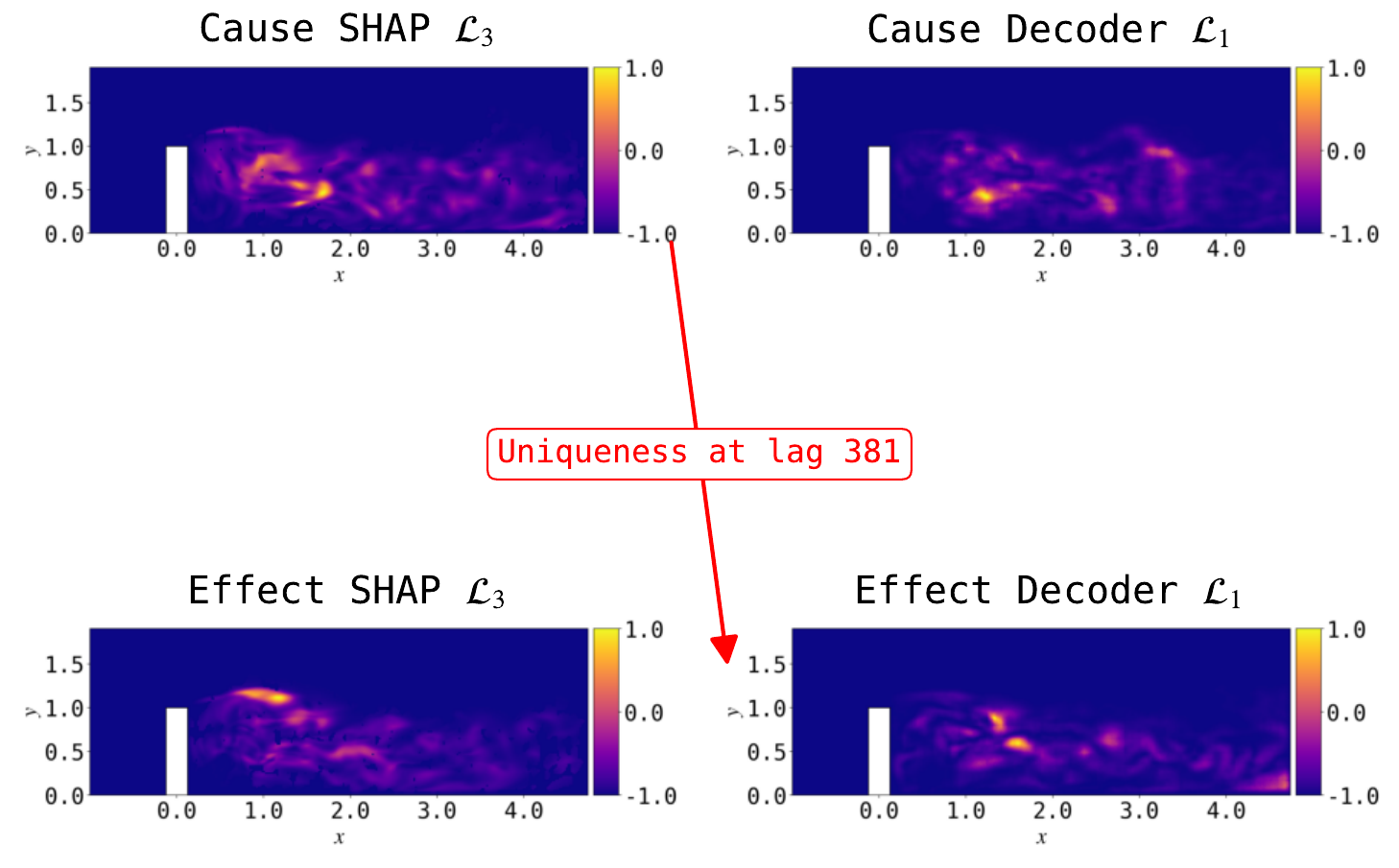}
    \caption{Cause and effect diagram for $\overline{STKE}_3$ (left) and $\overline{TKE}_1$ (right) at temporal window $T_1$ for causes and $T_3 + \text{lag}_3$ for effects.}
\label{SHAP_1}
\end{figure}

First, we analyze the unique contribution from $\mathcal{L}_3$ to $\mathcal{L}_1$ in Fig.~\ref{SHAP_1}. The mean energy field masked by the SHAP structures $\overline{STKE}_3$ and $\overline{TKE}_1$ at both temporal windows: cause and effect, are depicted. The $\overline{STKE}_3$ and $\overline{TKE}_1$ are investigated at $T_3$, producing temporal window, where the former carries the spatial locations important for encoding and the latter is the individual decodification of the studied variable.

\noindent When inspecting $\overline{STKE}_3$ one observes traces of the tip vortex being convected downstream into the mid-mid wake, achieving a parallel-like flow with the bottom wall. For this particular temporal window, $T_3$, the tip vortex that is being convected is on a positive regime when inspecting the stream-wise fluctuations. However, when observing $\overline{STKE}_3$ at the effect period, the concentration of energy is clearly at the tip of the obstacle, forming the tip-vortex dominating the flow, which is about to shed downstream.

\noindent On the other hand, the right plots of Fig.~\ref{SHAP_1} illustrate the de-codification of $\mathbf{\cal{L}}_1$, where the energy concentration arises at the near-low wake, which could be a consequence of the disappearance of the recirculation-bubble existing at the near-low section. At the effect temporal window, the energy concentrates at the high-mid wake, consequence of the shedding from the tip-vortex, and the mid-mid wake; where the energetic cluster is being convected downstream allowing the flow to produce the recirculation bubble once again on the conjugated regime.

\begin{figure}[H]
    \centering
    \begin{subfigure}{0.48\linewidth}
        \includegraphics[width=\linewidth]{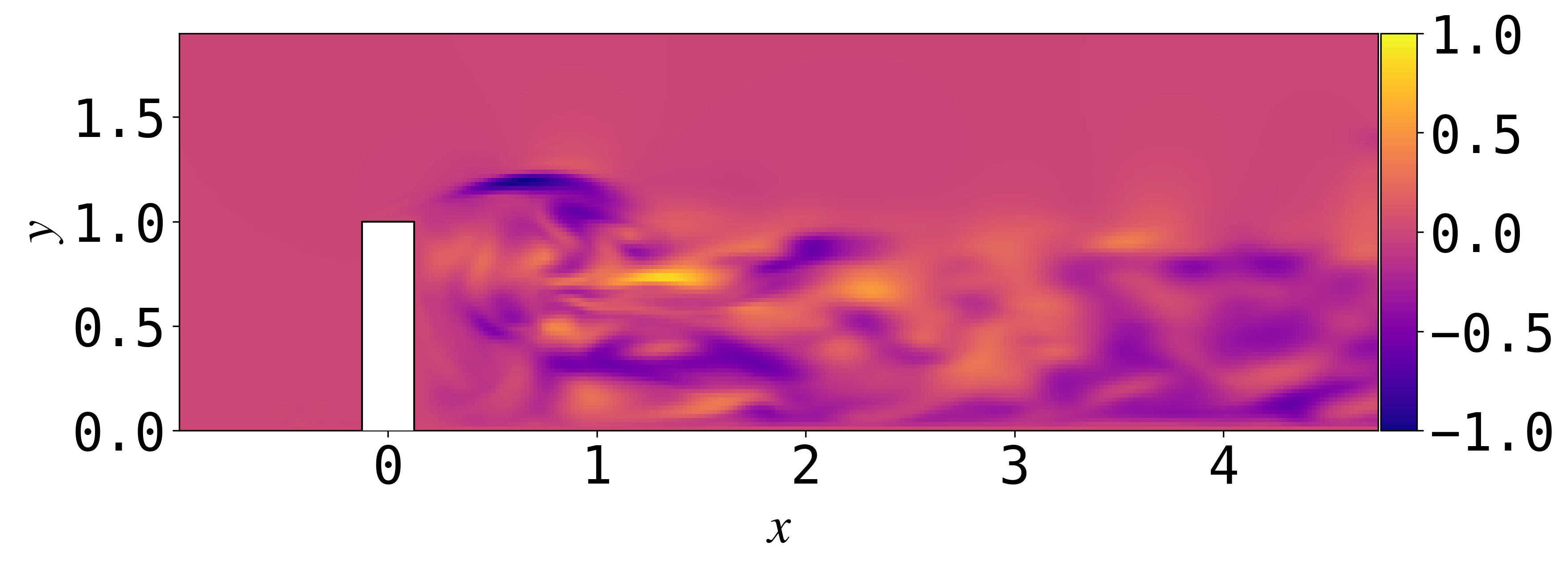}
        \caption*{(a) Stream-wise mean, cause window}
    \end{subfigure}
    \begin{subfigure}{0.48\linewidth}
        \includegraphics[width=\linewidth]{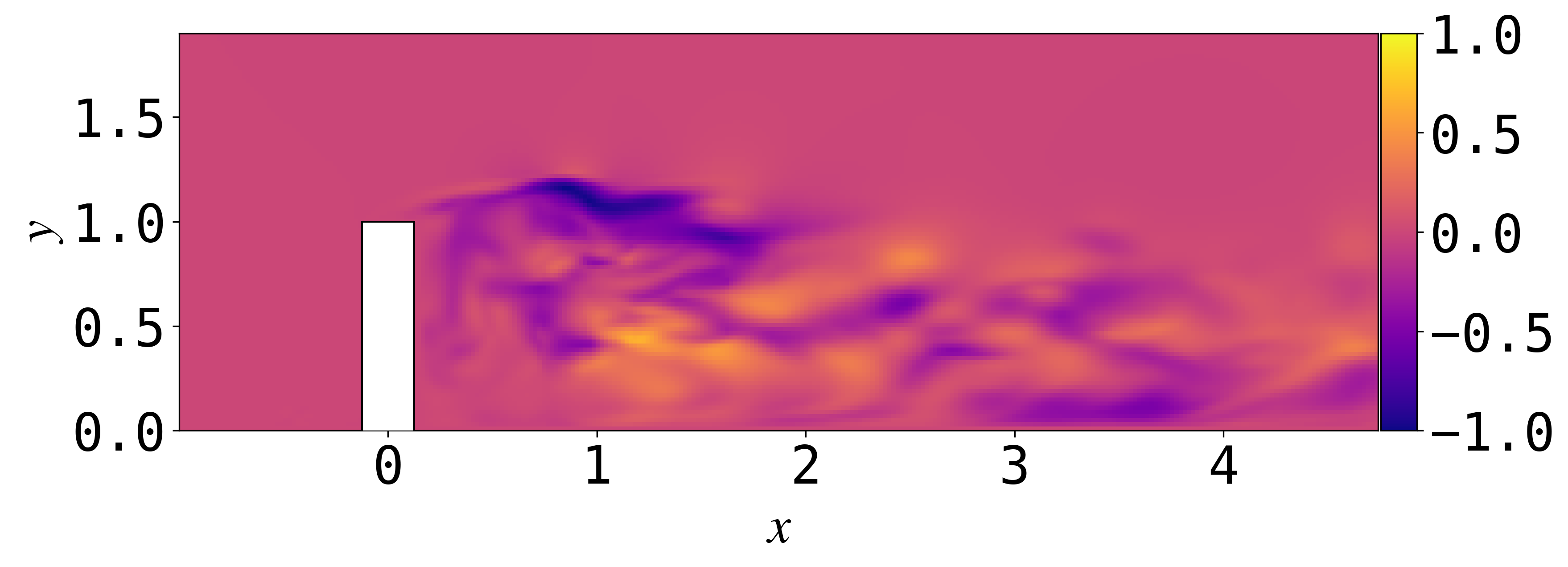}
        \caption*{(b) Stream-wise mean, effect window}
    \end{subfigure}
    \begin{subfigure}{0.48\linewidth}
        \includegraphics[width=\linewidth]{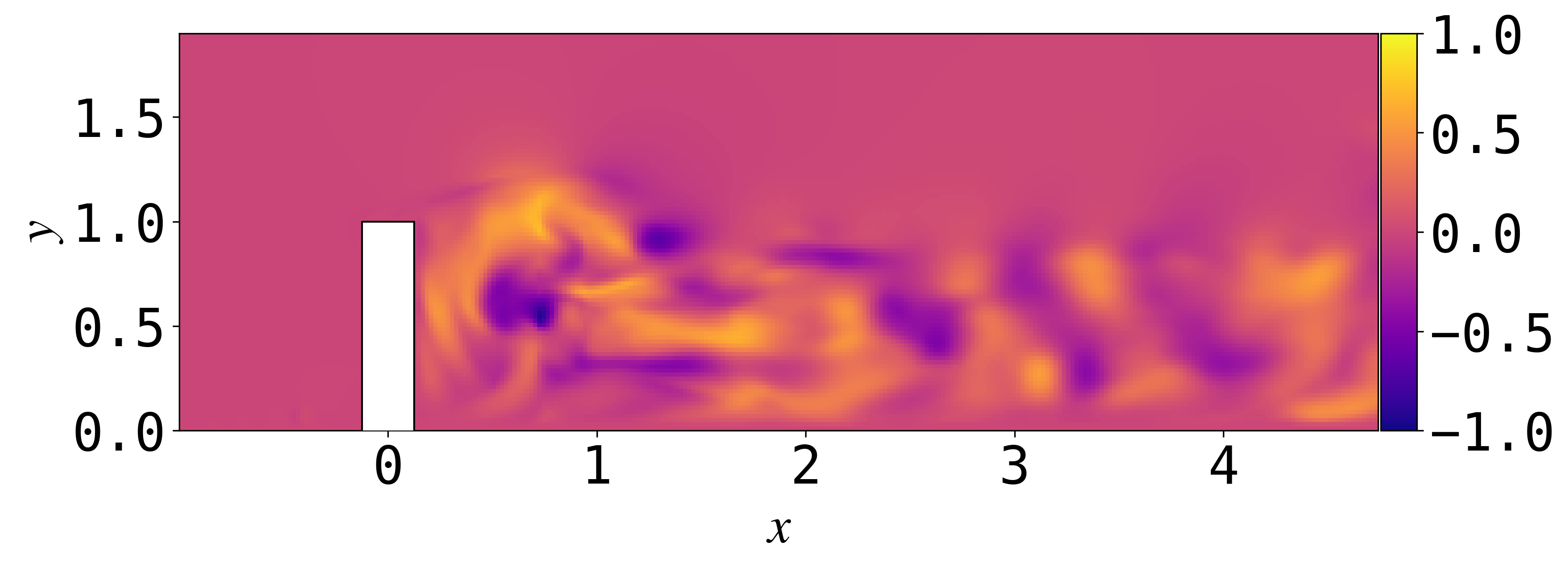}
        \caption*{(c) Vertical mean, cause window}
    \end{subfigure}
    \begin{subfigure}{0.48\linewidth}
        \includegraphics[width=\linewidth]{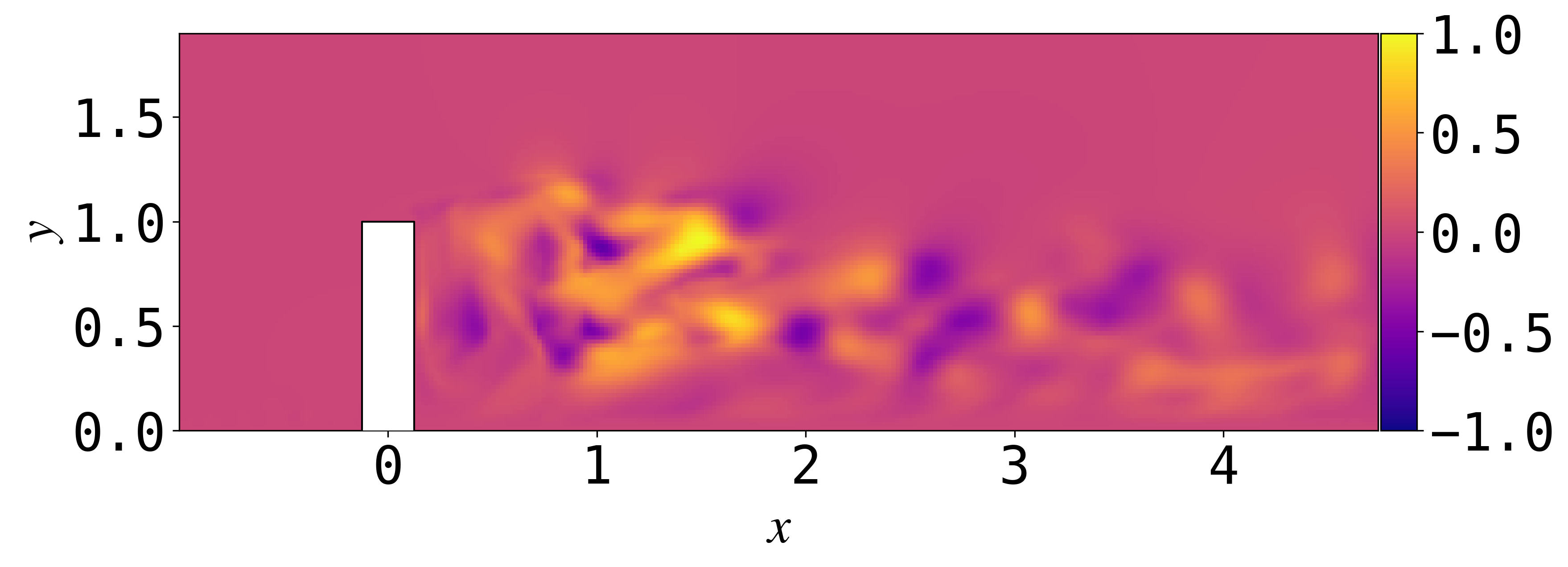}
        \caption*{(d) Vertical mean, effect window}
    \end{subfigure}
    \caption{Local fluctuations mean in the stream-wise (top row) and vertical (bottom row) components, computed over the cause window $T_3$ and the effect window $T_3 + \text{lag}_1$.}
    \label{fig:combined_fluctuations_1}
\end{figure}

\begin{figure}[H]
    \centering
    \includegraphics[width=1\linewidth]{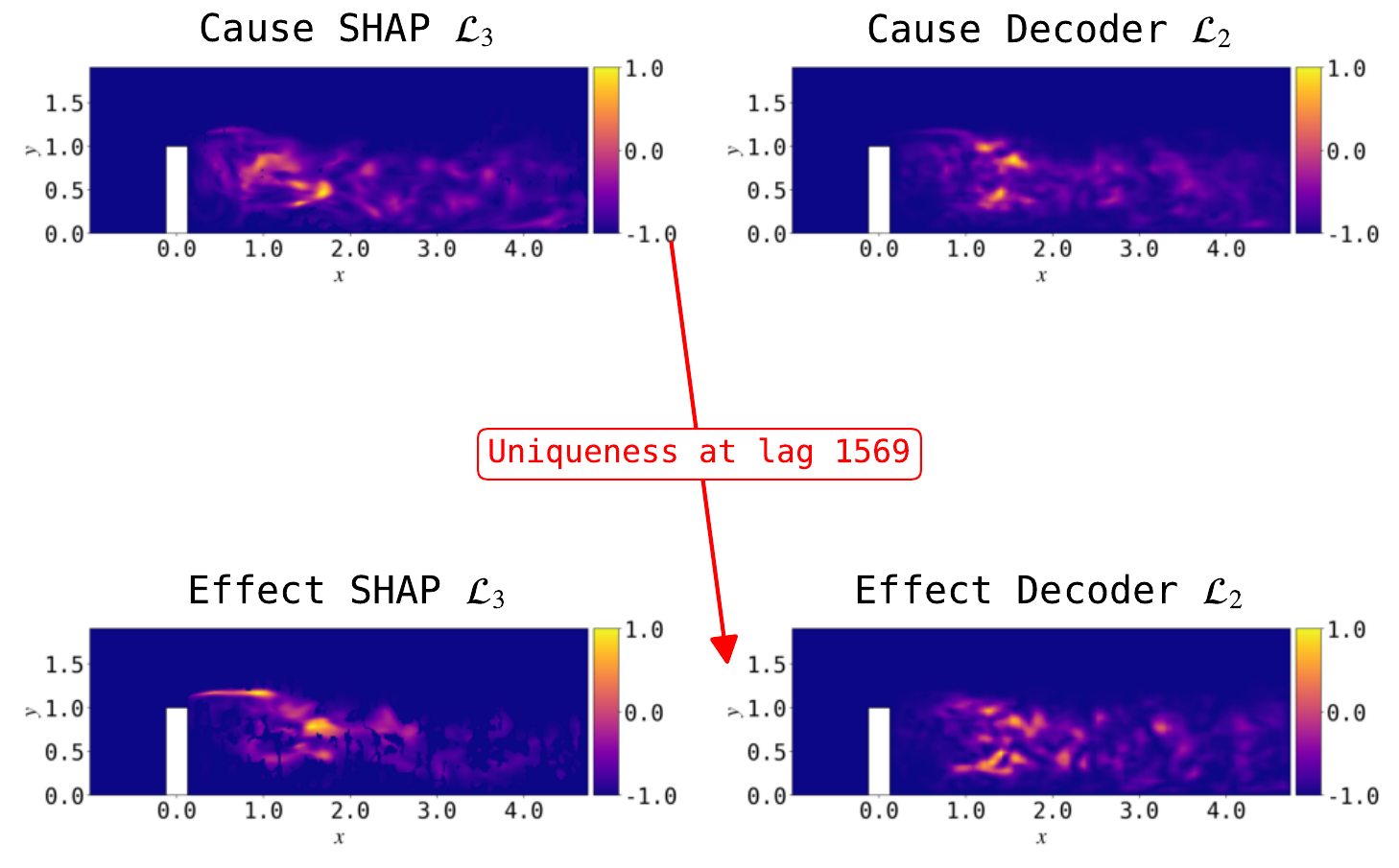}
    \caption{Cause and effect diagram for $\overline{STKE}_3$ (left) and $\overline{TKE}_2$ (right) at temporal window $T_3$ for causes and $T_3 + \text{lag}_2$ for effects.}
\label{SHAP_2}
\end{figure}

\noindent In Fig.~\ref{SHAP_2}, we investigate $\overline{STKE}_3$ and $\overline{TKE}_2$ at $T_3$, but this time with $lag_2$. Once again, we observe that $\overline{STKE}_3$ displays a convecting vortex structure aligned with the mid wake, representative of the quasi-parallel shedding state. At $T_3 + \text{lag}_2$, $\overline{TKE}_2$ localizes near the tip of the obstacle, identifying the formation of a new arch-vortex.

\begin{figure}[H]
    \centering
    \begin{subfigure}{0.48\linewidth}
        \includegraphics[width=\linewidth]{final_comp/u_streamwise_mean_window_0_14058_14108.png}
        \caption*{(a) Stream-wise mean, cause window}
    \end{subfigure}
    \begin{subfigure}{0.48\linewidth}
        \includegraphics[width=\linewidth]{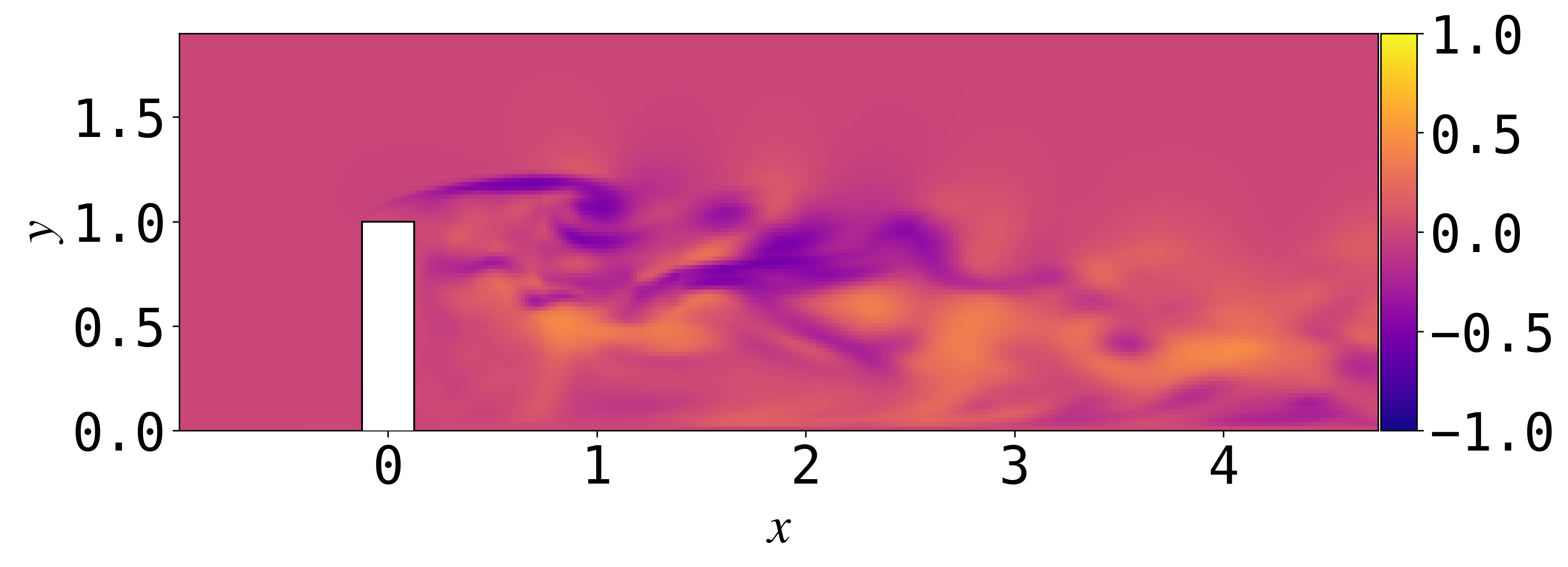}
        \caption*{(b) Stream-wise mean, effect window}
    \end{subfigure}
    \begin{subfigure}{0.48\linewidth}
        \includegraphics[width=\linewidth]{final_comp/u_vertical_mean_window_0_14058_14108.png}
        \caption*{(c) Vertical mean, cause window}
    \end{subfigure}
    \begin{subfigure}{0.48\linewidth}
        \includegraphics[width=\linewidth]{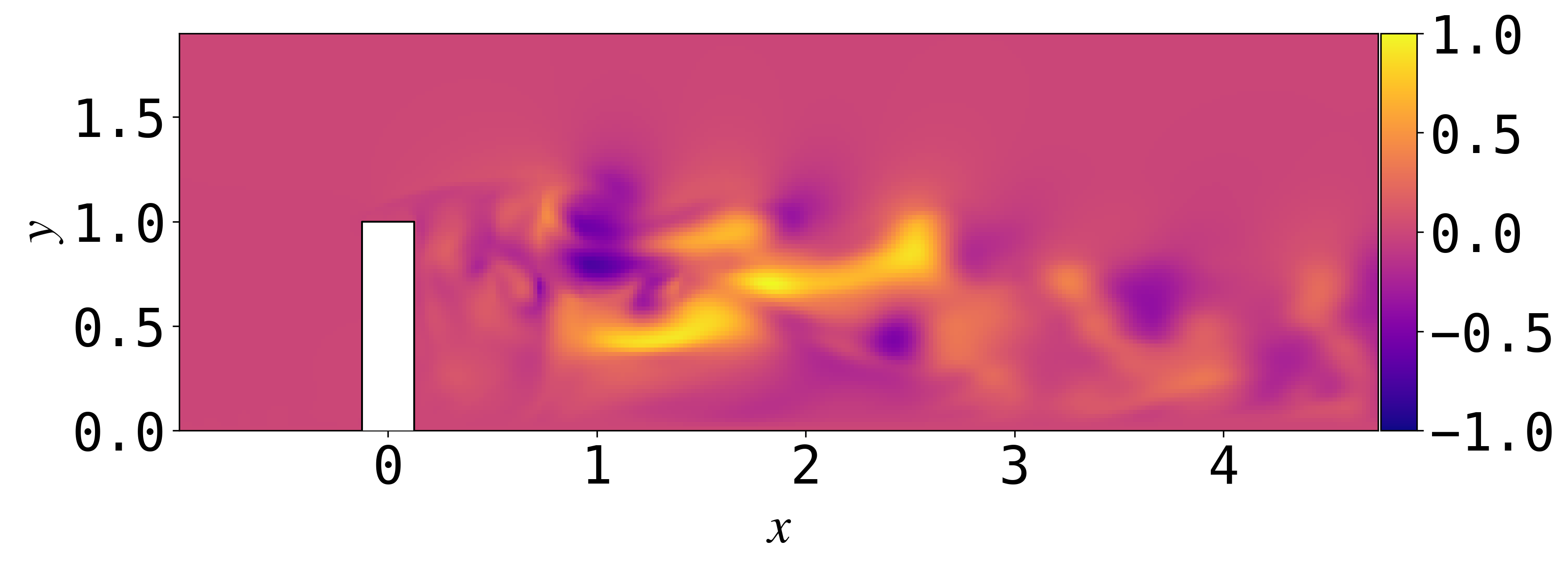}
        \caption*{(d) Vertical mean, effect window}
    \end{subfigure}
    \caption{Local fluctuations mean in the stream-wise and vertical components at $T_3$ (cause) and $T_3 + \text{lag}_2$ (effect).}
    \label{fig:combined_fluctuations_2}
\end{figure}

\noindent The fluctuation fields (Fig.~\ref{fig:combined_fluctuations_2}) demonstrate a transition from a parallel configuration at the cause window to a bursting, asymmetric wake pattern downstream. This reorganization from stable to shedding-like dynamics highlights the causality chain captured by the latent space.

\noindent Lastly, when studying the causal graph for $\mathbf{\cal{L}}_3$, we investigate the cause and effect of both $\overline{STKE}_1$ and $\overline{TKE}_3$. It is interesting to notice that the causal temporal window $T_1$ captures low-wake activity, while the effect window $T_1 + \text{lag}_3$ reveals enhanced upper-wake dynamics.

\begin{figure}[H]
    \centering
    \begin{subfigure}{0.48\linewidth}
        \includegraphics[width=\linewidth]{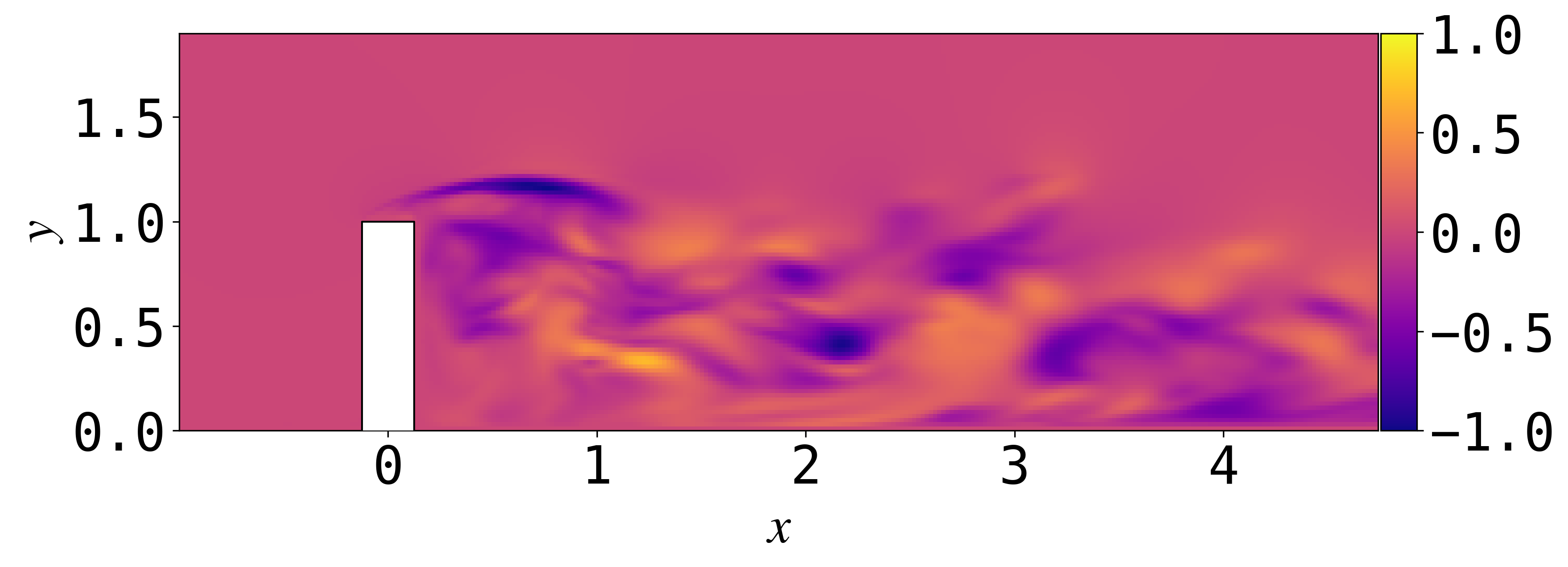}
        \caption*{(a) Stream-wise mean, cause window}
    \end{subfigure}
    \begin{subfigure}{0.48\linewidth}
        \includegraphics[width=\linewidth]{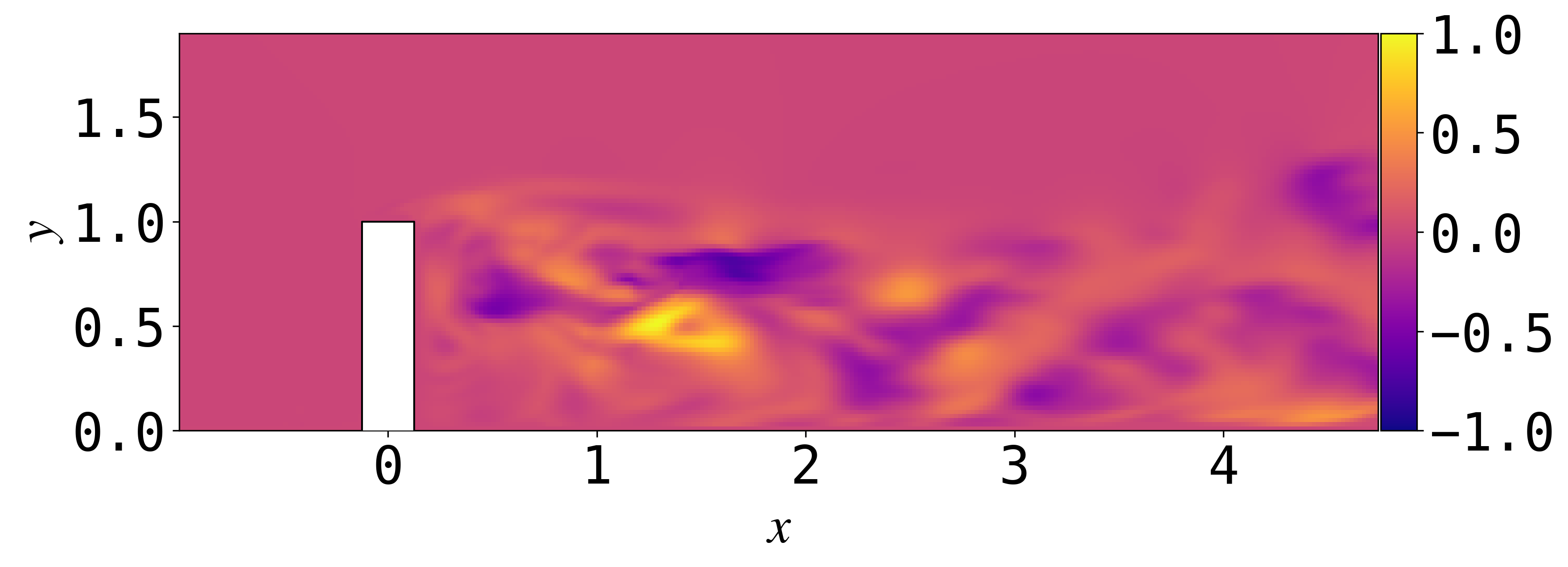}
        \caption*{(b) Stream-wise mean, effect window}
    \end{subfigure}
    \begin{subfigure}{0.48\linewidth}
        \includegraphics[width=\linewidth]{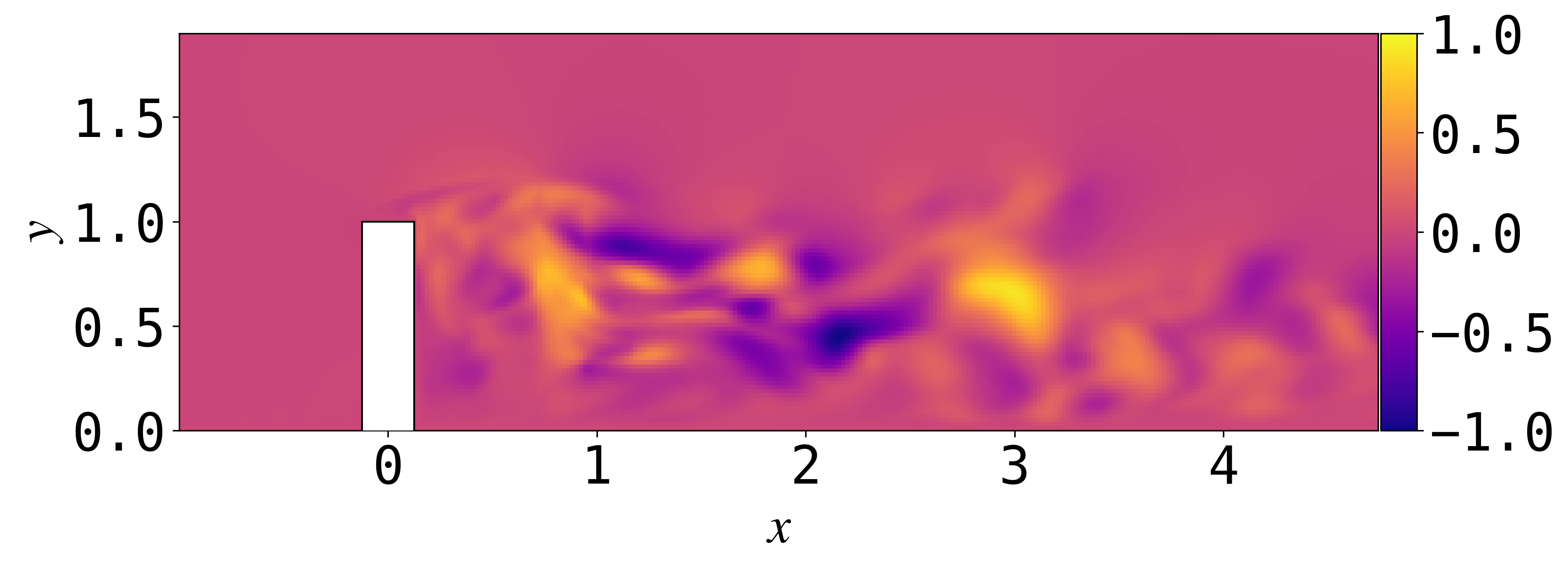}
        \caption*{(c) Vertical mean, cause window}
    \end{subfigure}
    \begin{subfigure}{0.48\linewidth}
        \includegraphics[width=\linewidth]{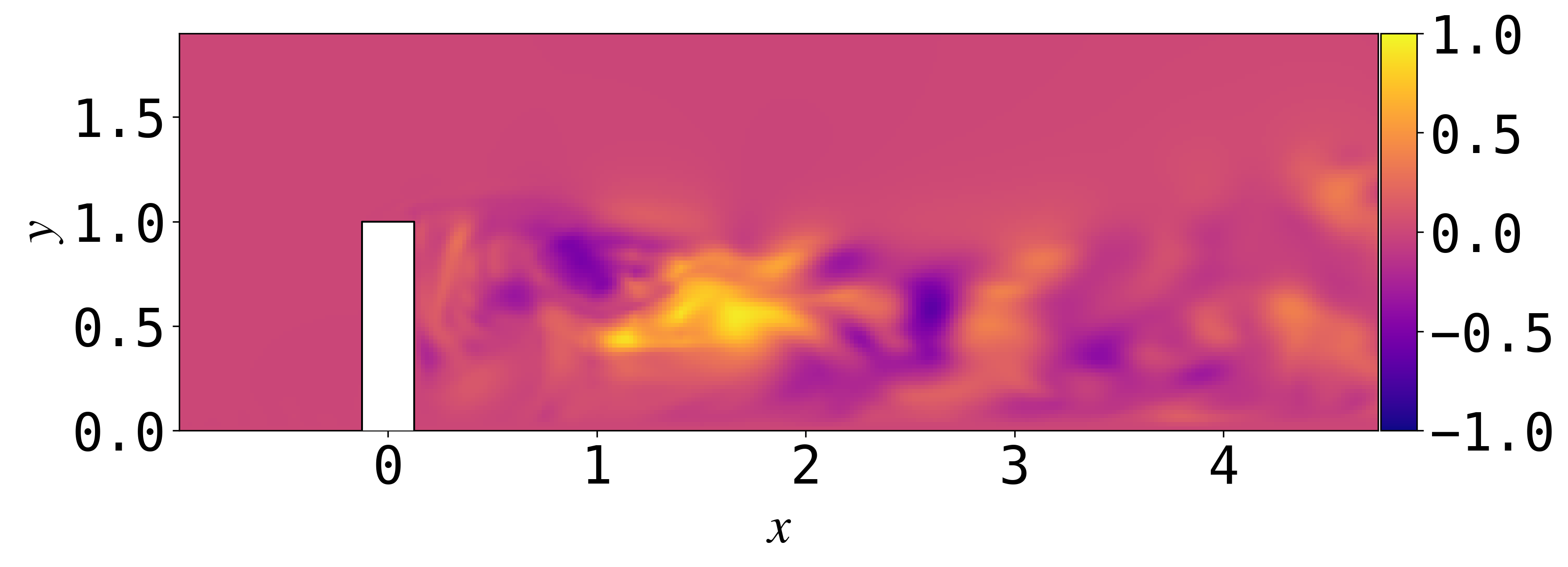}
        \caption*{(d) Vertical mean, effect window}
    \end{subfigure}
    \caption{Fluctuation means showing the transition from parallel-like in the low-wake to angle-like high-wake patterns for $\mathbf{\cal{L}}_3$.}
    \label{fig:combined_fluctuations_3}
\end{figure}

\noindent The fluctuation fields in Fig.~\ref{fig:combined_fluctuations_3} show that the parallel-like low-wake structures at the cause window transform into intense, angular, and upward propagating patterns at the effect window. This shift reflects a change from stable, convective base-wake behavior to a more energetic, possibly wall-influenced turbulent state. While traditionally associated with wall-bounded turbulence, recent studies suggest that burst- and streak-like dynamics may manifest in wake flows under certain geometric and Reynolds number conditions~\cite{Nekkanti2023JFM,Ghaemi2011JFM}.

\begin{figure}[H]
    \centering
    \includegraphics[width=1\linewidth]{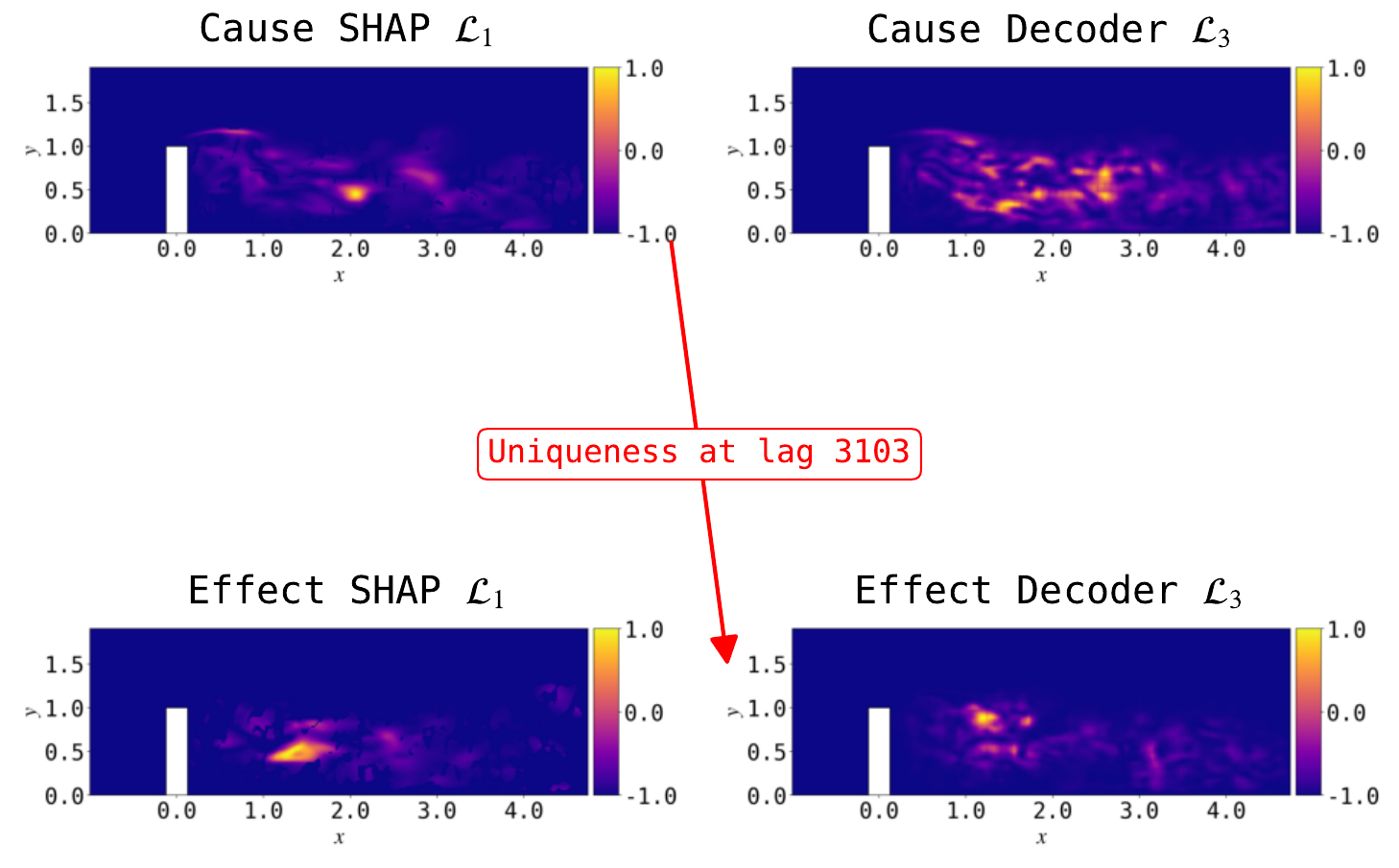}
    \caption{Cause and effect diagram for $\overline{STKE}_1$ (left) and $\overline{TKE}_3$ (right) at temporal window $T_1$ for causes and $T_1 + \text{lag}_3$ for effects.}
\label{SHAP_3}
\end{figure}

\paragraph{State analysis of unique causality:}

% \begin{figure}[h!]
%     \centering
%     \begin{subfigure}{0.30\linewidth}
%         \centering
%         \includegraphics[width=1\linewidth]{final_comp/obs_states/tbl_U1_3_states_ld2.pdf}
%     \end{subfigure}
%     \quad
%     \begin{subfigure}{0.30\linewidth}
%         \centering
%         \includegraphics[width=1\linewidth]{final_comp/obs_states/tbl_U3_2_states_ld2.pdf}
%     \end{subfigure}
%         \begin{subfigure}{0.30\linewidth}
%         \centering
%         \includegraphics[width=1\linewidth]{final_comp/obs_states/tbl_U3_1_states_ld2.pdf}
%     \end{subfigure}
%     \caption{From (left) to (right) the causal unique maps for $\mathcal{L}_1$,$\mathcal{L}_2$ and $\mathcal{L}_3$. We depict the source of the unique contribution on the $x$-axis while plotting the target on the $y$-axis.}
%     \label{causal_obs}
% \end{figure}\begin{figure}[H]
\begin{figure}
    \centering
    \begin{subfigure}{0.30\linewidth}
        \centering
        \includegraphics[width=1\linewidth]{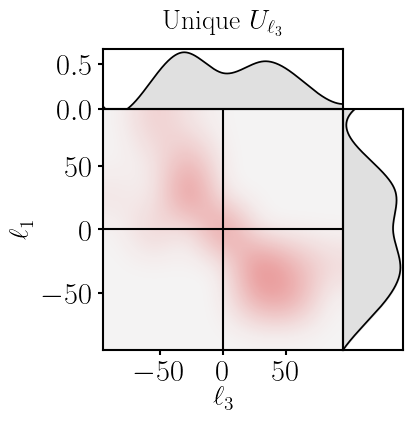}
    \end{subfigure}
    \quad
    \begin{subfigure}{0.30\linewidth}
        \centering
        \includegraphics[width=1\linewidth]{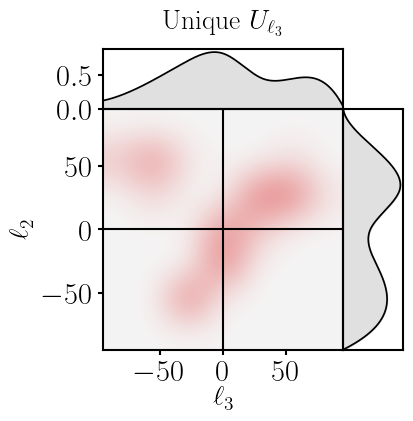}
    \end{subfigure}
        \begin{subfigure}{0.30\linewidth}
        \centering
        \includegraphics[width=1\linewidth]{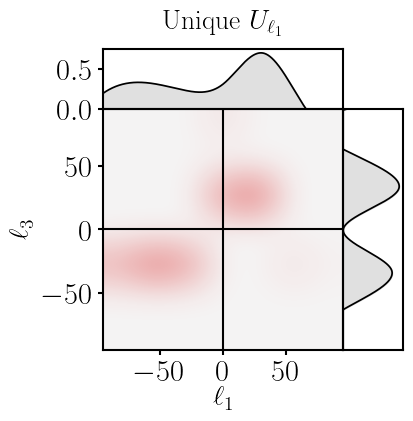}
    \end{subfigure}
    \caption{From (left) to (right) the causal unique maps for $\ell_1$,$\ell_2$ and $\ell_3$. We depict the source of the unique contribution on the $x$-axis while plotting the target on the $y$-axis.}
    \label{causal_obs}
\end{figure}
To conclude our causal analysis, we will join the SURD state analysis with the $STKE$ fields we already presented above. This study is complementary to the windows analysis, as now we will extract a temporal sequence for the target variable in the latent space, containing all time instances corresponding to the causal states. First, in Figure~\ref{causal_obs} we show the state-conditioned unique contributions $\Delta {I}^{U}_{i \rightarrow j}(\mathcal{L}_j^+ = \ell_j^+)$ from each source state $\ell_i \in \mathcal{L}_i$ to target states $\ell_j$ across all latent states, where the unique causal states are depicted in red. Notably, $\mathcal{L}_3$ acts as a strong source for both $\ell_1$ and $\ell_2$ at specific target states, consistent with the total SURD values in Fig.~\ref{histos}. These dominant interactions are localized in latent space and highlight directional flow of predictive information.

\begin{figure}[H]
    \centering
    \begin{subfigure}{0.70\linewidth}
        \centering
        \includegraphics[width=1\linewidth]{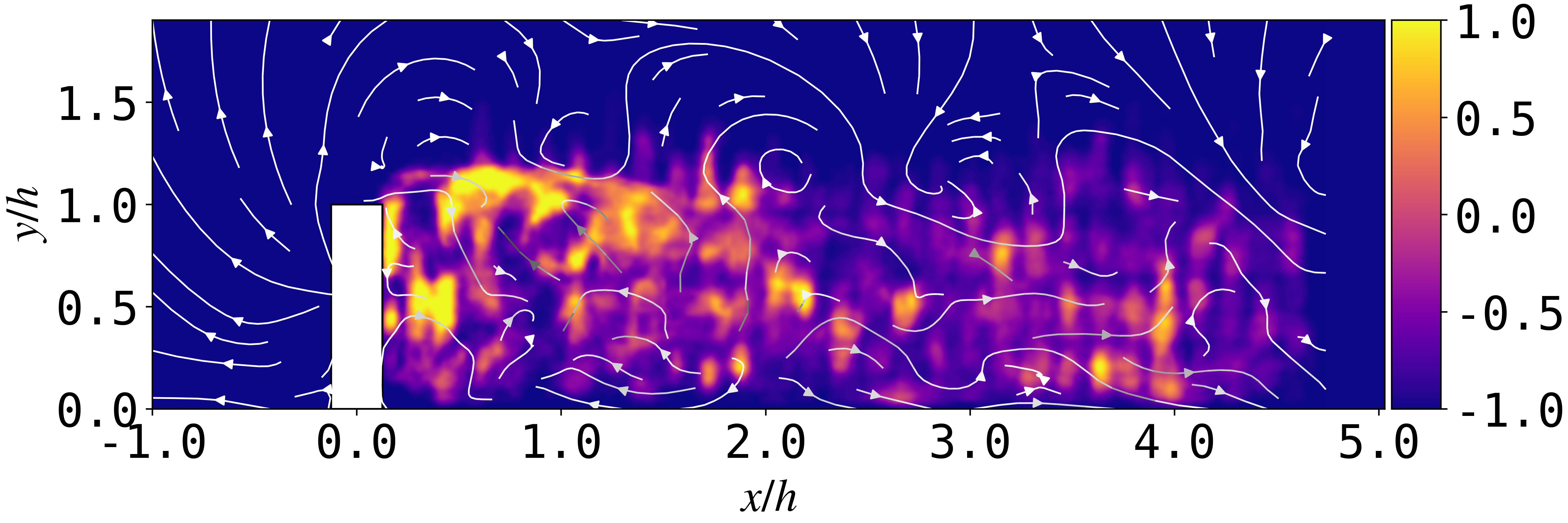
}
    \end{subfigure}
    \quad
    \begin{subfigure}{0.70\linewidth}
        \centering
        \includegraphics[width=1\linewidth]{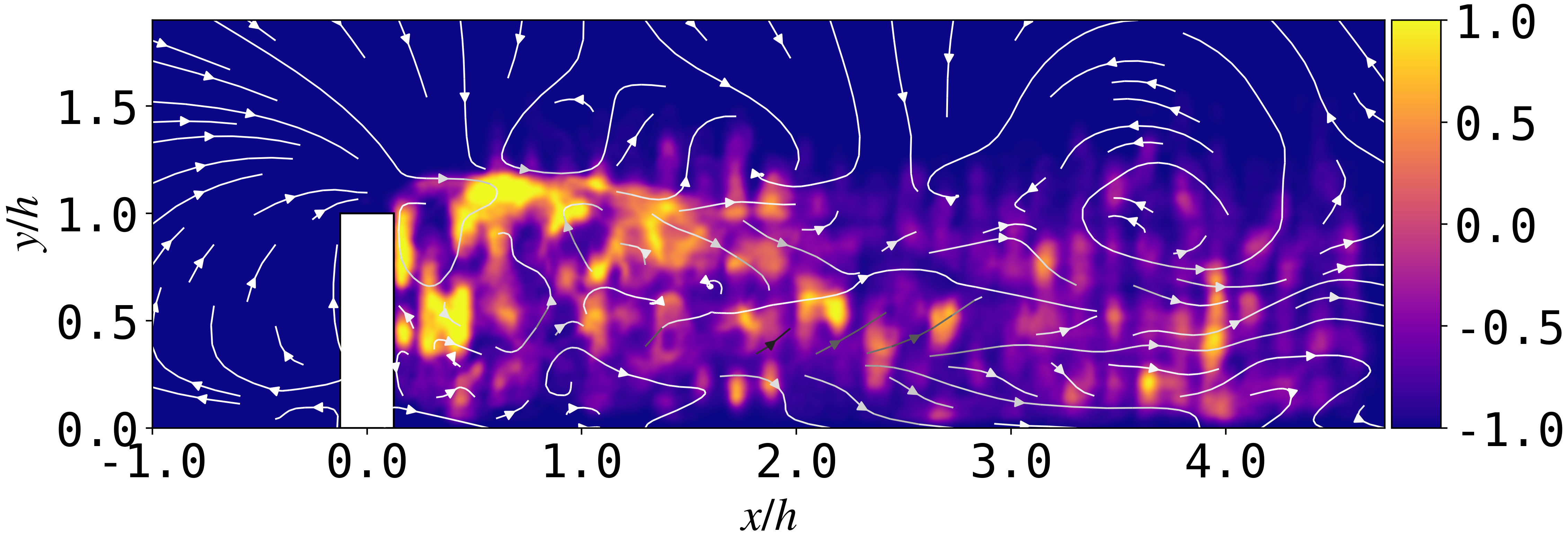}
    \end{subfigure}
        \begin{subfigure}{0.70\linewidth}
        \centering
        \includegraphics[width=1\linewidth]{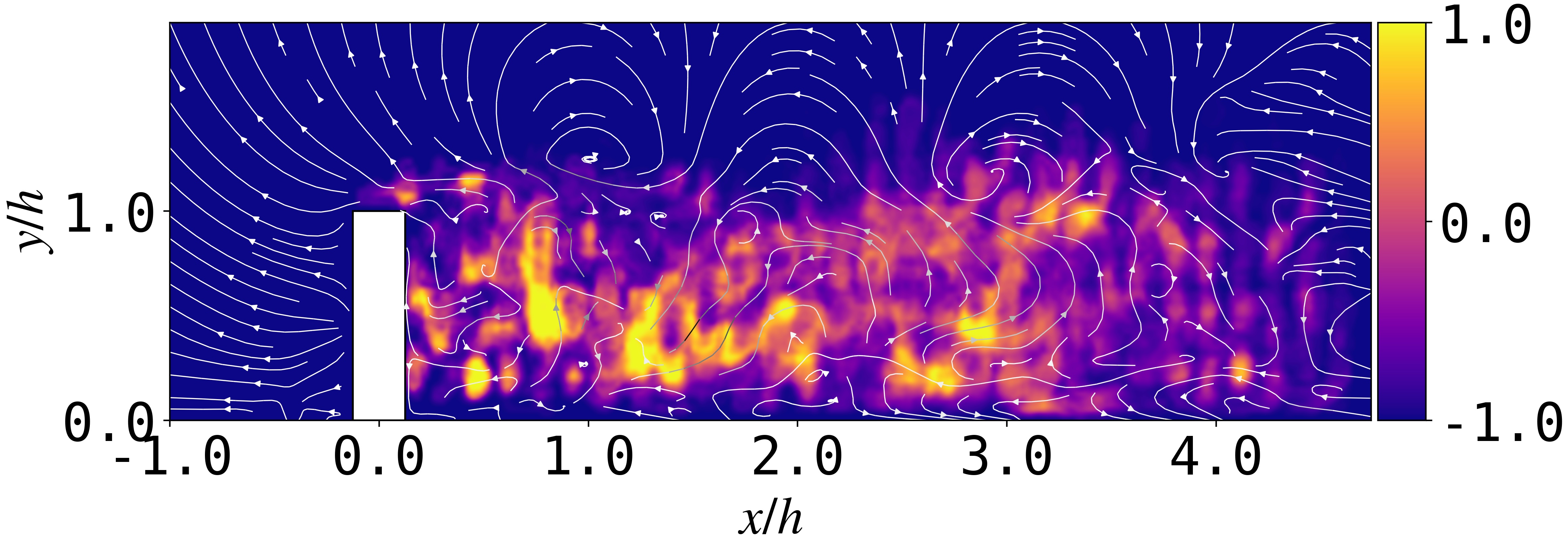}
    \end{subfigure}
    \caption{From (top) to (bottom) the ensemble mean turbulent kinetic energy ($STKE^{T_j}_i$) masked by $P_i(x,y,t)$ for sources $i = [3,3,1]$ respectively. $T_j$ is set as the causal temporal window for the states in Fig.~\ref{causal_obs} given the target $j$.}
    \label{causal_obs_un-s}
\end{figure}

To relate these events to physical space, Fig.~\ref{causal_obs_un-s} shows the local mean structured turbulent kinetic energy $\text{STKE}_i^T(x,y)$ for source latents $\ell_i = \{3,3,1\}$ over their respective most-informative time windows $T_j$, for a given target $j$, using SHAP-based structure masks $P_j(x,y,t)$. For $\ell_1 \rightarrow \ell_3$, structures concentrate in the near-bottom wake ($x/h\sim1.5$), suggesting base-vortex lift-up and near-wall events. The dynamic nature of $STKE^{T_3}_1$ is governed by wall structures being advected downstream, while merging its dynamics in the mid wake with the tip vortex. In $\ell_3 \rightarrow \ell_2$, the response emerges downstream near the mid-shear layer. One can observe the tip-vortex emerging at the top of the obstacle, while having circular like structure at $x \sim 0.5$ where the recirculation bubble emerges. It is also noticeable how most of the energy concentrates along the top-mid wake, specially near to the obstacle, as the flow depicted by the streamlines in $STKE_3^{T_2}$ assembles quasi-parallel to the bottom wall. Finally, the $\ell_3 \rightarrow \ell_1$ map shows top-wake concentration near $y/h\sim1$, likely linked to tip-shedding. It is also important to notice the similarities arising between $STKE_3^{T_2}$ and $STKE_3^{T_1}$, as both concentrate energetically around the tip of the obstacle and the recirculation zone behind the obstacle, specially capturing the tip-vortex shedding into the mid wake, clearly represented in Fig.~\ref{eneruni1} by the global mean $\overline{STKE_3}$. On the other hand, the main differences arise when inspecting the streamlines of the mid and far wake, where the turbulent nature of the wake gets more oscillatory and non-organized for $STKE_3^{T_1}$. These spatial patterns reinforce the interpretation of latent-level unique causality as arising from distinct regions involved in vortex–shear-layer interactions.
% \begin{table}[]
% \begin{tabular}{ll|lll}
%                        & Red & unique &  &  \\ \cline{1-3}
% \multicolumn{1}{l|}{1} & 431  & 1967  &  &  \\ \cline{1-3}
% \multicolumn{1}{l|}{2} & 164 & 3424  &  &  \\ \cline{1-3}
% \multicolumn{1}{l|}{3} & 116 & 2469 &  & 
% \end{tabular}
% \caption{unique}
% \end{table}

\section{Conclusions}
\label{conclusions}
In the present work we introduce a novel framework to study causality and fluid-flow mechanisms. We first use a $\beta$-VAE to compress the physical domain into a low-dimensional manifold, the latent space where we apply the mutual information implementation by SURD~\cite{surd} to study the causal dependencies between latent variables through information theory. Lastly, by the implementation of SHAP values through gradient SHAP~\cite{Cremades2024}, we can express the causal relations identified in the latent space in terms of important structures in the physical space. The objective of this method is to identify physical structures or mechanisms that govern the main dynamics of the flow for our given model. 

\noindent This framework would be extremely helpful to design control strategies, as one could not only control the latent space based on the analytical causality but also depending on the physical subdomains which contribute to the encoding of each $\mathbf{\cal{L}}_i$ for $i \in [1,2,3]$. Classically, control strategies have been developed based on ad-hoc assumptions regarding the most relevant structures on the flow. With this approach, the definition of importance becomes objective and based on SHAP values, which in practice are the most important structures to reconstruct the flow~\cite{beneitez2025improvingturbulencecontrolexplainable}.

\noindent On the other hand, one could go a step further and investigate the SHAP structures and the decoding of each latent variable independently as we did in \S.~\ref{study}, with the objective of understanding the mechanism governing the flow from a fundamental perspective. The coupling~\cite{SoleraRico2024} between $\beta$-VAEs and predictive models such as diffusion models~\cite{Guastoni2025Perspective} or transformers~\cite{sanchisagudo2023easy} has already been proposed in the literature. In these works, the compression and predictive capability of ML models are validated and implemented, however explainability and causality still remained unexplored.

\noindent On the other hand, the main findings of this work can be summarized as follows. 
Firstly, the \emph{windows analysis} revealed how unique causal dependencies unfold over the most informative lags. For $\mathbf{\cal{L}}_1$, its evolution is largely driven by $\mathbf{\cal{L}}_3$, with SHAP fields showing traces of the tip shear layer convecting downstream into the mid wake and reorganizing the low-wake recirculation bubble. For $\mathbf{\cal{L}}_2$, the dominant causal link from $\mathbf{\cal{L}}_3$ corresponded to a transition from a quasi-parallel upper shear layer into localized shedding near the obstacle tip, i.e.\ the initiation of a new vortex street. Lastly, in Figs.~\ref{SHAP_3},\ref{fig:combined_fluctuations_3}, it can be observed how the full tip-vortex shedding cycle is represented: the structures evolve from elongated streamwise-aligned patterns into angled features departing from the parallel organization above the wall. Together, the three latent variables capture complementary phases of the shedding process: the end of a previous cycle ($\mathbf{\cal{L}}_1$), the onset of tip-layer roll-up ($\mathbf{\cal{L}}_2$), and the complete development of the shedding event ($\mathbf{\cal{L}}_3$).

\noindent Secondly, the state analysis demonstrated that the three latent variables act jointly in physically meaningful conditions and that $\mathcal{L}_3$ arises as the co-founder of our reduced order model, as it has the capability of connecting different subdomains of the wake. Thus, predictive information is maximized not by individual latent contributions but by their joint action in the physical space, reflecting the cooperative dynamics by which tip-layer instabilities and spanwise shedding synchronize to organize the wake, making clear these transition by capturing the recirculation bubble behind the obstacle, present in $\overline{STKE^{T_1}}_3$ at Fig.~\ref{causal_obs_un-s}. Further studies need to be carried to understand better the dynamics of the flow, however with the analysis proposed around the X-CAL graph $\mathcal{G}_c$, we have a clear vision of the possible latent manifold ruling it. $\mathcal{L}_3$ is affected by $\mathcal{L}_1$ after approximately a full shedding cycle~\cite{vinuesa2015direct}, the inverse relation arises at half-cycle, while the relationship between $\mathcal{L}_2$ and $\mathcal{L}_3$ relies on the sustainment of the turbulence along the wake while the tip-vortex is being generated.

\noindent Overall, the causal-AI framework demonstrates that the latent variables systematically capture the dominant dynamics of the obstacle wake: the spanwise shedding, partially the base vortex and the 2D footprint of the tip vortex. This establishes a clear link between causal dependencies in the latent space and coherent structures in the physical domain. Extending the methodology to 3D datasets will make it possible to identify the full set of coherent motions (tip, spanwise, base and horseshoe), thereby strengthening the physical interpret-ability of causal analysis in turbulent wakes. Furthermore, we believe the presented framework has the capability of enhancing the understanding of closure~\cite{Eiximeno_Sanchis-Agudo_Miro_Rodriguez_Vinuesa_Lehmkuhl_2025} models when constructing reduced order models or accelerating already existing methods.
\section*{Acknowledgments}
R.V. acknowledges financial support from ERC grant no.‘2021-CoG-101043998, DEEPCONTROL’ and the EU Doctoral Network MODELAIR. ML model training and development was carried out using computational resources provided by the National Academic Infrastructure for Supercomputing in Sweden (NAISS).
\section*{References}
\bibliographystyle{apsrev4-2}
\bibliography{ref}

%========================  APPENDIX: SHAP as a symmetry tracker  ========================

%  
\end{document}

%% file: macros.tex
\let\bs\boldsymbol
\newcommand{\bq}{\bs{q}}        % short for boldsymbolq
\newcommand{\bi}{{\bs{i}}}        % short for boldsymboli
\newcommand{\bQ}{\bs{Q}}        % short for boldsymbolQ
\newcommand{\Is}{\tilde{\imath}}        % short for I specific

%% file: ref.bib
@article{POD_org,
  title={The structure of inhomogeneous turbulent flows},
  author={J.L. Lumley},
  journal={In: Yaglom, A.M. and Tartarsky, V.I., Eds., Atmospheric turbulence and radio wave propagation},
  volume={23},
  number={13},
  pages={166-178},
  year={1967}
}

@article{surd,
  author = {Mart{\'i}nez-S{\'a}nchez, {\'A}lvaro and Arranz, Gonzalo and Lozano-Dur{\'a}n, Adri{\'a}n},
  title = {Decomposing causality into its synergistic, unique, and redundant components},
  journal = {Nature Communications},
  year = {2024},
  volume = {15},
  number = {1},
  pages = {9296},
  doi = {10.1038/s41467-024-53373-4},
  url = {https://doi.org/10.1038/s41467-024-53373-4}
}

@inproceedings{Sundararajan2017,
  author    = {Sundararajan, Mukund and Taly, Ankur and Yan, Qiqi},
  title     = {Axiomatic Attribution for Deep Networks},
  booktitle = {Proceedings of the 34th International Conference on Machine Learning},
  series    = {Proceedings of Machine Learning Research},
  volume    = {70},
  pages     = {3319--3328},
  publisher = {PMLR},
  year      = {2017},
  url       = {https://proceedings.mlr.press/v70/sundararajan17a.html}
}

@article{SHAPley1953,
  author  = {Shapley, Lloyd S.},
  title   = {A value for n-person games},
  journal = {Contributions to the Theory of Games},
  volume  = {2},
  pages   = {307--317},
  year    = {1953},
  publisher = {Princeton University Press}
}

@article{Erion2021,
  author  = {Erion, Gabriel and Janizek, Joseph D. and Sturmfels, Pascal and Lundberg, Scott M. and Lee, Su-In},
  title   = {Improving performance of deep learning models with axiomatic attribution priors and expected gradients},
  journal = {Nature Machine Intelligence},
  year    = {2021},
  volume  = {3},
  number  = {7},
  pages   = {620--631},
  doi     = {10.1038/s42256-021-00343-w},
  url     = {https://doi.org/10.1038/s42256-021-00343-w}
}

@ARTICLE{shannon1948,
  author={Shannon, C. E.},
  journal={The Bell System Technical Journal}, 
  title={A mathematical theory of communication}, 
  year={1948},
  volume={27},
  number={3},
  pages={379-423},
  doi={10.1002/j.1538-7305.1948.tb01338.x}}

@article{DeWeese1999,
doi = {10.1088/0954-898X/10/4/303},
url = {https://dx.doi.org/10.1088/0954-898X/10/4/303},
year = {1999},
month = {nov},
publisher = {},
volume = {10},
number = {4},
pages = {325},
author = {Michael R DeWeese  and Markus Meister},
title = {How to measure the information gained from one symbol},
journal = {Network: Computation in Neural Systems}
}

@article{Jimenez_2018,
  author  = {Jimenez, Javier},
  title   = {Coherent structures in wall-bounded turbulence},
  journal = {Journal of Fluid Mechanics},
  year    = {2018},
  volume  = {842},
  pages   = {P1},
  doi     = {10.1017/jfm.2018.144}
}

@misc{states2025,
      title={Observational causality by states and interaction type for scientific discovery}, 
      author={{\'A}lvaro Mart{\'i}nez-S{\'a}nchez and Adri{\'a}n Lozano-Dur{\'a}n},
      year={2025},
      eprint={2505.10878},
      archivePrefix={arXiv},
      primaryClass={physics.data-an},
      url={https://arxiv.org/abs/2505.10878}, 
}

@article{SoleraRico2024,
  author = {Solera-Rico, Alberto and Sanmiguel Vila, Carlos and G{\'o}mez-L{\'o}pez, Miguel and Wang, Yuning and Almashjary, Abdulrahman and Dawson, Scott T. M. and Vinuesa, Ricardo},
  title = {\beta-Variational autoencoders and transformers for reduced-order modelling of fluid flows},
  journal = {Nature Communications},
  year = {2024},
  volume = {15},
  number = {1},
  pages = {1361},
  doi = {10.1038/s41467-024-45578-4},
  url = {https://doi.org/10.1038/s41467-024-45578-4}
}

@article{WANG2024109254,
	author = {Yuning Wang and Alberto Solera-Rico and Carlos {Sanmiguel Vila} and Ricardo Vinuesa},
	doi = {10.1016/j.ijheatfluidflow.2023.109254},
	issn = {0142-727X},
	journal = {International Journal of Heat and Fluid Flow},
	pages = {109254},
	title = {Towards optimal \ensuremath{\beta}-variational autoencoders combined with transformers for reduced-order modelling of turbulent flows},
	url = {https://www.sciencedirect.com/science/article/pii/S0142727X23001534},
	volume = {105},
	year = {2024}
}

@article{EIVAZI2022117038,
title = {Towards extraction of orthogonal and parsimonious non-linear modes from turbulent flows},
journal = {Expert Systems with Applications},
volume = {202},
pages = {117038},
year = {2022},
issn = {0957-4174},
doi = {https://doi.org/10.1016/j.eswa.2022.117038},
url = {https://www.sciencedirect.com/science/article/pii/S0957417422004535},
author = {Hamidreza Eivazi and Soledad {Le Clainche} and Sergio Hoyas and Ricardo Vinuesa}
}

@article{captumGradientShap,
  author  = {Kokhlikyan, Narine and Miglani, Vivek and Martin, Miguel and Wang, Edward and Alsallakh, Bilal and Reynolds, Jonathan and Melnikov, Alexander and Kliushkina, Natalia and Araya, Carlos and Yan, Siqi and Reblitz-Richardson, Orion},
  title   = {Captum: A unified and generic model interpretability library for {PyTorch}},
  journal = {arXiv preprint},
  eprint  = {2009.07896},
  archivePrefix = {arXiv},
  primaryClass  = {cs.LG},
  year    = {2020},
  url     = {https://arxiv.org/abs/2009.07896}
}

@article{cremades2024classically,
  title={Classically studied coherent structures only paint a partial picture of wall-bounded turbulence},
  author={Cremades, Andr{\'e}s and Hoyas, Sergio and Vinuesa, Ricardo},
  journal={Nature Communications},
  volume={16},
  number={1},
  pages={10189},
  year={2025},
  month={nov},
  publisher={Nature Publishing Group UK London},
  doi={10.1038/s41467-025-65199-9},
  url={https://doi.org/10.1038/s41467-025-65199-9}
}

@incollection{wiener1956,
  author    = {N. Wiener},
  title     = {The Theory of Prediction},
  booktitle = {Modern Mathematics for Engineers},
  year      = {1956},
  publisher = {McGraw-Hill},
  address   = {New York},
}

@article{kolmogorov1965,
  author    = {Andrei N. Kolmogorov},
  title     = {Three approaches to the quantitative definition of information},
  journal   = {Problems of Information Transmission},
  year      = {1965},
  volume    = {1},
  issue     = {1},
  pages     = {1-7},
}

@book{kolmogorovLegacy,
  title     = {The Kolmogorov Legacy in Physics},
  year      = {2004},
  editor    = {V. L. Ginzburg},
  publisher = {Springer},
  address   = {Berlin, Heidelberg},
}

@article{millionshchikov1970,
  author  = {V. Millionshchikov},
  title   = {On the theory of characteristic Lyapunov exponents},
  journal = {Mat. Zametki},
  year    = {1970},
  volume  = {7},
  pages   = {503--513},
}

@inproceedings{lundberg2017unified,
  title={A Unified Approach to Interpreting Model Predictions},
  author={Lundberg, Scott M and Lee, Su-In},
  booktitle={Advances in Neural Information Processing Systems},
  volume={30},
  year={2017},
  url={https://proceedings.neurips.cc/paper_files/paper/2017/file/8a20a8621978632d76c43dfd28b67767-Paper.pdf}
}

@inproceedings{sundararajan2017axiomatic,
  title={Axiomatic Attribution for Deep Networks},
  author={Sundararajan, Mukund and Taly, Ankur and Yan, Qiqi},
  booktitle={Proceedings of the 34th International Conference on Machine Learning},
  volume={70},
  pages={3319--3328},
  year={2017},
  organization={PMLR},
  url={https://proceedings.mlr.press/v70/sundararajan17a/sundararajan17a.pdf}
}

@inproceedings{zeiler2014visualizing,
  title={Visualizing and Understanding Convolutional Networks},
  author={Zeiler, Matthew D and Fergus, Rob},
  booktitle={European conference on computer vision},
  pages={818--833},
  year={2014},
  publisher={Springer},
  url={https://arxiv.org/abs/1311.2901}
}

@article{Cremades2024,
  author = {Cremades, Andr{\'e}s and Hoyas, Sergio and Deshpande, Rahul and Quintero, Pedro and Lellep, Martin and Lee, Will Junghoon and Monty, Jason P. and Hutchins, Nicholas and Linkmann, Moritz and Marusic, Ivan and Vinuesa, Ricardo},
  title = {Identifying regions of importance in wall-bounded turbulence through explainable deep learning},
  journal = {Nature Communications},
  year = {2024},
  volume = {15},
  number = {1},
  pages = {3864},
  doi = {10.1038/s41467-024-47954-6},
  url = {https://doi.org/10.1038/s41467-024-47954-6}
}

@article{sanchisagudo2023easy,
  title={Easy attention: A simple attention mechanism for temporal predictions with transformers},
  author={Sanchis-Agudo, Marcial and Wang, Yuning and Arnau, Roger and Guastoni, Luca and Lim, Jasmin and Duraisamy, Karthik and Vinuesa, Ricardo},
  journal={APL Computational Physics},
  volume={1},
  number={1},
  pages={016104},
  year={2025},
  publisher={AIP Publishing LLC}
}

@article{Sakamoto1986ArchtypeVF,
  title={Arch-type Vortex Formed Behind a Normal Plate Placed in Laminar Boundary Layer},
  author={Hiroshi Sakamoto and Hiroyuki Haniu},
  journal={Jsme International Journal Series B-fluids and Thermal Engineering},
  year={1986},
  volume={29},
  pages={2857-2862},
  url={https://api.semanticscholar.org/CorpusID:122014555}
}

@article{vinuesa2015direct,
  author    = {Vinuesa, Ricardo and Schlatter, Philipp and Malm, Johan and Mavriplis, Catherine and Henningson, Dan S.},
  title     = {Direct numerical simulation of the flow around a wall-mounted square cylinder under various inflow conditions},
  journal   = {Journal of Turbulence},
  year      = {2015},
  volume    = {16},
  number    = {6},
  pages     = {555--587},
  month     = {Mar},
  doi       = {10.1080/14685248.2014.989232},
  url       = {https://doi.org/10.1080/14685248.2014.989232},
  publisher = {Taylor \& Francis}
}

@article{Gery,
author = {Zampino, Gerardo and Atzori, Marco and Zea, Elias and Otero, Evelyn and Vinuesa, Ricardo},
year = {2025},
month = {03},
pages = {109672},
title = {Aspect-ratio effect on the wake of a wall-mounted square cylinder immersed in a turbulent boundary layer},
volume = {112},
journal = {International Journal of Heat and Fluid Flow},
doi = {10.1016/j.ijheatfluidflow.2024.109672}
}

@misc{beneitez2025improvingturbulencecontrolexplainable,
      title={Improving turbulence control through explainable deep learning}, 
      author={Miguel Beneitez and Andres Cremades and Luca Guastoni and Ricardo Vinuesa},
      year={2025},
      eprint={2504.02354},
      archivePrefix={arXiv},
      primaryClass={physics.flu-dyn},
      url={https://arxiv.org/abs/2504.02354}, 
}

@misc{vinuesa2025decodingcomplexitymachinelearning,
      title={Decoding complexity: how machine learning is redefining scientific discovery}, 
      author={Ricardo Vinuesa and Paola Cinnella and Jean Rabault and Hossein Azizpour and Stefan Bauer and Bingni W. Brunton and Arne Elofsson and Elias Jarlebring and Hedvig Kjellstrom and Stefano Markidis and David Marlevi and Javier Garcia-Martinez and Steven L. Brunton},
      year={2025},
      eprint={2405.04161},
      archivePrefix={arXiv},
      primaryClass={cs.LG},
      url={https://arxiv.org/abs/2405.04161}, 
}

@article{Schmid2010,
  title   = {Dynamic Mode Decomposition of Numerical and Experimental Data},
  author  = {Schmid, Peter J.},
  journal = {Journal of Fluid Mechanics},
  volume  = {656},
  pages   = {5--28},
  year    = {2010},
  doi     = {10.1017/S0022112010001217},
}

@article{Yousif2023,
  author  = {Yousif, Mustafa Z. and Yu, Linqi and Hoyas, Sergio and Vinuesa, Ricardo and Lim, HeeChang and et al.},
  title   = {A deep-learning approach for reconstructing 3D turbulent flows from 2D observation data},
  journal = {Scientific Reports},
  volume  = {13},
  pages   = {2529},
  year    = {2023},
  doi     = {10.1038/s41598-023-29525-9},
}

@article{Lorenz1963,
  author  = {Lorenz, Edward N.},
  title   = {Deterministic Nonperiodic Flow},
  journal = {Journal of the Atmospheric Sciences},
  volume  = {20},
  number  = {2},
  pages   = {130--141},
  year    = {1963},
  doi     = {10.1175/1520-0469(1963)020<0130:DNF>2.0.CO;2},
}

@incollection{Waleffe2009Exact,
  author       = {Waleffe, Fabian},
  title        = {Exact Coherent Structures in Turbulent Shear Flows},
  booktitle    = {Turbulence and Interactions},
  editor       = {Deville, Michel and L{\^e}, Thien-Hiep and Sagaut, Pierre},
  series       = {Notes on Numerical Fluid Mechanics and Multidisciplinary Design},
  volume       = {105},
  pages        = {139--158},
  publisher    = {Springer, Berlin, Heidelberg},
  year         = {2009},
  doi          = {10.1007/978-3-642-00262-5_7},
  isbn         = {978-3-642-00261-8 (print), 978-3-642-00262-5 (online)},
}

@article{WangZhou2009FiniteLength,
  author    = {Wang, H. F. and Zhou, Y.},
  title     = {The finite-length square cylinder near wake},
  journal   = {Journal of Fluid Mechanics},
  volume    = {638},
  pages     = {453--490},
  year      = {2009},
  doi       = {10.1017/S0022112009990693},
}

@article{Monnier2018BLM,
  author  = {Bruno Monnier and Sepehr A. Goudarzi and Ricardo Vinuesa and Candace Wark},
  title   = {Turbulent Structure of a Simplified Urban Fluid Flow Studied Through Stereoscopic Particle Image Velocimetry},
  journal = {Boundary-Layer Meteorology},
  year    = {2018},
  volume  = {166},
  number  = {2},
  pages   = {239--268},
  doi     = {10.1007/s10546-017-0303-9}
}

@article{Guastoni2025Perspective,
  title   = {A new perspective on the simulation of stochastic problems in fluid mechanics with diffusion models},
  author  = {Guastoni, Luca and Vinuesa, Ricardo},
  journal = {Nature Machine Intelligence},
  year    = {2025},
  volume  = {7},
  pages   = {816--817},
  month   = jun,
  doi     = {10.1038/s42256-025-01060-4},
  url     = {https://www.nature.com/articles/s42256-025-01060-4},
  note    = {News \& Views}
}

@article{Nekkanti2023JFM,
  author  = {Akhil Nekkanti and Sheel Nidhan and Oliver T. Schmidt and Sutanu Sarkar},
  title   = {Large-scale streaks in a turbulent bluff body wake},
  journal = {Journal of Fluid Mechanics},
  volume  = {974},
  pages   = {A47},
  year    = {2023},
  doi     = {10.1017/jfm.2023.783}
}

@article{Ghaemi2011JFM,
  author  = {Sina Ghaemi and Fulvio Scarano},
  title   = {Counter-hairpin vortices in the turbulent wake of a sharp trailing edge},
  journal = {Journal of Fluid Mechanics},
  volume  = {689},
  pages   = {317--356},
  year    = {2011},
  doi     = {10.1017/jfm.2011.431}
}

@article{Eiximeno_Sanchis-Agudo_Miro_Rodriguez_Vinuesa_Lehmkuhl_2025, title={On deep-learning-based closures for algebraic surrogate models of turbulent flows}, volume={1020}, DOI={10.1017/jfm.2025.10610}, journal={Journal of Fluid Mechanics}, author={Eiximeno, Benet and Sanchis-Agudo, Marcial and Mir{\'o}, Arnau and Rodriguez, Ivette and Vinuesa, Ricardo and Lehmkuhl, Oriol}, year={2025}, pages={A36}}
